\documentclass[aps,superscriptaddress,nofootinbib]{revtex4}

\usepackage{graphicx}
\usepackage{slashed}
\usepackage{amssymb,amsmath,amsbsy}

\usepackage{psfrag}
\usepackage{epsfig}

\def\be{\begin{equation}}
\def\ee{\end{equation}}

\def\m#1{m_{#1}}

\def\hpm{H^\pm}

\def\mhpm{\m{\hpm}}
\def\ltap{\;\centeron{\raise.35ex\hbox{$<$}}{\lower.65ex\hbox{$\sim$}}\;}
\def\gtap{\;\centeron{\raise.35ex\hbox{$>$}}{\lower.65ex\hbox{$\sim$}}\;}
\newcommand{\bea}{\begin{eqnarray}}
\newcommand{\eea}{\end{eqnarray}}

\begin{document}

\title{Wrong sign and symmetric limits and non-decoupling in 2HDMs}
\author{P.M.~Ferreira}
    \email[E-mail: ]{ferreira@cii.fc.ul.pt}
\affiliation{Instituto Superior de Engenharia de Lisboa - ISEL,
	1959-007 Lisboa, Portugal}
\affiliation{Centro de F\'{\i}sica Te\'{o}rica e Computacional,
    Faculdade de Ci\^{e}ncias,
    Universidade de Lisboa,
    Av.\ Prof.\ Gama Pinto 2,
    1649-003 Lisboa, Portugal}
\author{Renato Guedes}
    \email[E-mail: ]{renato@cii.fc.ul.pt}
\affiliation{Centro de F\'{\i}sica Te\'{o}rica e Computacional,
    Faculdade de Ci\^{e}ncias,
    Universidade de Lisboa,
    Av.\ Prof.\ Gama Pinto 2,
    1649-003 Lisboa, Portugal}
\author{Marco O. P. Sampaio,} 
    \email[E-mail: ]{msampaio@ua.pt}
\affiliation{Departamento de F\'\i sica da Universidade de Aveiro and I3N \\ 
Campus de Santiago, 3810-183 Aveiro, Portugal} 
\author{Rui Santos}
    \email[E-mail: ]{rsantos@cii.fc.ul.pt}
\affiliation{Instituto Superior de Engenharia de Lisboa - ISEL,
	1959-007 Lisboa, Portugal}
\affiliation{Centro de F\'{\i}sica Te\'{o}rica e Computacional,
    Faculdade de Ci\^{e}ncias,
    Universidade de Lisboa,
    Av.\ Prof.\ Gama Pinto 2,
    1649-003 Lisboa, Portugal}

\date{\today}

\begin{abstract}
We analyse the possibility that, in two Higgs doublet models, one or more of the Higgs couplings
to fermions or to gauge bosons change sign, relative to the respective Higgs Standard Model couplings. 
Possible sign changes in the coupling of a neutral scalar to charged ones are also discussed. These \textit{wrong signs} can
have important physical consequences, manifesting themselves in Higgs production via gluon fusion or Higgs decay into two gluons or into two photons. We consider all possible
wrong sign scenarios, and also the \textit{symmetric limit}, in all possible Yukawa implementations of the two Higgs doublet model, 
in two different possibilities: the observed Higgs boson is the lightest CP-even scalar,
or the heaviest one. We also analyse thoroughly the impact of the currently available LHC data on such scenarios. With all 8 TeV data analysed, all wrong sign scenarios are allowed in all Yukawa types, even at the 
1$\sigma$ level. However, we will show that B-physics constraints are crucial in excluding the possibility of wrong sign scenarios
in the case where $\tan \beta$ is below $1$. We will also discuss the future prospects for probing
the wrong sign scenarios at the next LHC run. Finally we will present a scenario where the alignment limit
could be excluded due to non-decoupling in the case where the heavy CP-even Higgs is the 
one discovered at the LHC. 
\end{abstract}

\maketitle

\section{Introduction}
\label{sec:intro}

The Large Hadron Collider (LHC) has confirmed the existence of a Higgs boson~\cite{ATLASHiggs, CMSHiggs} 
compatible with the one predicted by the Standard Model (SM). The Higgs couplings to fermions and gauge bosons are well within
the expected SM couplings. In addition, no extra scalar particles were found, leaving us with a theory that, at the present scale, is indeed very close to the SM. All extensions of the SM are therefore being pushed
to some kind of SM limit. Such is the case of the simplest extensions of the scalar sector like the ones obtained by
simply adding a complex singlet or a complex doublet to the SM field content, the latter designated by 
the two-Higgs doublet model (2HDM).

In a previous work~\cite{Ferreira:2014naa} we have discussed  the interesting possibility of a sign change in one of the Higgs Yukawa couplings.
There, we have defined the wrong sign scenario to be such that a sign change occurs in one of the Yukawa couplings relative to the Higgs coupling to $VV$ ($V=W^\pm$ or $Z$).
The LHC data analysed so far does not allow to differentiate between scenarios where a sign change in one of the Yukawa couplings occurs
 (see e.g.~Refs.~\cite{Espinosa:2012im, Falkowski:2013dza,Belanger:2013xza, Cheung:2014noa}). These studies were performed taking into account the
 measured properties of the SM-like Higgs.

In this work we will discuss all the possible sign changes in the Higgs couplings to fermions and to massive gauge bosons. 
The various wrong sign scenarios will all have in common the property that they are physically meaningful, that is, each of them can in principle be probed experimentally and distinguished from the limit where the model resembles the SM. In contrast to the wrong sign scenarios are the cases where all Higgs couplings to other SM particles change sign (while no significant difference occurs in the Higgs self-couplings). Since we will be interested in probing sign changes through loop induced vertices (which receive contributions from several couplings to SM particles and can only change if relative sign changes occur) the latter are not considered.

The study will be performed in the framework of the softly-broken $\mathbb{Z}_2$ symmetric and CP-conserving 
2HDM. The 2HDM is the simplest model that can provide wrong sign scenarios as defined above, since adding instead a (simpler) singlet field implies that the shift of the Higgs couplings to the other SM particles is the same for all such couplings. The 2HDM contains a decoupling limit and an alignment limit. In the exact decoupling limit~\cite{Gunion:2002zf}
the theory is the SM 
while in the alignment limit the SM-like Higgs boson couplings to the SM
particles are exactly the SM ones. However, the coupling structure of the 2HDM further allows 
for a change in the sign of the tree-level couplings to fermions and to massive gauge bosons. This sign change can affect 
both the $hgg$ and the $h \gamma \gamma$ effective couplings which are one-loop generated. We will examine two different 
wrong sign scenarios each associated with one of the two CP-even states of the 2HDM ($h$ or $H$) being identified 
with the scalar state that has already been found at the LHC (by convention $m_h < m_H$). In both cases there is an 
associated alignment limit where the tree-level Higgs couplings to the SM particles are equal to the SM ones.
Furthermore, each scenario also contains wrong sign limits, some of which are still compatible with current data. 
We will discuss in detail all the wrong sign scenarios - the ones that are already excluded or highly disfavoured, 
and those that can be probed at the upcoming runs of the LHC and at a future International Linear Collider (ILC). 

Finally we will discuss a very interesting feature  of the scenario where the heavy CP-even scalar
is identified with the SM Higgs. In fact, because there are two light states the theory does not decouple.
This non-decoupling nature of the heavy scenario will be discussed with the presentation of a situation 
where, although in the alignment limit, a given scenario could be excluded with a precise measurement of the signal rate $\mu_{\gamma \gamma}$.

We will adopt a twofold approach in our analysis. On one hand, we present the currently allowed parameter space regarding 
the wrong sign scenarios using all experimental data analysed so far. On the other hand, in order to make predictions related to a
 future increase in the precision of the measured rates, we will analyse the consequences of forcing such rates to 
be within 20, 10 or 5\% of the SM prediction. In doing this we will not separate the LHC production mechanisms
($gg\to h$, $b\bar{b}\to h$, Vector Boson Fusion (VBF), $Vh$ associated production and $t\bar{t}h$ associated
production); that is, we sum over all production mechanisms in computing the cross section.  

This paper is organized as follows. In Section~\ref{sec:model}, we describe the 2HDM and the constraints imposed
by theoretical and phenomenological considerations including the most recent LHC data. In Section~\ref{sec:ws}
we discuss the possible wrong sign limits for the 2HDM. In particular we will discuss the case where the
 heaviest CP-even scalar is the SM Higgs boson. In Section \ref{sec:res} we analyse in detail the different
wrong sign scenarios in view of present and future LHC data. In Section~\ref{sec:non-dec} we discuss the non-decoupling
nature of the heavy Higgs scenario. Finally in Section~\ref{sec:sym} we define and discuss the \textit{symmetric limit}
of 2HDMs. Our conclusions are presented in Section\ref{sec:conc}.


\section{The CP-conserving 2HDM}
\label{sec:model}

The two-Higgs double model (2HDM) is an extension of the SM where an extra complex scalar doublet is added to the field content of the SM while keeping its gauge symmetry.  It was first proposed by T.D. Lee~\cite{Lee:1973iz} as a means to explain the matter-antimatter asymmetry (see Refs.~\cite{hhg, Branco:2011iw} for a detailed description of the model).
With two doublet fields (denoted henceforth $\Phi_1$ and $\Phi_2$) the most general Yukawa Lagrangian gives rise to tree-level (Higgs-mediated)  
flavour-changing neutral currents (FCNC) which are severely constrained by experimental data. 

A natural way~\cite{GWP} of avoiding FCNCs is to impose an extra symmetry on the scalar potential.
We choose to impose a $\mathbb{Z}_2$ symmetry such that  the potential is invariant under 
$\Phi_1 \rightarrow \Phi_1$, $\Phi_2 \rightarrow - \Phi_2$ (i.e the doublets are $\mathbb{Z}_2$-even and $\mathbb{Z}_2$-odd respectively). The symmetry is extended to the Yukawa sector such that a fermion of a given charge couples only to one doublet. 
There are four possible independent coupling choices for the Yukawa Lagrangian~\cite{Barger:1989fj}.
In the literature two of the models have been named type I and type II and the other two have been changing
names over the years.  We shall call them type Flipped (F)  and type Lepton Specific (LS) (also called Y and X~\cite{Aoki:2009ha}, respectively). The different Yukawa types are built such that: only $\Phi_2$ couples to all fermions (type I); 
or $\Phi_2$ couples to up-type quarks and $\Phi_1$ couples to 
down-type quarks and leptons (type II); or $\Phi_2$ couples to up-type quarks and 
to leptons and $\Phi_1$ couples to down-type quarks (type F); or finally $\Phi_2$ couples to all 
quarks and $\Phi_1$ couples to leptons (type LS). 

The scalar potential in a softly broken $\mathbb{Z}_2$ symmetric 2HDM can be written as
\begin{align*}
V(\Phi_1,\Phi_2) =& m^2_1 \Phi^{\dagger}_1\Phi_1+m^2_2
\Phi^{\dagger}_2\Phi_2 - (m^2_{12} \Phi^{\dagger}_1\Phi_2+{\mathrm{h.c.}
}) +\frac{1}{2} \lambda_1 (\Phi^{\dagger}_1\Phi_1)^2 +\frac{1}{2}
\lambda_2 (\Phi^{\dagger}_2\Phi_2)^2\nonumber \\ 
+& \lambda_3
(\Phi^{\dagger}_1\Phi_1)(\Phi^{\dagger}_2\Phi_2) + \lambda_4
(\Phi^{\dagger}_1\Phi_2)(\Phi^{\dagger}_2\Phi_1) + \frac{1}{2}
\lambda_5[(\Phi^{\dagger}_1\Phi_2)^2+{\mathrm{h.c.}}] ~, \label{higgspot}
\end{align*}
where $\Phi_i$, $i=1,2$ are complex SU(2) doublets. We choose all parameters and the vacuum expectations values to be real.
This leads to an 8-parameter CP-conserving potential and we take as free parameters the four masses, the rotation
angle in the CP-even sector, $\alpha$, the ratio of the vacuum expectation
values,  $\tan\beta=v_2/v_1$, and the soft breaking parameter $m_{12}^2$.
Without loss of generality, we choose the conventions $0\leq\beta\leq \pi/2$ and $- \pi/2 \leq  \alpha \leq \pi/2$.

It is also instructive for our study to re-call how the two physical CP-even eigenstates,  $h$ and $H$, relate to the original field fluctuations (before diagonalisation) which determine the coupling to other SM particles. If we denote them by $h_i$ (for each $\Phi_i$ respectively), then in our convention
\begin{equation}
\left(\begin{array}{c}
h_{1}\\
h_{2}
\end{array}\right)=\left(\begin{array}{cc}
\cos\alpha & -\sin\alpha\\
\sin\alpha & \cos\alpha
\end{array}\right)\left(\begin{array}{c}
H\\
h
\end{array}\right)\; .
\end{equation} 
Now we can find a map between the couplings of the (already observed) Higgs in the scenario where it is $h$ (the lightest CP-even state), to the case where it is instead
$H$ (the heaviest CP-even state). Due to our convention for the range of $\alpha$ one has to be careful. One can check that the correct map (which also preserves our convention) is 
\begin{equation}
\label{eq:h2H}
\alpha \rightarrow \alpha -{\rm sign}(\alpha)\frac{\pi}{2} \; .
\end{equation}
Thus all expressions later obtained in the discussion of the wrong sign scenario can be transposed from the case where $h$ is the observed Higgs boson to the case where it is $H$ by using this map. Eq.~\eqref{eq:h2H} will later explain some sign flips in our results. Nevertheless, it is clear that the experimental constraints have different effects on the allowed parameter space for each scenario (light or heavy), because the various limits on new (yet to observe) scalars are not uniform in mass (thus the allowed parameter space of one scenario {\em cannot} be obtained by applying this map to the data points allowed in the other scenario).

\subsection{Theoretical and experimental constraints}

The constraints to impose on the CP-conserving 2HDM models originate from two sources (for a recent review see~Ref.~\cite{Arhrib:2013oia}): i) consistency with theoretical principles/conditions and ii)  consistency with experimental data. 
Regarding the theoretical constraints it is well known, at tree level, that once a CP-conserving minimum of the potential is chosen,
no additional minima that spontaneously break the electric charge and/or CP symmetry
exist~\cite{vacstab}. Furthermore we demand that the CP-conserving minimum is the global one~\cite{Barroso:2013awa}, that the potential is bounded from below~\cite{vac1} and that tree-level unitarity~\cite{unitarity} is obeyed. 

Regarding the consistency with experimental data we impose various conditions. We require the model to satisfy electroweak precision constraints~\cite{Peskin:1991sw,STHiggs,lepewwg,gfitter1}, i.e. that the $S,T,U$ variables~\cite{Peskin:1991sw} predicted by the model are within the $95\%$ ellipsoid centred on the best fit point to the electroweak data. There are also indirect constrains originating from loop processes that involve charged Higgs bosons, which depend on $\tan\beta$ through the charged Higgs coupling to fermions.  They originate
mainly from  $B$ physics observables~\cite{BB, Deschamps:2009rh} and from the $R_b\equiv\Gamma(Z\to b\bar{b})/\Gamma(Z\to{\rm hadrons})$~\cite{Ztobb} measurement. 
They give rise to the best bound on the charged Higgs mass in a type II model which yields $\mhpm \gtrsim 340$ GeV almost independently of $\tan \beta$.

LEP searches based on $e^+ e^- \to H^+ H^-$~\cite{Abbiendi:2013hk} and recent LHC results~\cite{ATLASICHEP, CMSICHEP} based on
$pp \to \bar t \, t (\to H^+ \bar b ) $ constrain the mass of the charged Higgs to be above $O(100)$ GeV, depending on the model type.
Finally, we should note that there is a 3.4~$\sigma$ discrepancy
between the  value of $\overline{B}\to D^{(*)}\tau^-\overline{\nu}_\tau$  measured by
the {B\sc{a}B\sc{ar}} collaboration~\cite{Lees:2012xj} and the corresponding SM prediction. 
If confirmed, this observation would exclude both the SM and the versions of the 2HDM considered in this work.

So far we have described mostly pre-LHC bounds. 
The parameter space of the 2HDM is already very constrained by the LHC results~\cite{manyrefs}.
We will now briefly re-analyse these results to find the parameter space still allowed after the 8~TeV run. We have used the~\textsc{ScannerS}~\cite{Coimbra:2013qq} program interfaced with \textsc{SusHi}~\cite{Harlander:2012pb} for the 
$pp (gg+bb) \to h$  production process at NNLO and \textsc{Hdecay}~\cite{Djouadi:1997yw, Harlander:2013qxa} for all 2HDM decays. 
The numbers were cross-checked with \textsc{HIGLU}~\cite{Spira:1995mt} and \textsc{2HDMC}~\cite{Eriksson:2009ws}.
The remaining Higgs production cross sections, VBF, associated production (with a $Z$ or $W$) and $t \bar t h$ were taken from~\cite{LHCHiggs} at NLO. 
SM electroweak corrections were not considered in any production process because the 2HDM electroweak corrections can be significantly different. All $95\%$ C.L. exclusion limits, obtained experimentally from the non-observation of new scalars in experimental searches at colliders,  were applied using \textsc{HiggsBounds}~\cite{Bechtle:2013wla}. Consistency with the observed signals of the Higgs boson at the LHC was tested with~\textsc{HiggsSignals}~\cite{Bechtle:2013xfa}, which computes a probability for the model point to fit all known signal data\footnote{Later we will show results for points which are consistent within a $3\sigma$, $2\sigma$ or $1\sigma$ probability, for example.}. The theoretical constraints associated with vacuum stability and tree level unitarity are inbuilt in the~\textsc{ScannerS} code for any model, whereas specific functions were developed for the 2HDM to test electroweak precision observables and B-physics observables (all constrained to be within $95\%$ of the best fit values as discussed above).

We will use the standard definition of signal strength
\begin{equation}
\mu^h_f \, = \, \frac{\sigma \, {\rm BR} (h \to
  f)}{\sigma^{\scriptscriptstyle {\rm SM}} \, {\rm BR^{\scriptscriptstyle{\rm SM}}} (h \to f)}
\label{eg-rg}
\end{equation}
where $\sigma$ is the Higgs production cross section and ${\rm BR} (h \to f)$ is
the branching ratio of the decay into some given final state $f$;  $\sigma^{\scriptscriptstyle {\rm {SM}}}$
and ${\rm BR^{\scriptscriptstyle {\rm SM}}}(h \to f)$ are the expected values for the same quantities in the SM. 
In the following sections we will also make predictions for the next LHC run at 13 TeV. In these predictions
we will not use the present LHC data (but will use all other constraints) but instead we will ask that the rates $\mu^h_f$
for the final states $f=WW$, $ZZ$, $\gamma \gamma$ and $\tau^+ \tau^ -$ to be within $20$, $10$ or $5$ \% of
the SM predictions.

We also define
\begin{equation}
\kappa_i^2=\frac{\Gamma^{\scriptscriptstyle {\rm 2HDM}}  (h \to i)}{\Gamma^{\scriptscriptstyle {\rm SM}} (h \to i)}
\end{equation}
which at tree-level is just the ratio of the couplings $\kappa_i=g_{i}^{\scriptscriptstyle {\rm 2HDM}}  /g_{i}^{\scriptscriptstyle {\rm SM}} $.
Taking the  $h W^+ W^-$ coupling as an example, we write
\begin{equation}
\kappa_{W}^2= \frac{\Gamma^{\scriptscriptstyle {\rm 2HDM}}  (h \to W^+ W^-)}{\Gamma^{\scriptscriptstyle {\rm SM}}  (h \to W^+ W^-)}=
\left( \frac{g_{\scriptscriptstyle  h W^+ W^-}^{\scriptscriptstyle {\rm 2HDM}} }{g_{ \scriptscriptstyle h W^+ W^-}^{\scriptscriptstyle {\rm SM}}} \right)^2=\sin^2 (\beta - \alpha)
\end{equation}
and the last equality only holds for Leading Order (LO) widths. Obviously, because the decays $h \to \gamma \gamma$
and $h \to gg$ are one-loop processes at LO, $\kappa_\gamma$ or $\kappa_g$ can only be calculated by the ratio of the 2HDM width
to the respective SM width. Unless otherwise stated, the theoretical values of $\kappa_F$ (where $F$ is a fermion) and $\kappa_V$
(where $V$ is a massive vector boson) refer to LO widths. Note that while $\kappa_F$ and $\kappa_V$ can be either positive or negative,
$\kappa_\gamma$ and $\kappa_g$ are strictly positive.
These definitions for the couplings $\kappa$ coincide with the definitions used by the experimental
groups at the LHC~\cite{rec}, at leading order.  We shall also make the simplifying assumption (which holds in the 
SM and in the 2HDM under consideration) that all down-type [up-type] fermion final states are governed by the same $\kappa_D$ [$\kappa_U$].

\subsection{Allowed parameter space after the 8 TeV LHC and the wrong sign limit}

In this section we discuss some important features of the allowed parameter space of the models, after imposing all theoretical and experimental constraints mentioned above. Unless stated otherwise, we have set one of the CP-even eigenstates to a mass of 125.9~GeV and left all other masses free to run over an interval. We will refer to the case where $m_h=125.9$~GeV as the light Higgs scenario, and to the case where $m_H=125.9$~GeV as the heavy Higgs scenario. For all scans, before applying the constraints, we allow $0.1<\tan\beta<50$, $|\alpha|<\pi/2$ and\footnote{For the scans we present it turns out that the combination of the global minimum condition with the other constraints implies $m_{12}^2>0$ in practice.}   
$- (900~{\rm GeV})^2<m_{12}^2 <(900~{\rm GeV})^2$. All (eigenstate) masses are free in the range [50,1000]~GeV, but we have also imposed that the masses of the other neutral scalars are away from the Higgs mass $125.9$~GeV by more than 5~GeV. These conditions apply to all scans. Any other extra condition (such as lower or upper bounds imposed on masses) will be specified for each scan.

\begin{figure}
\centering
\mbox{\includegraphics[width=0.34\linewidth,clip=true]{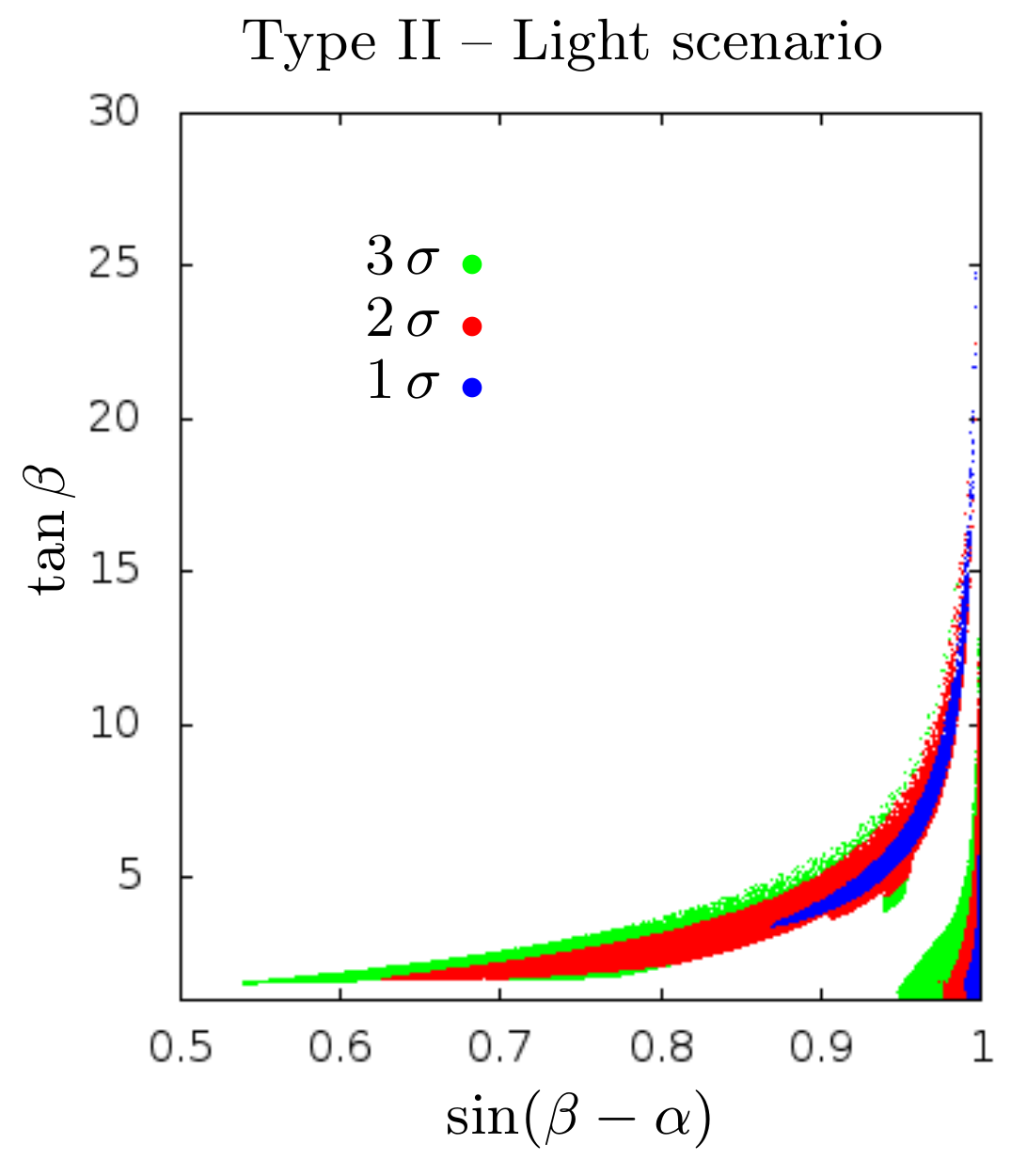}\hspace{0.017\linewidth}\includegraphics[width=0.313\linewidth,clip=true,trim= 25 0 0 0]{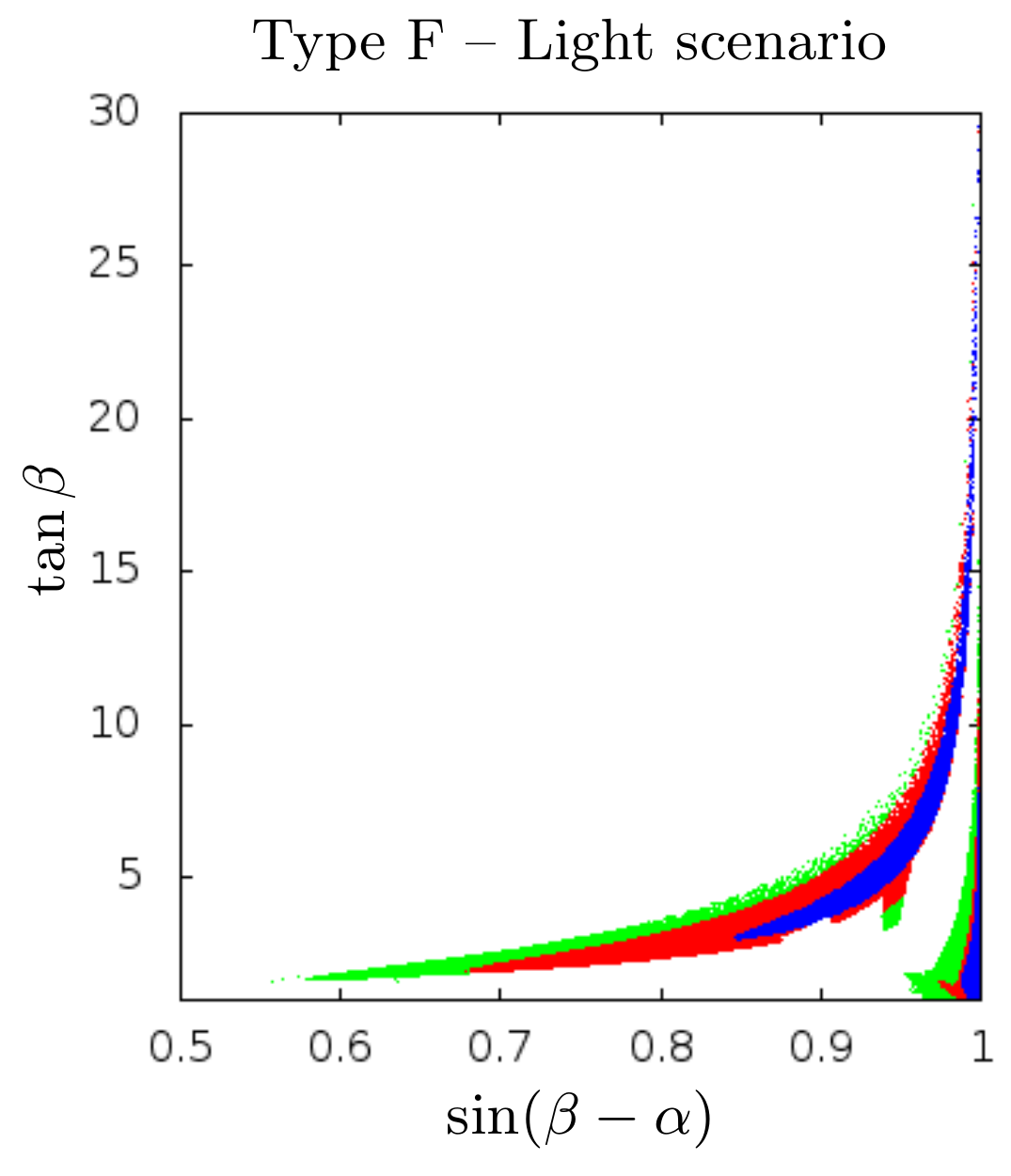}\hspace{0.017\linewidth}\includegraphics[width=0.313\linewidth,clip=true,trim= 25 0 0 0]{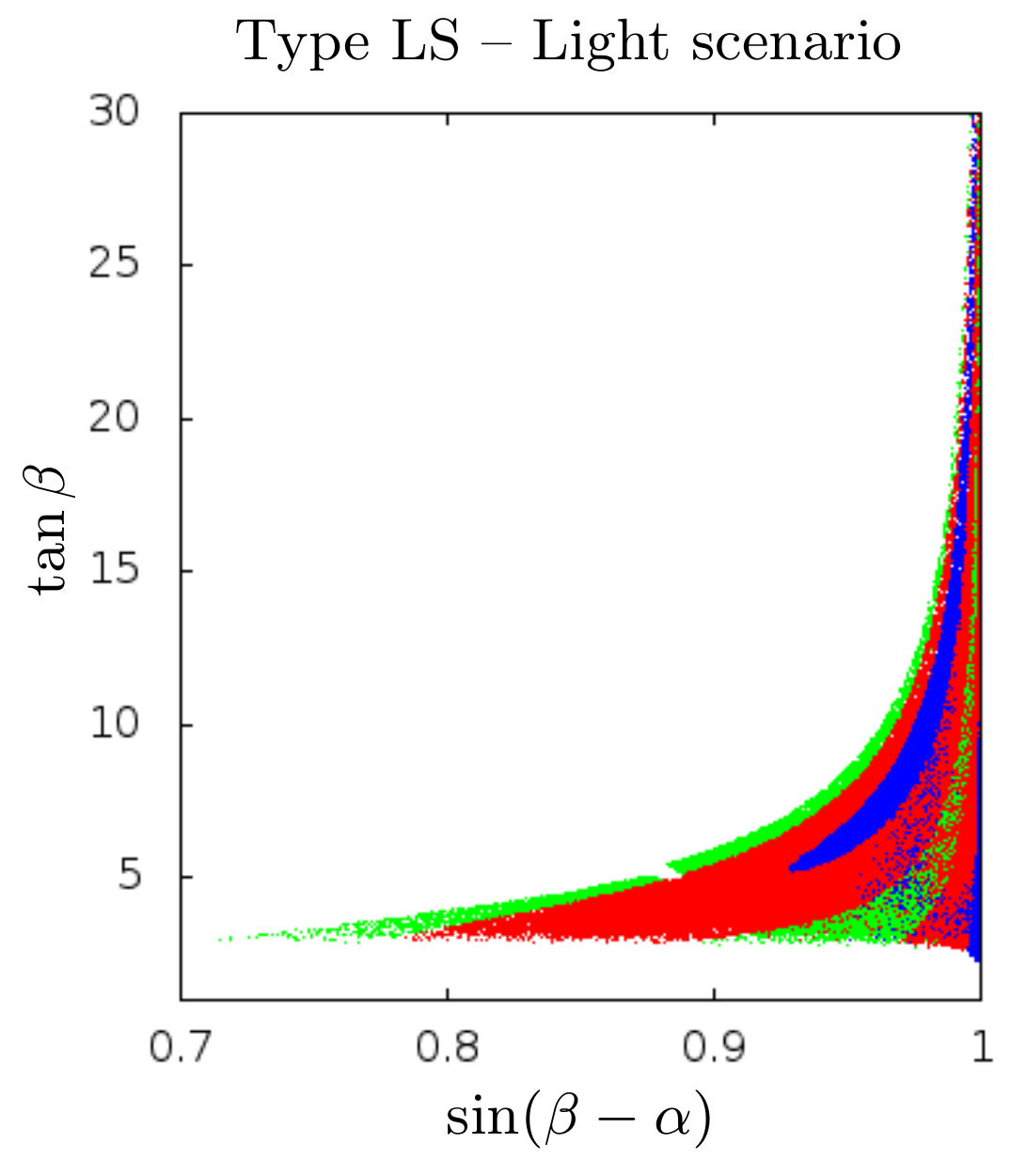}} \vspace{1mm}\\
\mbox{\includegraphics[width=0.34\linewidth,clip=true]{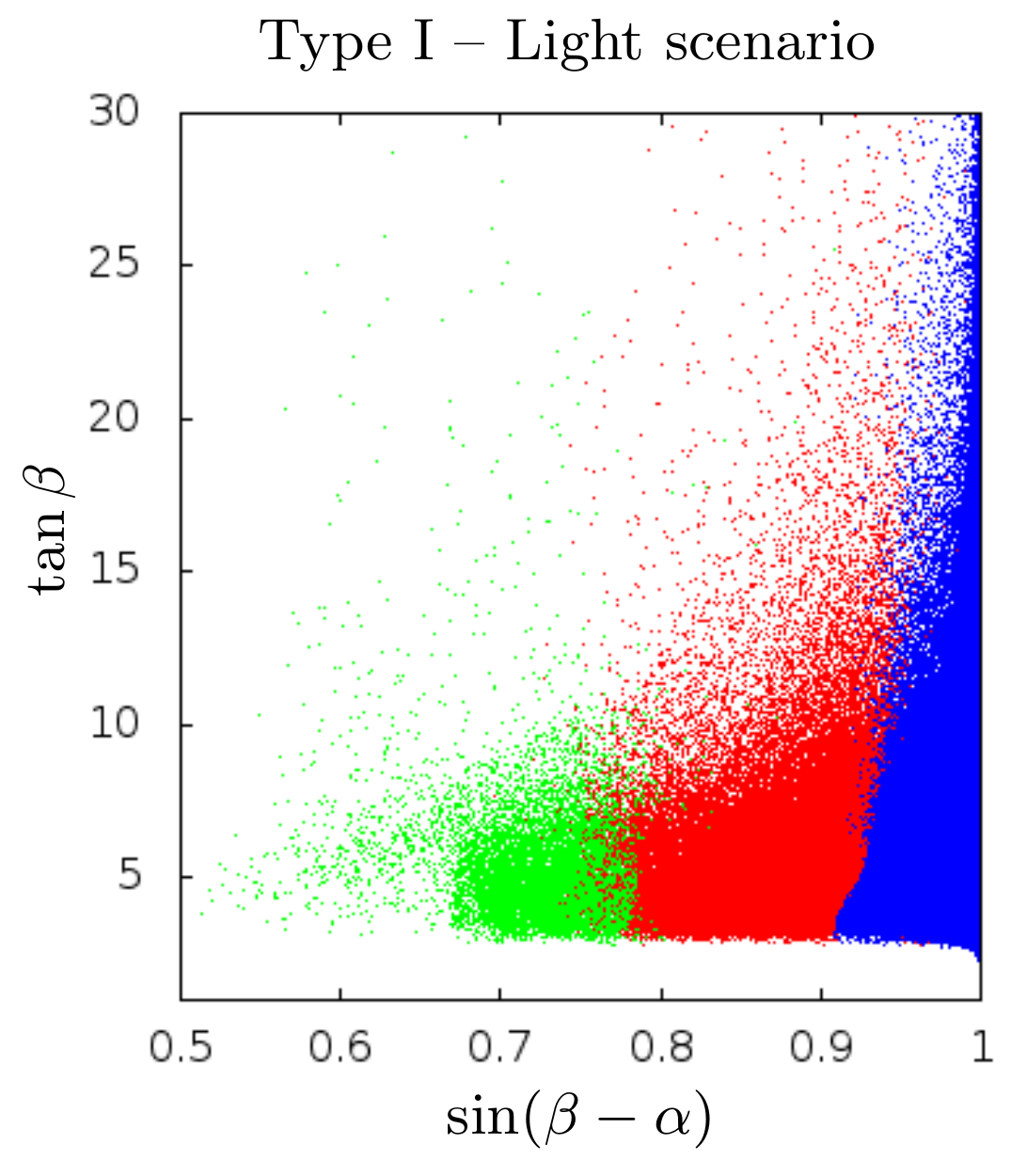}\hspace{0.017\linewidth}\includegraphics[width=0.313\linewidth,clip=true,trim= 25 0 0 0]{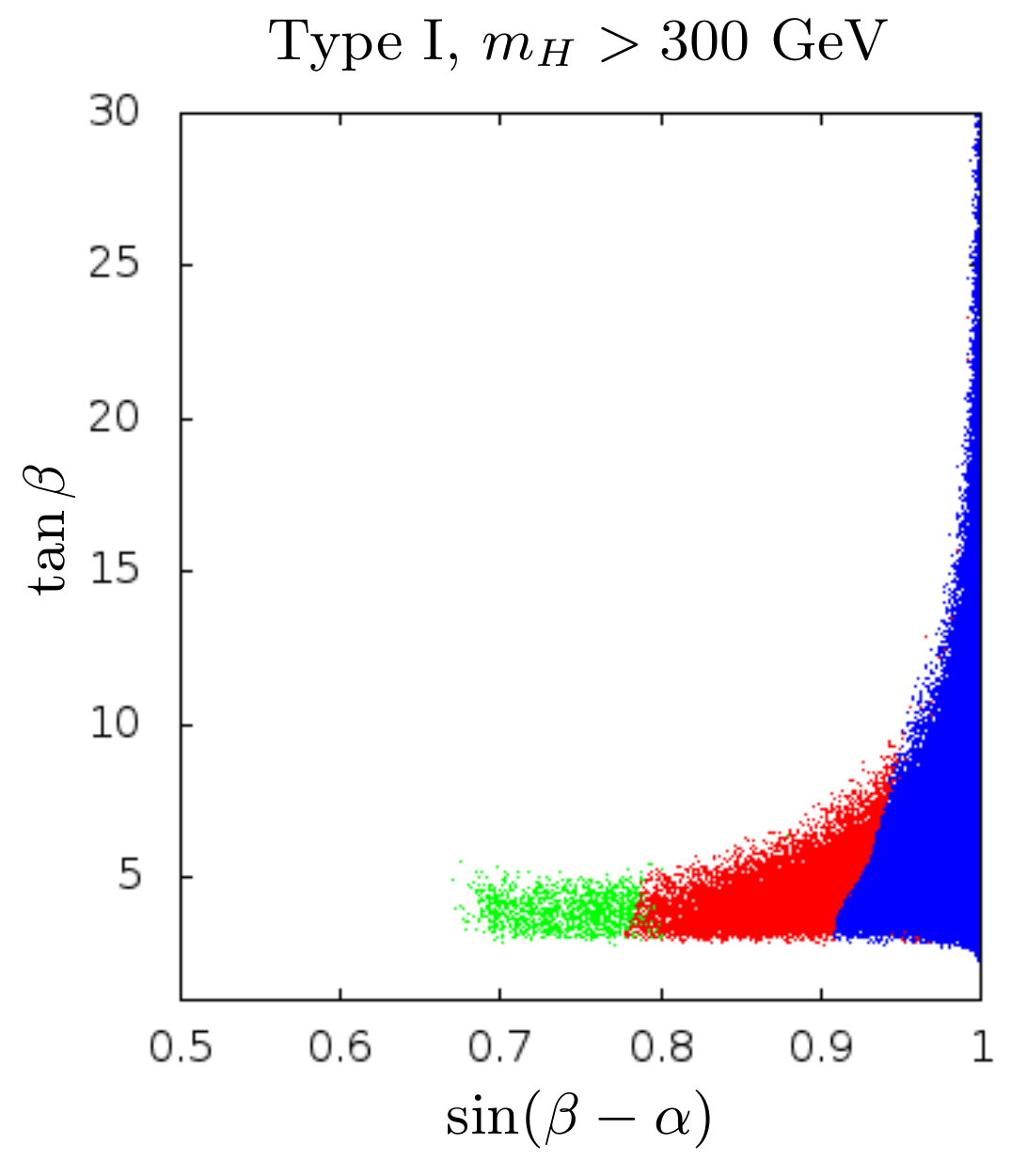}\hspace{0.017\linewidth}\includegraphics[width=0.313\linewidth,clip=true,trim= 25 0 0 0]{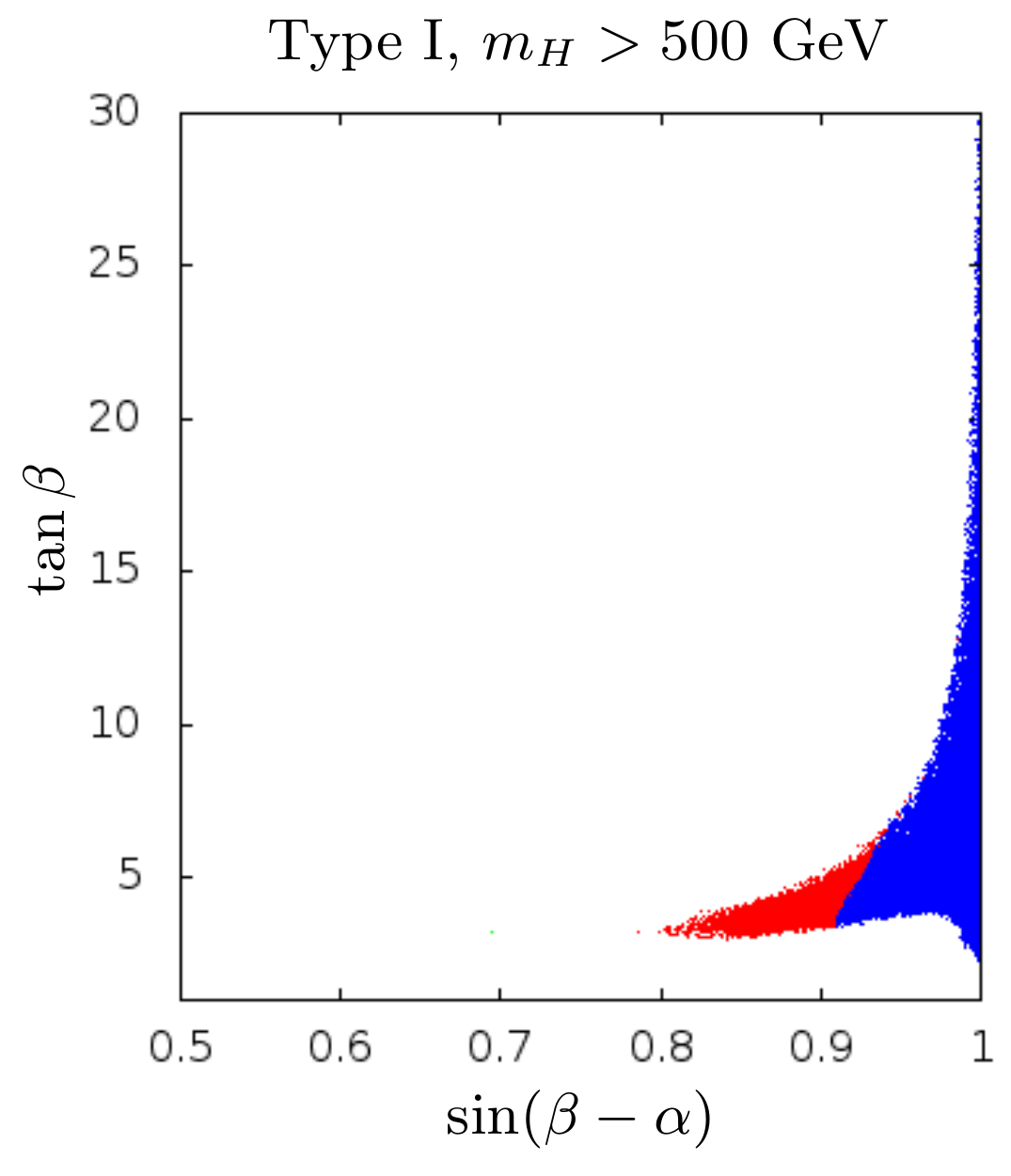}}
\caption{{\em Light Higgs scenario}: Allowed parameter space in the $\tan\beta$ vs $\sin(\beta-\alpha)$ plane after the LHC 8~TeV run for the Light Higgs scenario for each model type. Points have been accepted according to their p-value being within a number of standard deviations as show in the key (see top left panel).  We have imposed that $m_A>m_h+5$~GeV.   
}
\label{fig:LHC8_1}
\end{figure}
Let us start by discussing the light Higgs scenario where $m_h=125.9$~GeV. In figure~\ref{fig:LHC8_1} we present the allowed parameter space projected on the $\tan\beta$ \textit{vs} $\sin(\beta-\alpha)$ plane with all constraints applied. We have also imposed that the CP-odd scalar mass obeys $m_A> m_h+5~{\rm GeV}$.
We have accepted points that explain the observed Higgs signals with a fit probability within $3\sigma$, and also represent on top of these the points that survive at $2\sigma$ and $1\sigma$. 
 For the top left and middle panels (Types II and F), the allowed parameter space is centred around two lines.
 One is $\sin (\beta - \alpha) = 1$, the alignment limit where all Higgs couplings to fermions and massive gauge bosons are exactly the SM ones. The other one is $\sin (\beta + \alpha) = 1$ which we have
 called the \textit{wrong sign limit}\footnote{Note that in our convention the band of points associated with this line appears for $\alpha>0$.} in~\cite{Ferreira:2014naa}, to be discussed in detail
 in the next section. The two plots are very similar, both in the allowed range for $\sin (\beta - \alpha) $
 and in that large values of $\tan \beta$ are excluded except for $\sin (\beta - \alpha) $ very close
 to $1$. In order to understand the shape of the curves let us consider the approximation where the production occurs only via $gg$ while the 
 the total width is dominated by $h \to b \bar b$. As shown in~\cite{Ferreira:2012nv} , this approximation can be written
\begin{equation}
\mu_{VV} \approx \frac{\sin^2 (\beta - \alpha) }{\tan^2\beta \, \tan^2 \alpha}
\label{mutII}
\end{equation}
and by imposing $0.8< \mu_{VV} <1.2$ one reproduces figure~\ref{fig:LHC8_1} for types II and F with remarkable accuracy, as shown in~\cite{Fontes:2014tga}.  
Hence, the bounds on $\mu_{VV}$ alone, can explain not only the shape but also the numerical values presented in 
the plots in figure~\ref{fig:LHC8_1} for types II and F.  
Furthermore,  the b-loop contribution in $gg \to h$ and $bb \to h$ grows with
\begin{equation}
\frac{\sin^2\alpha}{\cos^2\beta}=(\sin (\beta - \alpha) -\cos (\beta - \alpha) \tan\beta)^2,
\label{hddhuuIIFS}
\end{equation}
which is exactly $\tan^2\beta$ when $\sin (\beta - \alpha)=0$ but even for, say $\sin (\beta - \alpha)=0.8$,
taking $\tan \beta = 10$ we get an enhancement factor of $27$ relative to the respective SM contribution.
As $\sin (\beta - \alpha)$ approaches $1$, the 2HDM lightest Higgs branching ratios (BRs) to SM particles do not differ much
from the values of the respective SM Higgs decays. Therefore, the inclusion of the b-loops would just
confirm the exclusion of the high $\tan \beta$ region except close to the alignment limit. 

In the top right and bottom panels of figure~\ref{fig:LHC8_1} we show the allowed parameter space for type LS and type I. Let us focus first on the top right and bottom left panels (for which no extra cut is present). We start by observing that there is
no $\tan \beta$ enhancement in the Higgs production cross section. In fact, the Higgs couplings to both up-type and 
down-type quarks are the same and the SM cross section for $gg+bb \to h$ is just multiplied by the factor 
\begin{equation}
\frac{\cos^2\alpha}{\sin^2\beta}=(\sin (\beta - \alpha) + \cos (\beta - \alpha) \cot \beta)^2,
\label{hddhuuILS}
\end{equation}
that could only be large for $\tan \beta \ll 1$, which is forbidden by B-physics constraints. For type I, considering the limit where the production occurs only via $gg$ while the total width is dominated by $h \to b \bar b$ (similarly to type II and F), we obtain
\begin{equation}
\mu_{VV} \approx \sin^2 (\beta - \alpha) \, .
\label{mutI}
\end{equation}
We conclude that the result for type I is a bound on $\sin (\beta - \alpha)$ which is  almost independent of $\tan \beta$. In fact, except for the Higgs self-couplings, the type I 2HDM is similar to the model obtained by adding a singlet
to the SM, because if $\tan \beta \gg 1$ (using equation \ref{hddhuuILS})
\begin{equation}
\kappa_F \approx \kappa_V = \sin (\beta - \alpha) \, .
\label{mutIsinglet}
\end{equation}
Hence, only constraints related to the shape of the potential (such as the ones arising from vacuum stability and perturbative unitarity) can introduce some $\tan\beta$ dependence.

In the case of type LS, a similar approximation needs to take into account both $\Gamma (h \to b \bar b)$
and $\Gamma (h \to \tau^+ \tau^-)$. In fact, if we take say $\sin (\beta - \alpha) = 0.8$ the two widths
are equal for $\tan \beta \approx 6$. The value of $\tan \beta$ for which the widths cross grows
with $\sin (\beta - \alpha) $ and above the crossing value $h \to \tau^+ \tau^-$ dominates. Therefore,
depending on the values of $\sin (\beta - \alpha) $ and $\tan \beta$ we either have an approximate expression
for $\mu_{VV}$ that is closer to type II (when $h \to \tau^+ \tau^-$ dominates - equation~\eqref{mutII}) or to type I 
 (when $h \to b \bar b$ dominates - equation~\eqref{mutI}). We can also write an approximate expression for
 $\mu_{VV}$ by considering as dominant the sum of the two widths $\Gamma (h \to b \bar b) + \Gamma (h \to \tau^+ \tau^-)$,
\begin{equation}
\mu_{VV} \approx \frac{10 \,(m_b^2/m_\tau^2) \,  \sin^2 (\beta - \alpha) }{9 \,(m_b^2/m_\tau^2) +\tan^2\beta \, \tan^2 \alpha} \; .
\label{mutLS}
\end{equation}
Finally, also the measurement of $pp \to h \to \tau^ + \tau^ -$ affects considerably more
the parameter space of type LS than that of type I~\cite{Arhrib:2011wc}. As this decay becomes more
important with growing $\tan \beta$ the exclusion region increases in the large $\tan \beta$ region.

\begin{figure}
\centering
\hspace{0.27\linewidth}\includegraphics[width=0.65\linewidth,clip=true]{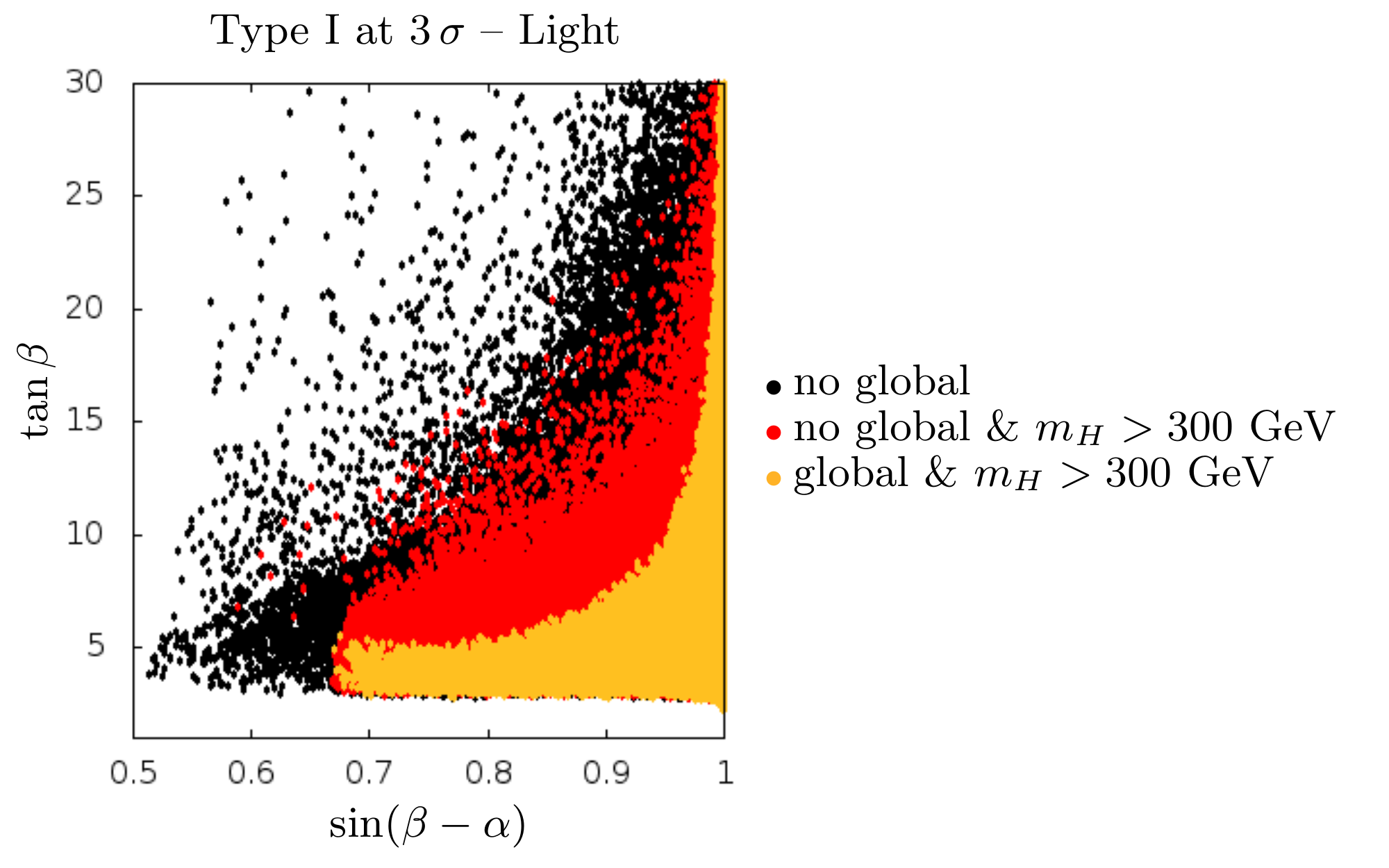}
\caption{{\em Light Higgs scenario}: Combined effect of cutting points for which the minimum is not the global one, and cutting the mass of the Heavy Higgs. For the black points the scan is as before except that the global minimum condition was lifted. For the other layers the mass cut and the global condition are re-introduced in turn as indicated in the key.
}
\label{fig:LHC8_global}
\end{figure}

In the bottom panels for type I, we also present (middle and right) the effect of placing a cut on the heavy Higgs mass, $m_H$. 
It is quite remarkable the effect that this cut (combined with the constraints) has on the allowed parameter space for type I (which is otherwise almost independent of $\tan \beta$). This behaviour is mainly related to the theoretical constraints imposed on the
2HDM including the discriminant that forces the model to be in the global minimum~\cite{Barroso:2013awa} at tree-level. 
To see the effect of the latter we show in figure~\ref{fig:LHC8_global}:  in black, points for which the global minimum conditions was lifted; 
in red, the subset of the black points that survive the mass cut $m_H>300$~GeV at $3\sigma$; and in yellow, the subset which survives the mass cut and the global minimum condition.
One should note that the global minimum condition does not play in general a major role in constraining the parameter space. Indeed this condition 
does not change the allowed regions for the other models, once the LHC constraints are imposed. Moreover, as discussed
in~\cite{Barroso:2013awa}, the theory can still be viable in a local minimum provided that the tunnelling time to the global one
is larger than the age of the universe.

\begin{figure}
\centering
\mbox{\includegraphics[width=0.45\linewidth,clip=true]{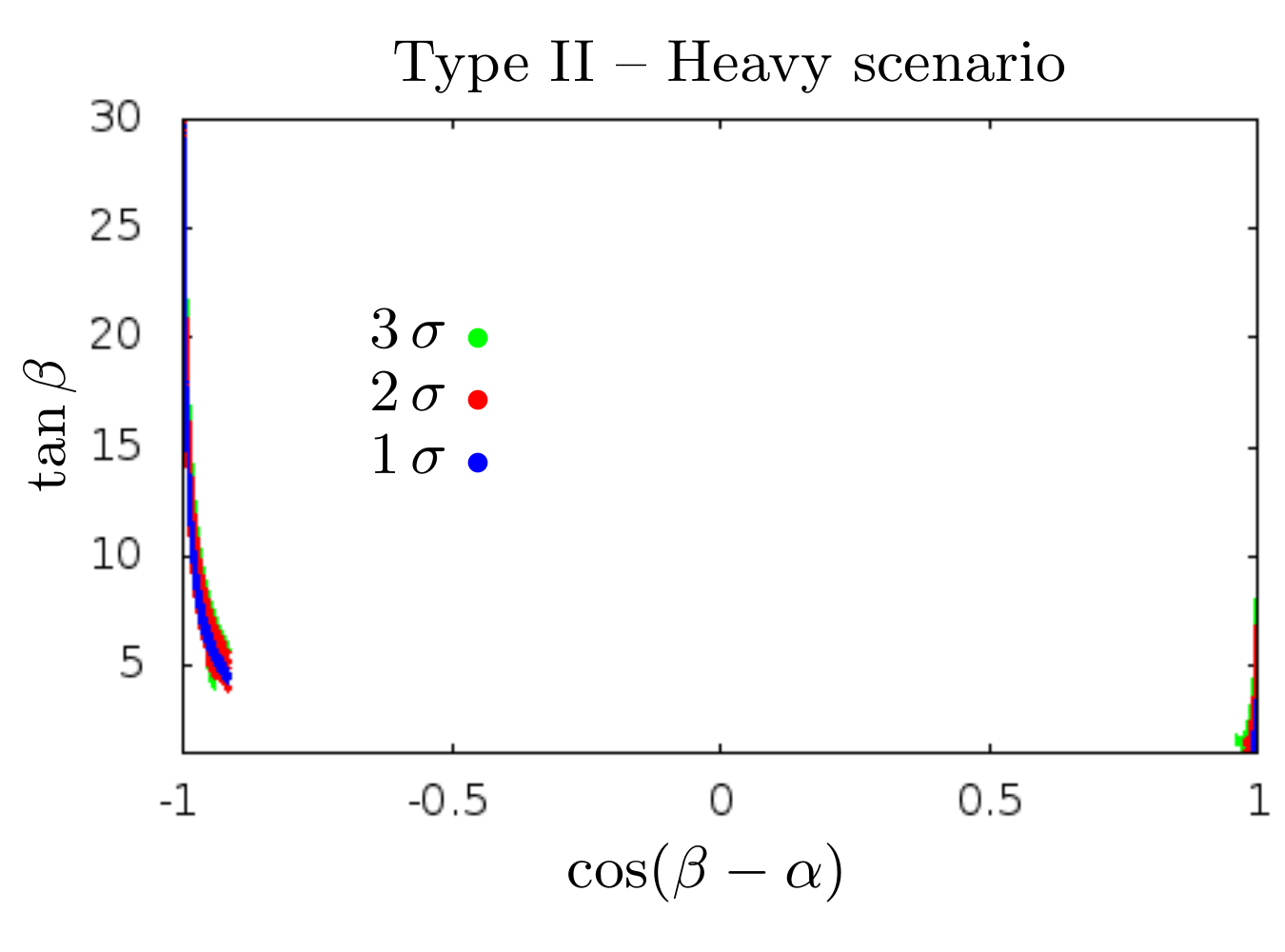}\hspace{6mm}\includegraphics[width=0.45\linewidth,clip=true]{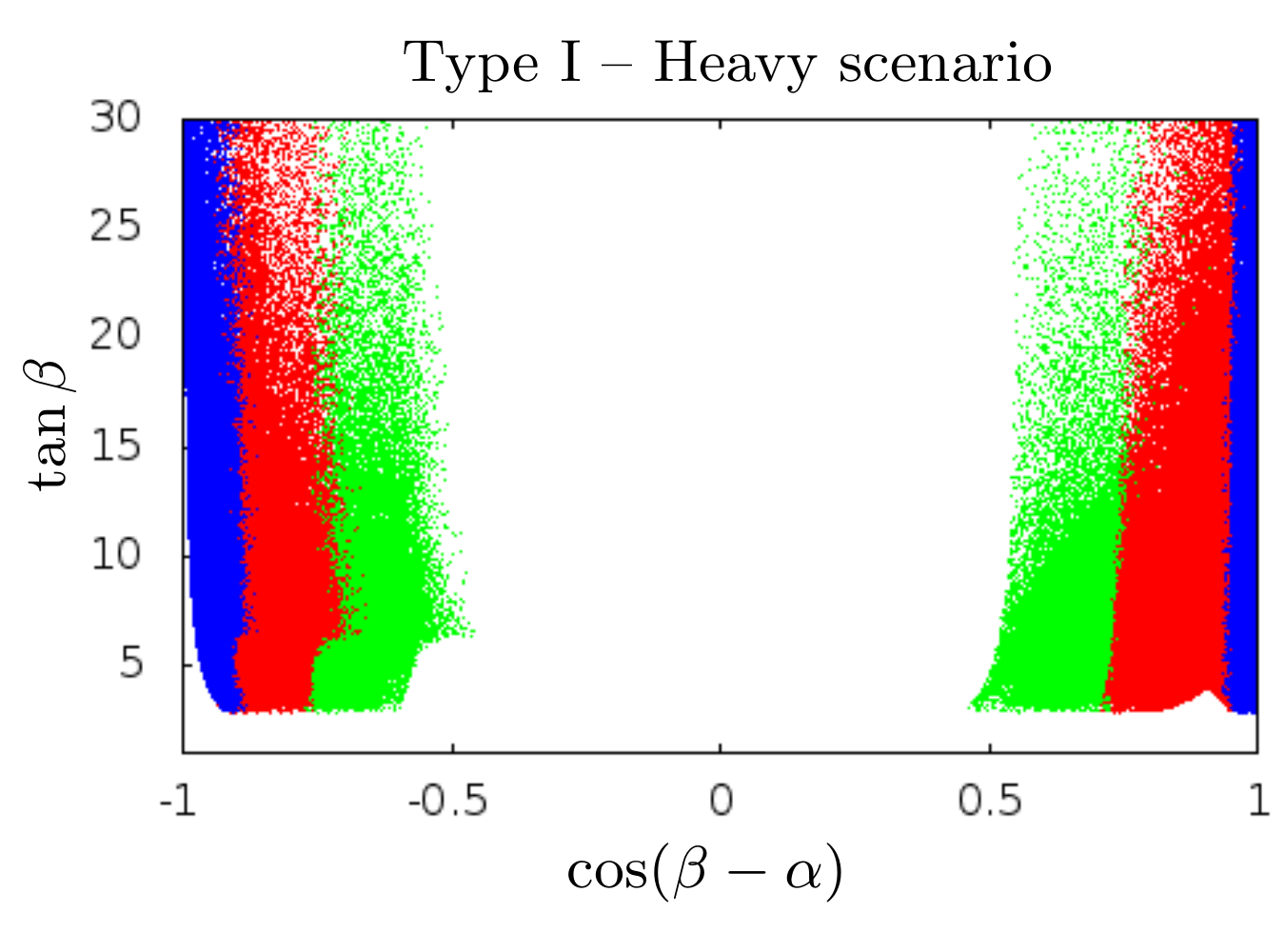}}
\mbox{\includegraphics[width=0.45\linewidth,clip=true]{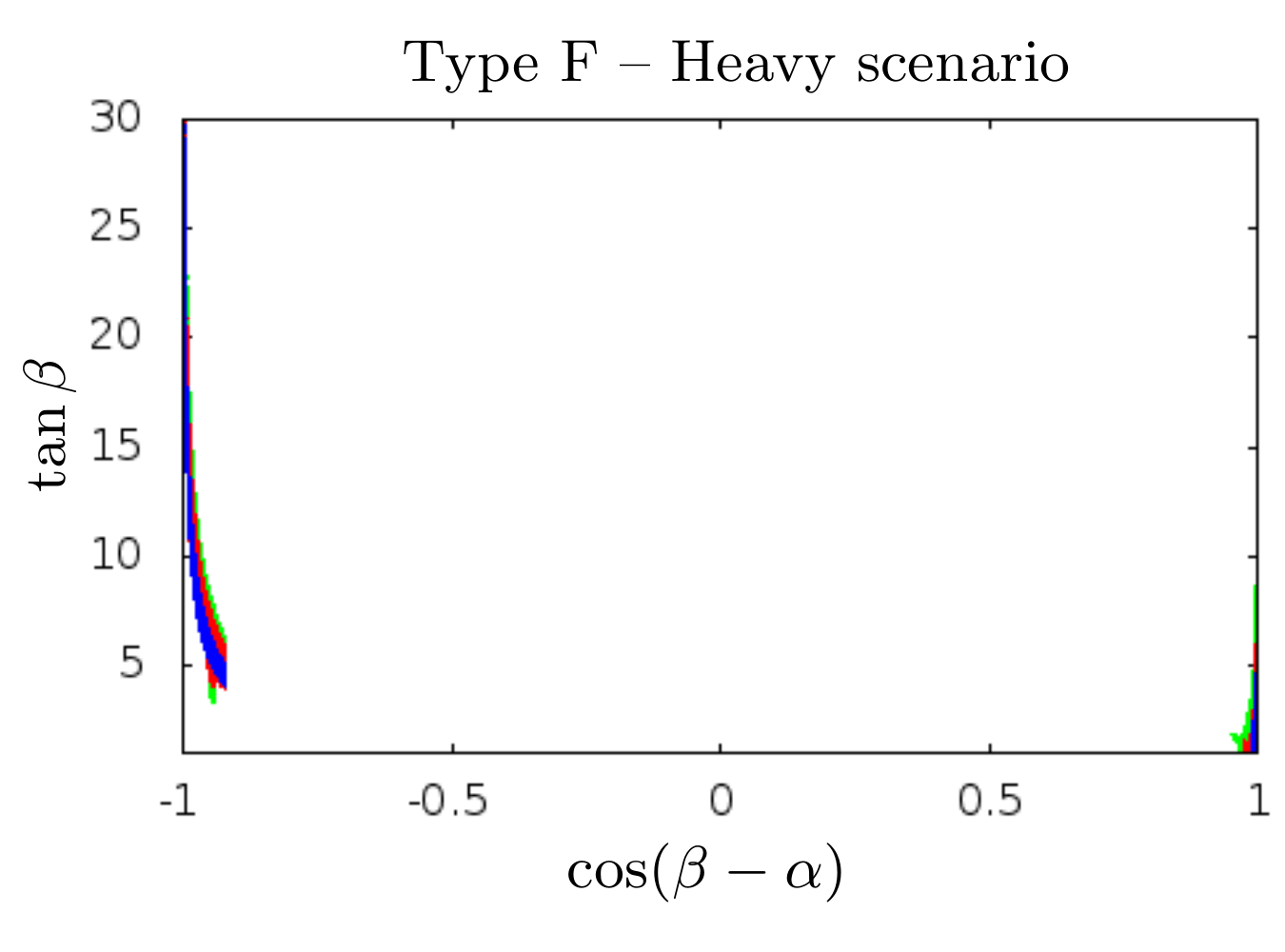}\hspace{6mm}\includegraphics[width=0.45\linewidth,clip=true]{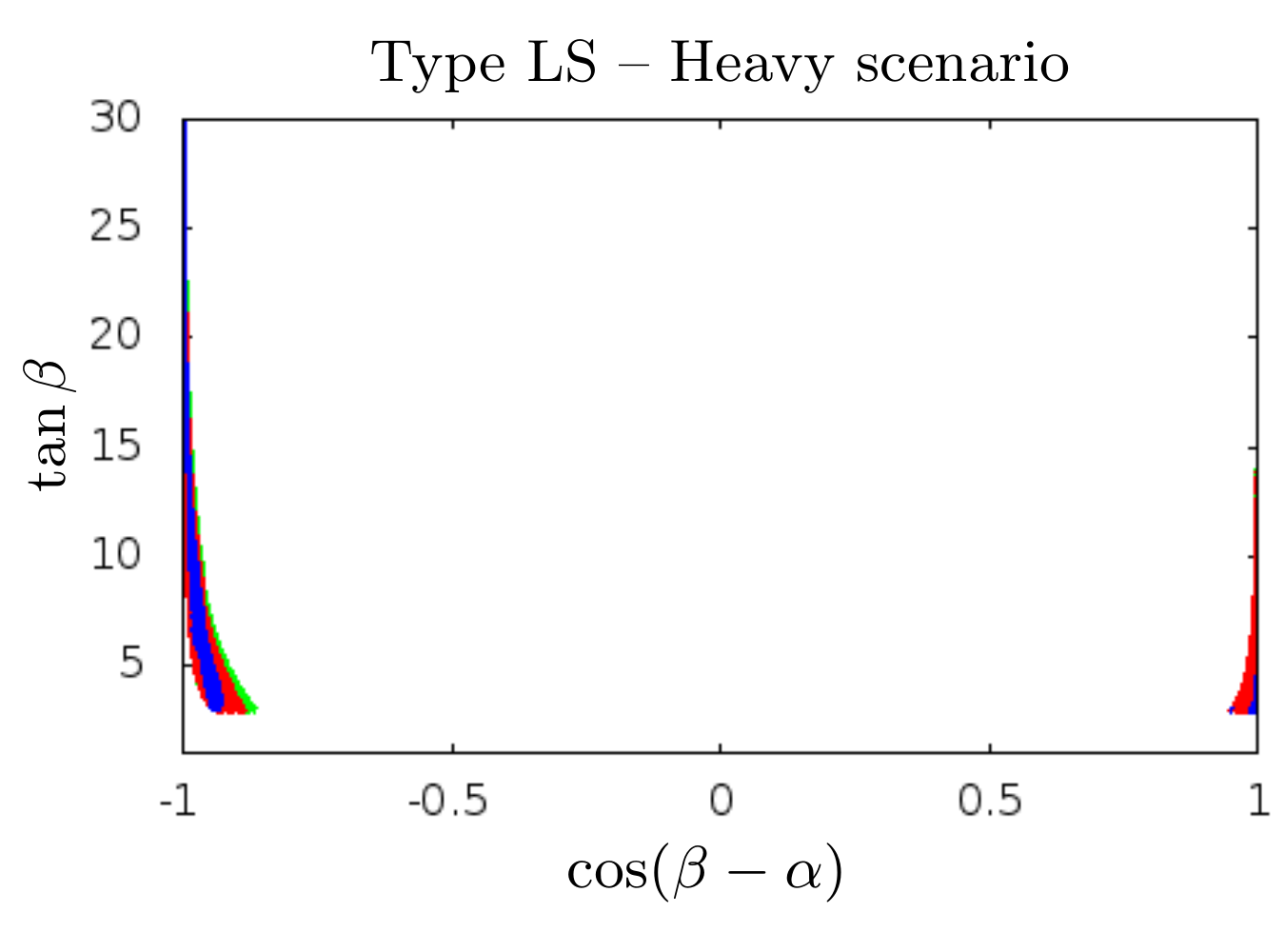}}
\caption{{\em Heavy Higgs scenario}: Allowed parameter space in the $\tan\beta$ vs $\cos(\beta-\alpha)$ plane after the LHC 8~TeV run for the Heavy Higgs scenario for each model type. Points have been accepted according to their p-value being within a number of standard deviations as show in the key (see top right panel).  
}
\label{fig:LHC8_2}
\end{figure}
The second scenario we consider is the heavy Higgs scenario of figure~\ref{fig:LHC8_2}, where $m_H = 125.9~{\rm GeV}$. In this case we have also imposed the lightest CP-even scalar mass, $m_h$ to be varied in the range $70$ to $120~{\rm GeV}$. This somewhat short range for $m_h$ was chosen mainly because we want to disallow the decay $H \to hh$. If allowed it would be the main decay channel and the model would have to be very fine-tuned
(taking $g_{Hhh} \approx 0$)~\cite{heavyh} for $H$ to still be the SM-like Higgs found at the LHC.
This is justified because our main goal is to compare the alignment limit with the different wrong sign scenarios for the heavy Higgs case (to be discussed in detail in the next sections).
One should stress that all collider bounds were taken into account, including the LEP bound on a light scalar coming from $e^+ e^- \to Zh$.

 It is straightforward to show (using the map of Eq.~\eqref{eq:h2H}) that all arguments that were used above to explain the excluded (large) $\tan\beta$ regions in the light scenario still hold with the replacements
\begin{equation}
\begin{cases}
\sin(\beta-\alpha)\rightarrow {\rm sign}(\alpha)\cos(\beta-\alpha) & \\
\cos(\beta-\alpha)\rightarrow -{\rm sign}(\alpha)\sin(\beta-\alpha)&
\end{cases}
\end{equation} 
in Eqs.~\eqref{hddhuuIIFS} and~\eqref{hddhuuILS}. This also explains why the two bands that are still allowed for each model type, now appear separated, on the left ($\cos(\beta-\alpha)\rightarrow -1$) and on the right ($\cos(\beta-\alpha)\rightarrow 1$) respectively.  Similarly the line equivalent to the $\sin(\beta+\alpha)=1$  line was mapped to $\cos(\beta+\alpha)=1$ (now with $\alpha<0$) (left band of each plot in figure~\ref{fig:LHC8_2}).
After discussing in detail all the possible wrong sign scenarios we will return to the discussion of figure~\ref{fig:LHC8_2} in Sect.~\ref{subsec:heavy}.


\section{The wrong sign limits of the CP-conserving 2HDM}
\label{sec:ws}

In this section we will classify the possible sign changes that can occur (for each scenario) in the Higgs couplings to fermions and massive gauge bosons, relative to the corresponding SM Higgs couplings. 

\subsection{The light Higgs scenario}
We start by discussing the scenario where $m_h = 125.9$ GeV.
As discussed in the previous section,  in this scenario, models II and F have two disjunct allowed regions. One corresponding to the alignment limit and the 
other one centred around the line $\sin(\beta +\alpha)=1$. With our conventions, the latter corresponds to the situation 
where the Higgs coupling to down-type
quarks changes sign relative to the SM, while couplings to up-type quarks and massive gauge bosons are the SM ones.
This is the wrong sign limit~\cite{Ferreira:2014naa} (see also~\cite{Dumont:2014wha, Fontes:2014tga,Ferreira:2014sld} for the CP-conserving 2HDM
and \cite{Fontes:2014xva} for the complex 2HDM) and it is imposed 
only at tree-level. The wrong sign scenarios were first studied in the context of the 2HDM in~\cite{Ginzburg:2001ss, GKO}.

We will now analyse the limit $\sin (\beta + \alpha) = 1$  for all Yukawa types. The main goal is to understand if a sign change in a given Higgs coupling can be measured at the LHC, being therefore distinguishable from the alignment limit. 
The Yukawa coupling signs for the different model types, when $\sin (\beta + \alpha) = 1$, are shown in Table~\ref{tab:couplings}.
\begin{table}[h!]
\begin{center}
\begin{tabular}{cccccccccccccccccccc}
\hline
& & Type I  & & Type II & & Type F & & Type LS & \\
\hline
$\kappa_U$   & & $+1$  & &  $+1$ & &  $+1$ & &  $+1$ & \\
$\kappa_D$   & &  $+1$  & &  $- 1$ & &  $- 1$ & &  $+1$ & \\
$\kappa_L$    & &  $+1$  & &  $- 1$ & & $+1$ & & $- 1$ & \\
\hline
\end{tabular}
\end{center}
\caption{Lightest Higgs Yukawa couplings in models I, II, F and LS in the limit $\sin(\beta+\alpha)=1$.
\label{tab:couplings}}
\end{table}
In order to probe sign changes in the Higgs couplings we need processes where interference occurs.
The best way to probe a sign change in the Yukawa sector is to use the effective $hgg$ vertex. The amplitude for both 
the $gg \to h$ production process and the $h \to gg$ decay is the sum of two contributions (considering only the third generation), 
one with a top-quark loop and the other one with a bottom-quark loop. Therefore if $\kappa_U \kappa_D < 0$
the interference term changes sign relative to the SM, so, in principle, the signal rates may be substantially different from the SM value as to allow for a discrimination between the 2HDM and the SM. In fact, focusing on the types II and F, for which this wrong sign scenario may occur, the ratio between the two LO
widths in the exact limit $\sin (\beta + \alpha) = 1$ is 
\begin{equation}
\frac{\Gamma^{\scriptscriptstyle {\rm 2HDM}} (h \to gg)_{{\rm LO}}} {\Gamma^{\scriptscriptstyle {\rm SM}} (h \to gg)_{{\rm LO}}}  = 1.27  \qquad  (\sin (\beta + \alpha) = 1)\, .
\label{eg-ga}
\end{equation}
As discussed in~\cite{Ferreira:2014naa}, this interference effect, almost 30\% relative to the SM, is not so strong in the $gg \to h$  production process,
which is the main Higgs production mode at the LHC.  In fact, in contrast with the LO result,
\begin{equation}
\frac{\sigma^{\scriptscriptstyle {\rm 2HDM}}  (gg \to h)_{{\rm LO}}} {\sigma^{\scriptscriptstyle {\rm SM}} (gg \to h)_{{\rm LO}}}  \approx
\frac{\Gamma^{\scriptscriptstyle {\rm 2HDM}} (h \to gg)_{{\rm LO}}} {\Gamma^{\scriptscriptstyle {\rm SM}} (h \to gg)_{{\rm LO}}}  \approx 1.27  \qquad  (\sin (\beta + \alpha) = 1)\,,
\label{eg-ga2}
\end{equation}
at NNLO in the limit of $\sin (\beta + \alpha) = 1$, we have
\begin{equation}
\frac{\sigma^{\scriptscriptstyle {\rm 2HDM}}  (gg \to h)_{{\rm NNLO}}} {\sigma^{\scriptscriptstyle {\rm SM}} (gg \to h)_{{\rm NNLO}}}  \approx  1.12  \qquad  (\sin (\beta + \alpha) = 1)\,,
\label{eg-ga3}
\end{equation}
while the ratio of the partial widths of $h \to gg$ does not suffer any significant change in going from LO to NNLO. 
In order to test the stability of the ratio~\eqref{eg-ga3}, we have performed the calculation with two PDF sets,
 MSTW2008nnlo68cl.LHgrid~\cite{Martin:2009iq} and CT10nnlo.LHgrid~\cite{Gao:2013xoa}
and we have varied the factorization and renormalization scales (taken equal) from $m_h/4$ to $m_h$ (all tests were performed with HIGLU.). The maximal variation was 
with the scales and it ranged from $1.122$ to $1.130$, that is, below $1 \%$.
For a center-of-mass energy of 14 TeV, the maximal variation was from $1.107$ to $1.120$, about a $1 \%$ variation.
Therefore, the ratio is stable and $\kappa_g$ can in principle be used to distinguish between the two scenarios in model types II and F if measured
with enough accuracy. However, the difference in the values of~\eqref{eg-ga} and of~\eqref{eg-ga3} is one of the reasons why this scenario
is not yet excluded at the LHC. In fact, a wrong sign cross section about 30\%  above the SM one, would probably have already been excluded.
However the enhancement of about 12 \% is not enough to exclude this scenario at present energies.  

When $\sin (\beta + \alpha) = 1$, the tree-level coupling to massive gauge bosons can be written as
\begin{equation}
\kappa_V = \sin(\beta - \alpha) =\frac{\tan^2 \beta -1}{\tan^2 \beta +1}  \, \, .
\label{eq:eq1}
\end{equation}
Therefore, there are two distinct regimes regarding the sign of $\kappa_V$: when $\tan \beta > 1$, $\kappa_V>0$,  while if  $\tan \beta < 1$, $\kappa_V<0$.
Note that when  $\kappa_V>0$,  $\tan \beta \gg 1$ implies $\kappa_V \approx 1$; on the contrary, if $\kappa_V <0$ because  $\tan \beta \ll 1$ is disallowed,
 $\kappa_V$ can never reach the alignment limit.
In fact, even for very small $\tan \beta$, say $0.5$, we would get $\kappa_V = -0.6$ and therefore a value of $\kappa_V^2$ quite far from $1$.
We will come back to this point later.

The other effective vertex with interference being measured at the LHC is the $h\gamma\gamma$ coupling. In this case, besides the fermion loops
we have the W-loop and also the charged Higgs loop contribution, where a new vertex, $g_{h H^\pm H^\mp}$,  comes into play.  In the notation of~\cite{Ferreira:2014naa},
in the wrong sign limit, the coupling $g_{h H^\pm H^\mp}$ takes the form~\cite{Ginzburg:2001ss}
\begin{equation}
g_{h H^\pm H^\mp}  = - \frac{\tan^2 \beta -1}{\tan^2 \beta +1} \, \frac{2 m_{H^\pm}^2 - m_h^2}{v^2} =- \kappa_V \, \frac{2 m_{H^\pm}^2 - m_h^2}{v^2}    \,.
\label{eq:gHclight}
\end{equation}
Hence, as discussed in~\cite{Ferreira:2014naa}, when $\kappa_V  >0$ the charged Higgs contribution approaches a constant (negative) value and reduces
the value of $\Gamma (h \to \gamma \gamma)$. 
However, when $\kappa_V  < 0 $, 
this contribution is positive and can be very close to zero (it is exactly zero when $\tan \beta =1$). Therefore, $\Gamma (h \to \gamma \gamma)$ 
is no longer reduced by the charged Higgs loop contribution when $\kappa_V  < 0 $. 

By examining Table~\ref{tab:couplings} we can now enumerate the wrong sign scenarios that could in principle be probed in
each model. This is shown in Table~\ref{tab:wrongsignlight}, where $\kappa_L$ was left out because there is no relevant interference 
term contributing to either $\kappa_g$ or  $\kappa_\gamma$.
\begin{table}[h!]
\begin{center}
\begin{tabular}{cccccccccccccccccccc}
\hline
& & Type I  & & Type II & & Type F & & Type LS & \\
\hline
$\tan \beta  > 1$  & & No  & &  $\kappa_D \, \kappa_V <0$ & &  $\kappa_D \, \kappa_V <0$ & &  No & \\
                              & &       & &  $\kappa_D \, \kappa_U <0$ & &  $\kappa_D \, \kappa_U <0$ & &       & \\
                              & &       & &                                               & &                                                & &       & \\
$\tan \beta  < 1$  & &  $\kappa_U \, \kappa_V <0$  & & $\kappa_U \, \kappa_V <0$ & & $\kappa_U \, \kappa_V <0$& &  $\kappa_U \, \kappa_V <0$ & \\
                              & &                                                 & & $\kappa_D \, \kappa_U <0$ & &   $\kappa_D \, \kappa_U <0$  & &   & \\
\hline
\end{tabular}
\end{center}
\caption{Possible wrong sign scenarios in the four Yukawa types for the lightest CP-even Higgs
\label{tab:wrongsignlight}}
\end{table}
In conclusion, the wrong sign scenario can be defined as either $\kappa_D \, \kappa_V <0$ (for $\tan\beta>1$) or $\kappa_U \, \kappa_V <0$ (for $\tan\beta<1$). We can further have (for types II and F) $\kappa_D \, \kappa_U <0$, in which case both $\kappa_g$ and $\kappa_\gamma$ are affected, otherwise, if 
$\kappa_D \, \kappa_U >0$ (and $\kappa_V <0$) , only $\kappa_\gamma$ is affected.

\subsection{The heavy Higgs scenario}

In the scenario where we set the heaviest CP-even state, $H$, to be the Higgs, i.e. $m_H=125.9$ GeV, the alignment limit is obtained by setting $\cos (\beta - \alpha)=1$.
The Higgs couplings to fermions and massive gauge bosons are $\kappa_F^H = \kappa_V^H = 1$.
In this scenario the wrong sign limit is obtained when $\cos (\beta + \alpha)=1$. The Yukawa couplings for 
the different model types, in the limit $\cos (\beta + \alpha)=1$, are shown in Table~\ref{tab:couplingsH}.
\begin{table}[h!]
\begin{center}
\begin{tabular}{cccccccccccccccccccc}
\hline
& & Type I  & & Type II & & Type F & & Type LS & \\
\hline
$\kappa_U$   & & $-1$  & &  $-1$ & &  $-1$ & &  $-1$ & \\
$\kappa_D$   & &  $-1$  & &  $+ 1$ & &  $+ 1$ & &  $-1$ & \\
$\kappa_L$    & &  $-1$  & &  $+ 1$ & & $-1$ & & $+ 1$ & \\
\hline
\end{tabular}
\end{center}
\caption{Heavy Higgs Yukawa couplings in models I, II, F and LS in the limit $\cos (\beta+\alpha)=1$.
\label{tab:couplingsH}}
\end{table}
The Yukawa couplings have all changed sign relative to lightest Higgs scenario. It is clear that it is again type II
and type F that can be distinguished from the corresponding alignment limit in $\kappa_g$. It should be noted, however, that in each of the corresponding wrong sign scenarios (heavy Higgs scenario and last section's light Higgs scenario) the value of $\kappa_g$ is exactly the same. 

As for the Higgs coupling to massive gauge bosons, when $\cos (\beta + \alpha) = 1$, it is given by
\begin{equation}
\kappa_V^H = \cos(\beta - \alpha) =-\frac{\tan^2 \beta -1}{\tan^2 \beta +1}  \, .
\label{eq:KvH}
\end{equation}
Again we have two distinct regimes regarding the sign of $\kappa_V^H$: when $\tan \beta > 1$, $\kappa_V^H<0$ while if  $\tan \beta < 1$, $\kappa_V^H>0$.
Therefore, also the sign of $\kappa_V^H$ is reversed relative to that of $\kappa_V$. As will be discussed later in detail, the case $\tan \beta <1$ is already excluded and 
it will not be further mentioned. Finally, the charged Higgs coupling for $\cos (\beta+\alpha)=1$ is
\begin{equation}
g_{H H^\pm H^\mp}  =  \frac{\tan^2 \beta -1}{\tan^2 \beta +1} \, \frac{2 m_{H^\pm}^2 - m_H^2}{v^2} =- \kappa_V^H \, \frac{2 m_{H^\pm}^2 - m_H^2}{v^2}   \, , \qquad  \cos (\beta+\alpha)=1
\label{eq:gHcHeavy}
\end{equation}
while in the heavy Higgs alignment limit we obtain
\begin{equation}
g_{H H^\pm H^\mp}  = - \, \frac{2 m_{H^\pm}^2 + m_H^2- 2 M^2}{v^2} \, ,  \qquad  \cos (\beta-\alpha)=1  \, ,
\end{equation}
where $M^2=m_{12}^2/(\sin \beta \, \cos \beta)$.

Therefore, there is a simultaneous change of sign in the Higgs couplings to massive gauge bosons and in the Yukawa couplings relative to the lightest 
Higgs case. That is, the wrong sign scenarios are exactly the same as the ones for the lightest Higgs case. 
This is true even for the charged Higgs coupling to the heavy Higgs. As we will see, no major difference is found in the results regarding the wrong sign limits relative to the 
light Higgs scenario.

\section{The present status and the future of the different wrong sign scenarios}
\label{sec:res}

Throughout this section we will use (as we did in our previous work~\cite{Ferreira:2014naa}) the expected errors for the 14 TeV LHC, Tables~1-20 of~\cite{Dawson:2013bba}, as a reference. 
The quoted expected errors for $\kappa_g$ based on fittings are $6$--$8\%$ for $L=300$ fb$^{-1}$ and $3$--$5\%$ for $L=3000$ fb$^{-1}$.
The predicted accuracy for $\kappa_\gamma$ is $5$--$7\%$ for an integrated luminosity of $L=300$ fb$^{-1}$ and $2$--$5\%$ for $L=3000$ fb$^{-1}$.
For comparison, the predicted accuracy at the International Linear Collider can be found in~\cite{Ono:2012ah, Asner:2013psa}.
%


\subsection{Light Higgs scenario for $ \rm \tan \beta >1$}
This scenario was discussed in detail in a previous work~\cite{Ferreira:2014naa}. There, we have analysed the case where the lightest Higgs
is the SM one in a type II model. We have forced all rates to be within $20$, $10$ and $5  \%$ of the SM predictions. 
We have concluded 
that measurements of either $\kappa_g$ and $\kappa_\gamma$ with $5  \%$ accuracy would enable us to distinguish between the
alignment limit and the wrong sign scenario. In this section we show the status of  $\kappa_g$ and $\kappa_\gamma$ at the end of the 8 TeV run. 
\begin{figure}
\centering
\mbox{\includegraphics[width=0.45\linewidth,clip=true]{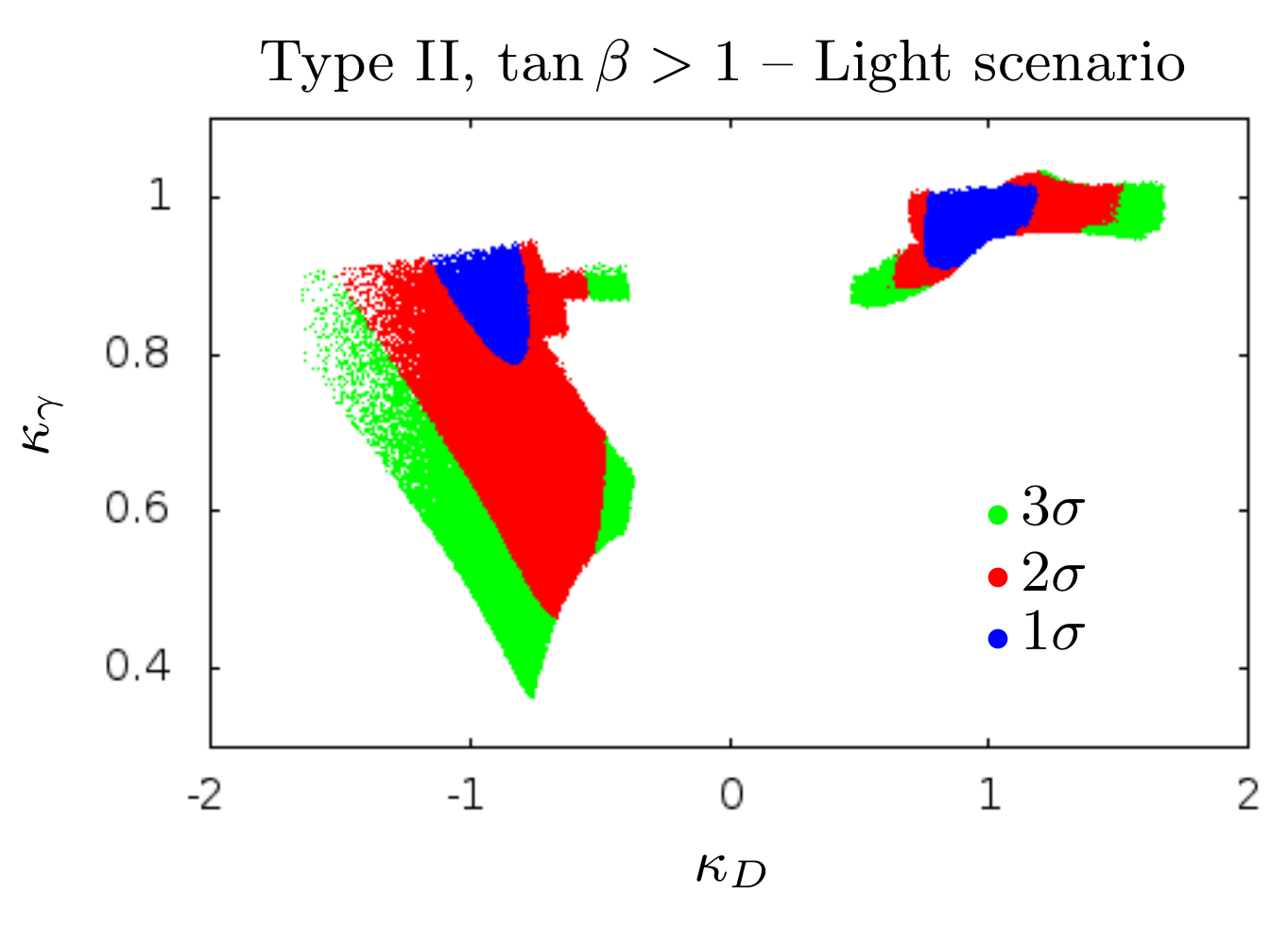}\hspace{6mm}\includegraphics[width=0.45\linewidth,clip=true]{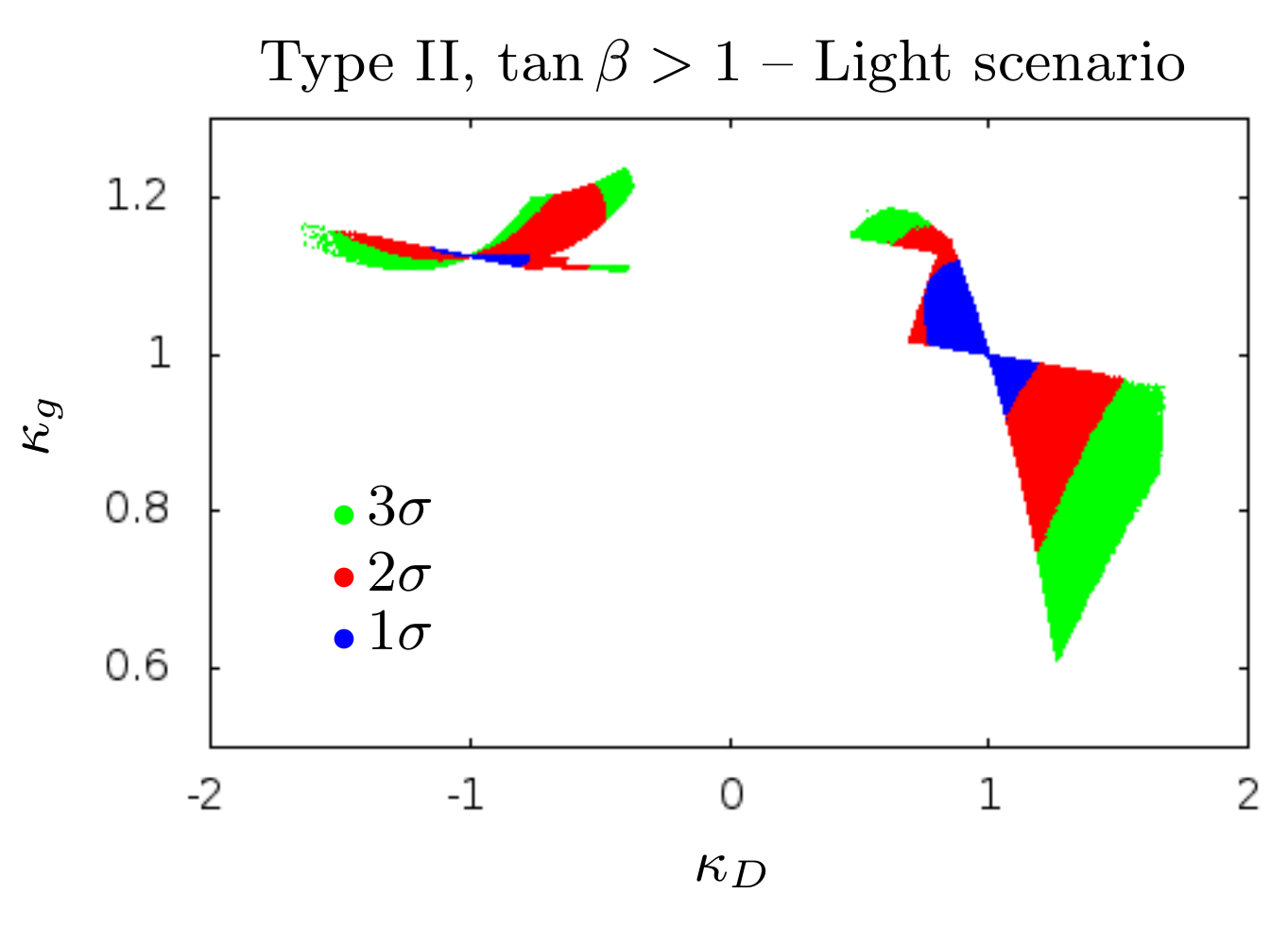}}
\caption{Left: $\kappa_\gamma$ as a function of $\kappa_D$; right: $\kappa_g$ as a function of $\kappa_D$.
 All points are for type II with $\tan \beta > 1$ and have passed both the pre-LHC constraints 
and the 7/8 TeV LHC Higgs data at 1$\sigma$ (blue), 2$\sigma$ (red) and 3$\sigma$ (green).}
\label{fig:f01}
\end{figure}
In the left panel of figure~\ref{fig:f01}
we show $\kappa_\gamma$ as a function of $\kappa_D$ and in the right panel we can see $\kappa_g$ as a function of $\kappa_D$.
 All points are for type II with $\tan \beta > 1$ and have passed both the pre-LHC constraints 
and the 7/8 TeV LHC Higgs data at 1$\sigma$ (blue), 2$\sigma$ (red) and 3$\sigma$ (green).
It is clear that, at the end of the 8 TeV run, the wrong sign scenario is still allowed at 1$\sigma$. This was expected
because, as discussed in~\cite{Ferreira:2014naa}, we need the 13/14 TeV LHC with at least an integrated luminosity
of 300 $fb^{-1}$ to exclude this particular wrong sign scenario. Finally, although not exactly the same, the plots 
for type F look very similar, and there is no point in showing them here. As discussed in~\cite{Ferreira:2014naa} there is no wrong sign scenario in types I and LS
for $\tan \beta > 1$.

\subsection{Light Higgs scenario in the low $\tan \beta$ regime} 
We will now analyse the lightest Higgs wrong sign scenarios for $\tan \beta < 1$ for the type II model. As discussed earlier, B-physics constraints and $R_b$
force $\tan \beta > O(1)$ although values of $\tan \beta$ slightly smaller than 1 are still allowed depending on the charged
Higgs mass. We will come back to this point later. 

In order to understand if the different wrong sign scenarios are still allowed for $\tan \beta < 1$ after the 8 TeV LHC, we have first generated a separate set of points with the same experimental and theoretical constraints applied as before, but where we have turned off the $R_b$ and B-physics constraints and the experimental constraints coming from colliders (LHC, Tevatron and LEP). In addition we also impose $0.5 < \tan \beta < 1$ which is the region of interest that we will discuss below in the full sample with all constraints turned on.

\begin{figure}
\centering
\includegraphics[width=0.35\linewidth]{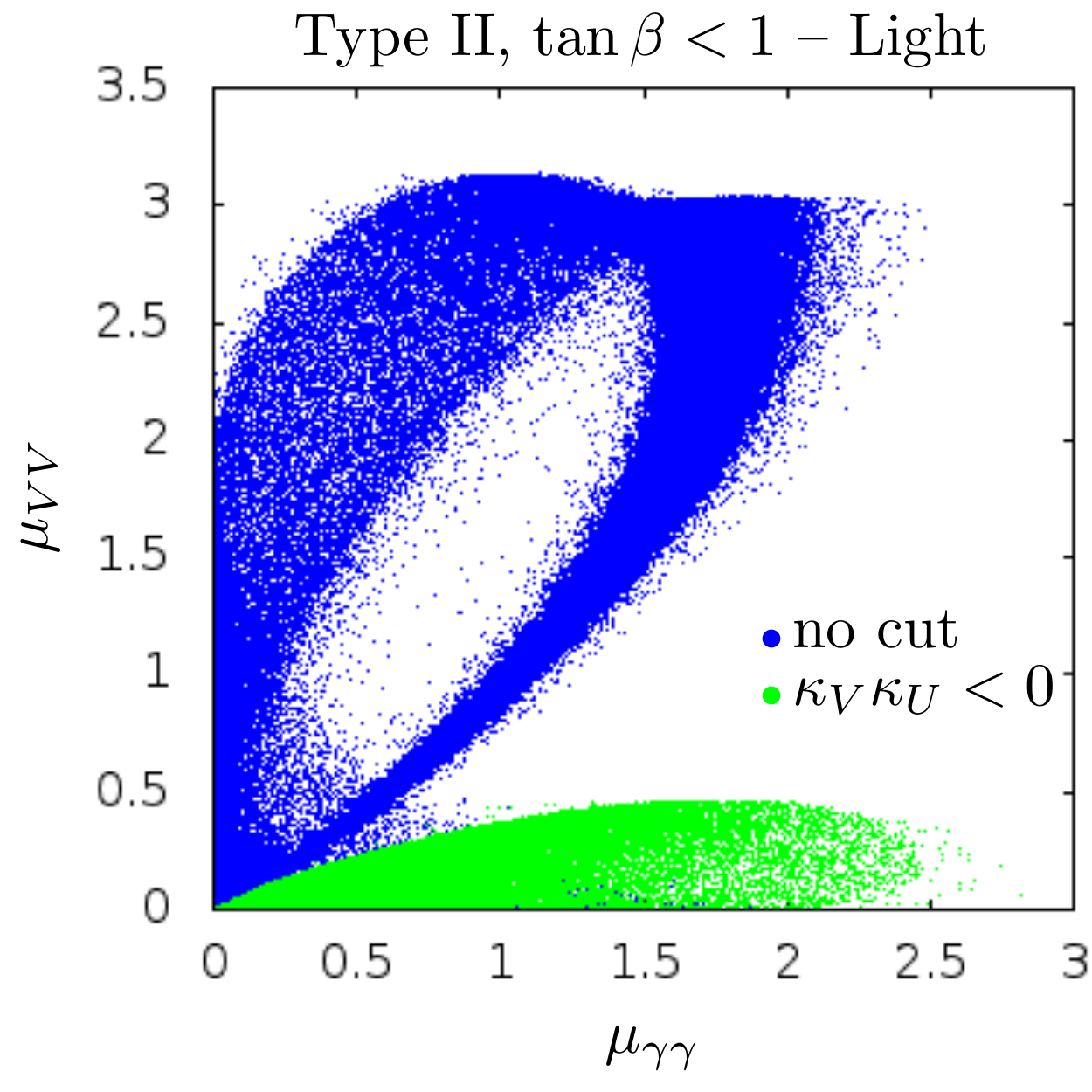}
\hspace{0.1\linewidth}
\includegraphics[width=0.35\linewidth]{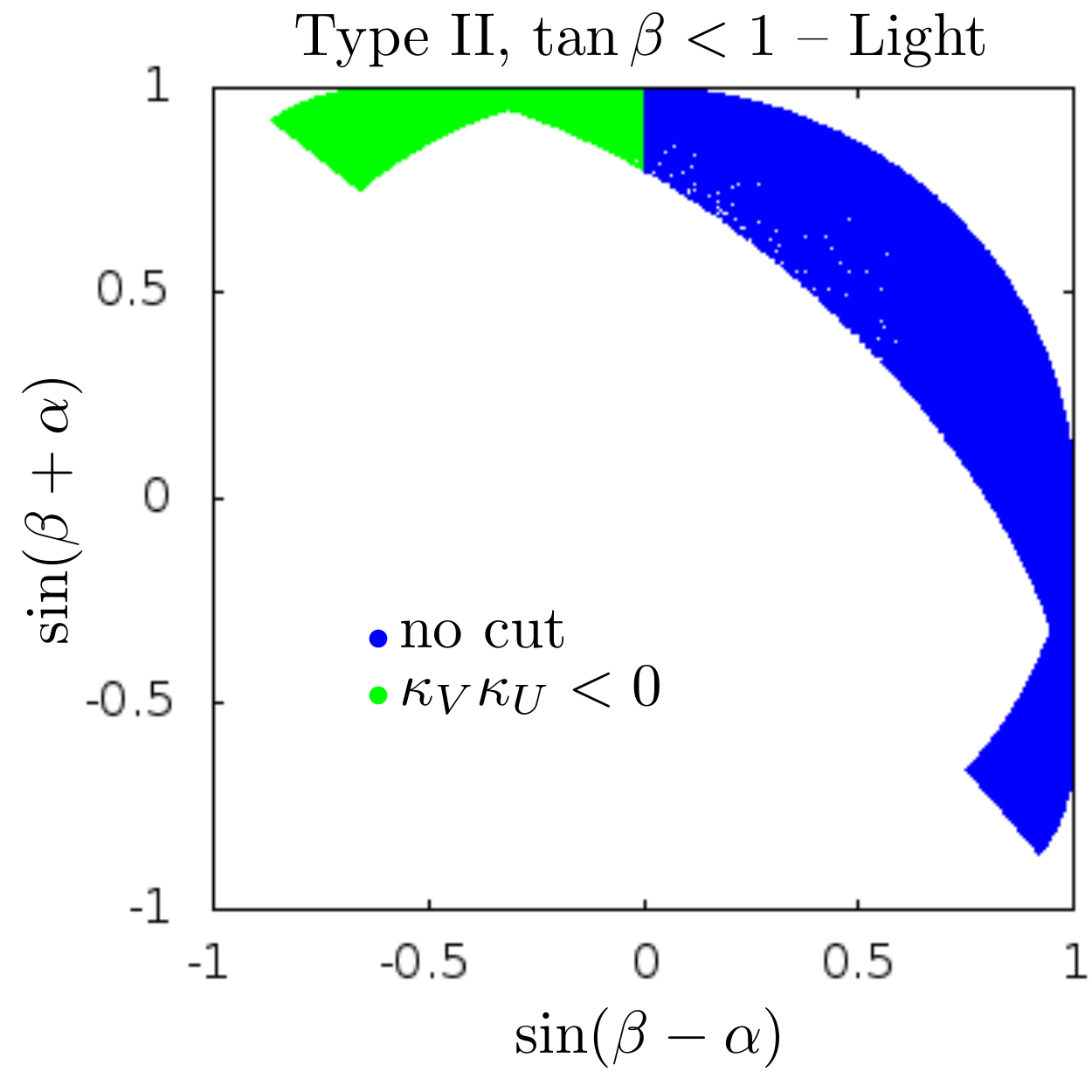}
\caption{Left: $\mu_{VV}$ as a function of $\mu_{\gamma \gamma}$ in the wrong sign scenario for $\tan \beta < 1$ and $\kappa_V \, \kappa_U <0$; right: $\sin  (\beta - \alpha)$ as a function of $\sin  (\beta + \alpha) $ for the same scenario.}
\label{fig:f1}
\end{figure}
\begin{figure}
\centering
\includegraphics[width=0.35\linewidth]{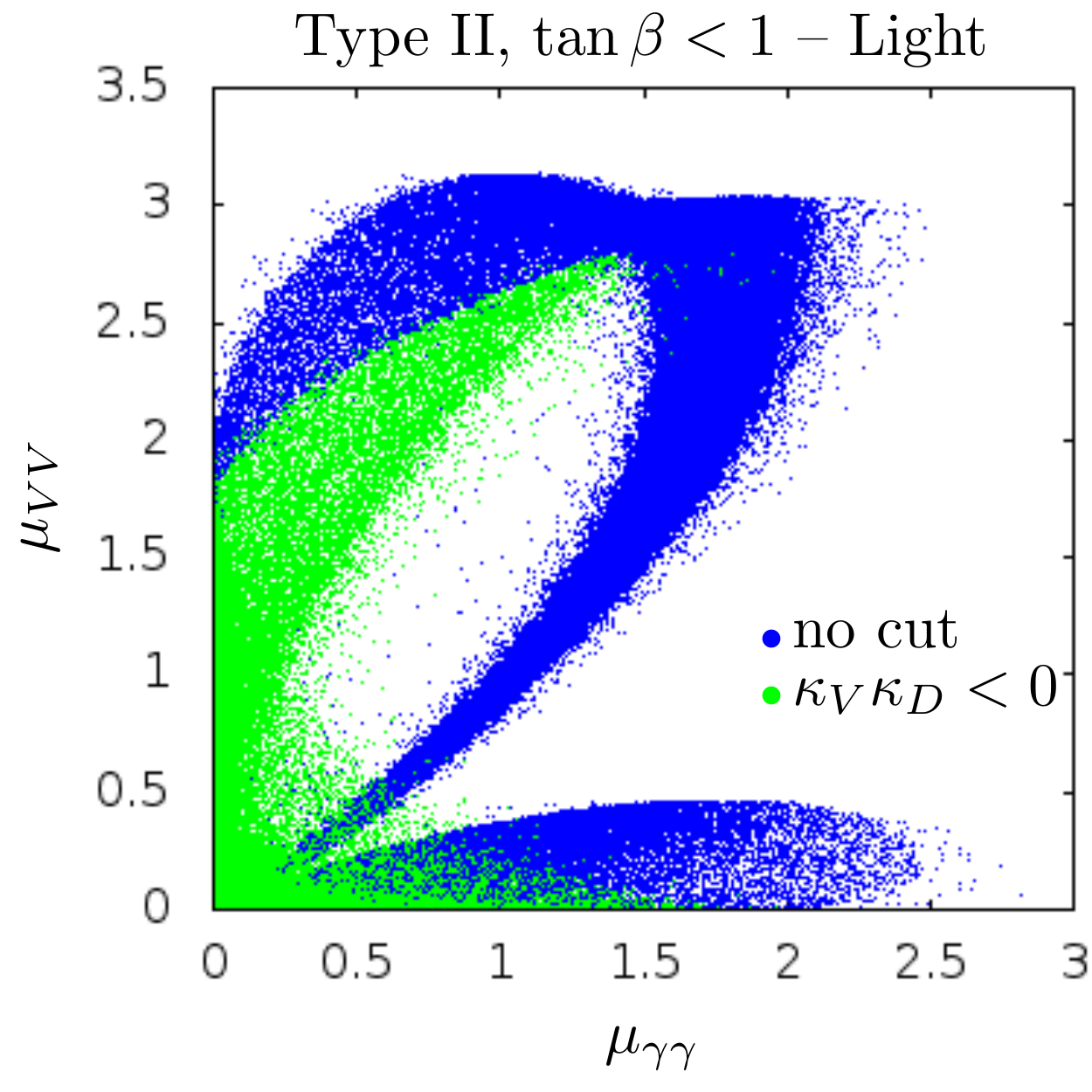}
\hspace{0.1\linewidth}
\includegraphics[width=0.35\linewidth]{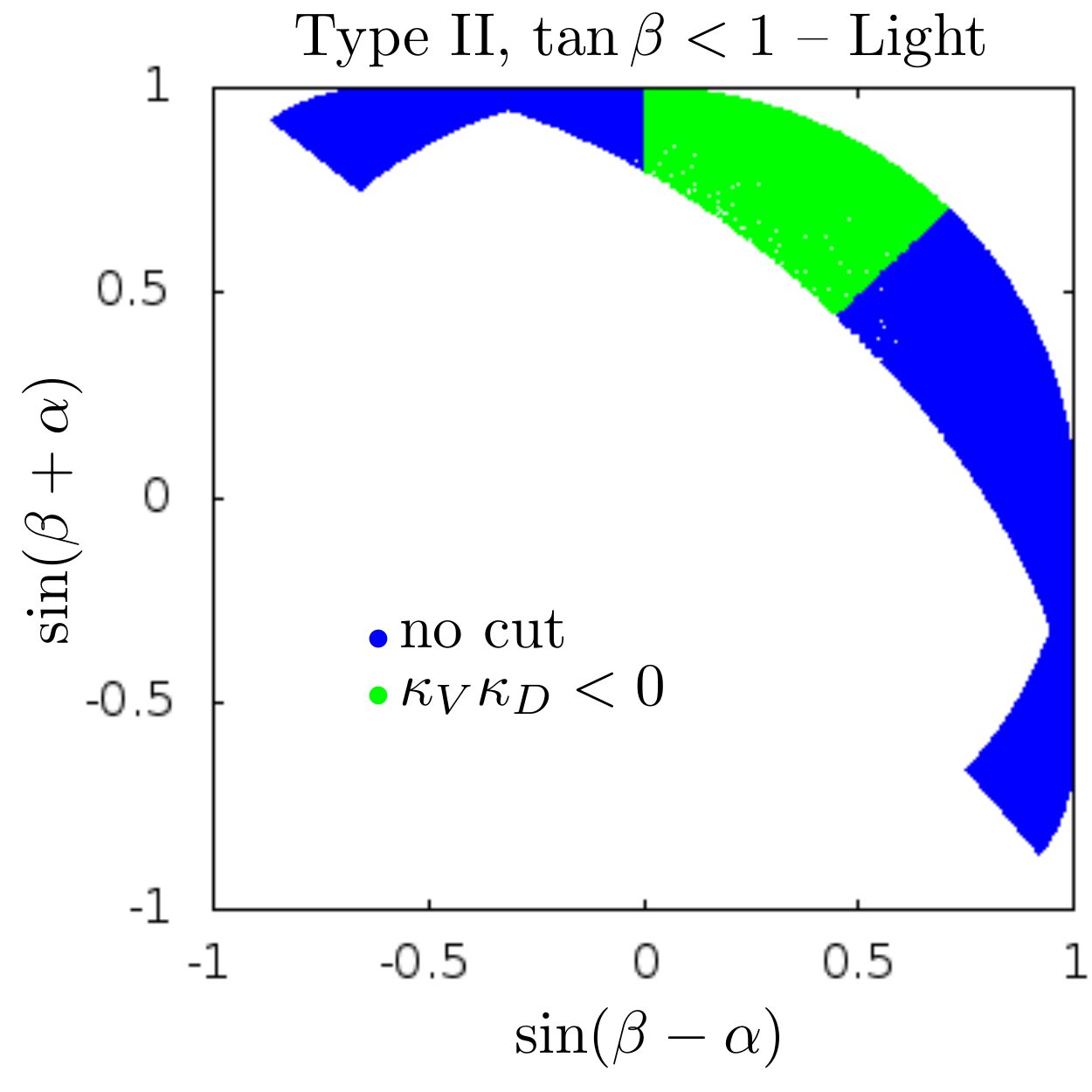}
\caption{Left: $\mu_{VV}$ as a function of $\mu_{\gamma \gamma}$ in the wrong sign scenario for $\tan \beta < 1$ for $\kappa_V \, \kappa_U <0$; right: $\sin  (\beta + \alpha) $
as a function of $\sin  (\beta - \alpha) $ for the same scenario.}
\label{fig:f2}
\end{figure}
In figure~\ref{fig:f1}, left panel, we plot $\mu_{VV}$ as a function of $\mu_{\gamma \gamma}$  for type II without (blue) and with (green) the cut $\kappa_V \, \kappa_U <0$. It is clear that the wrong sign-scenario $\kappa_V \, \kappa_U <0$ will be very constrained if the values of $\mu_{VV}$ measured at the LHC are taken into account. In fact whatever the rates for $\mu_{\gamma \gamma}$ and $\mu_{FF}$,  $\mu_{VV}$ is always well below $1$ when $\kappa_V \, \kappa_U <0$ in this region of small $\tan\beta$.

Due to our conventions for the angles,  $\kappa_U$ is always positive and therefore the  region where $\kappa_V \, \kappa_U <0$ corresponds to $\sin (\beta - \alpha) <0$. 
This is shown in the right panel of figure~\ref{fig:f1} where $\sin  (\beta + \alpha) $
as a function of $\sin  (\beta - \alpha) $ is presented for the same scenario. Not only the green points are all in the  $\sin (\beta - \alpha) <0$ region but it is clear that when 
$\sin (\beta+\alpha) \approx 1$ the allowed values of $\sin (\beta - \alpha)$ are quite far from $1$, forcing $\mu_{VV}$ to be well below $1$ and thus contradicting the LHC results.

The second possibility is to have $\kappa_V \, \kappa_D <0$. In figure~\ref{fig:f2}, left panel, we plot $\mu_{VV}$ as a function of $\mu_{\gamma \gamma}$ using the same color key but now the green points correspond to the cut  $\kappa_V \, \kappa_D <0$.  The latter are distributed around two regions. The first one is similar to the one in  figure~\ref{fig:f1} and corresponds to small values of $\sin (\beta - \alpha)$.
The second region corresponds to larger values of $\sin (\beta - \alpha)$. The $W$ and top loops give the largest contribution to $h \to \gamma \gamma$ and interfere destructively.
When $\kappa_V$ is reduced (although larger than in the previous scenario, it is always below $\approx 0.7$), because $\tan \beta < 1$, $\kappa_U$ is enhanced 
and the amplitude is reduced (taking $\kappa_V=0.7$ and $\tan \beta =0.7$, $\kappa_U \approx 1.7$).
Hence, it is foreseeable that both scenarios are already excluded by the LHC data analysed so far. 
Clearly, the scenario of low $\tan \beta$ in the case of $\kappa_V \, \kappa_D < 0$ is indeed excluded
as was shown in~\cite{Ferreira:2014naa}. In fact, only for large $\tan \beta$ does $\sin (\beta -\alpha)$
approaches $\sin (\beta +\alpha)$ thus leading to values of the rates closer to the SM ones. Therefore,
this scenario is allowed only for large values of $\tan \beta$.

\begin{figure}
\centering
\includegraphics[width=0.35\linewidth]{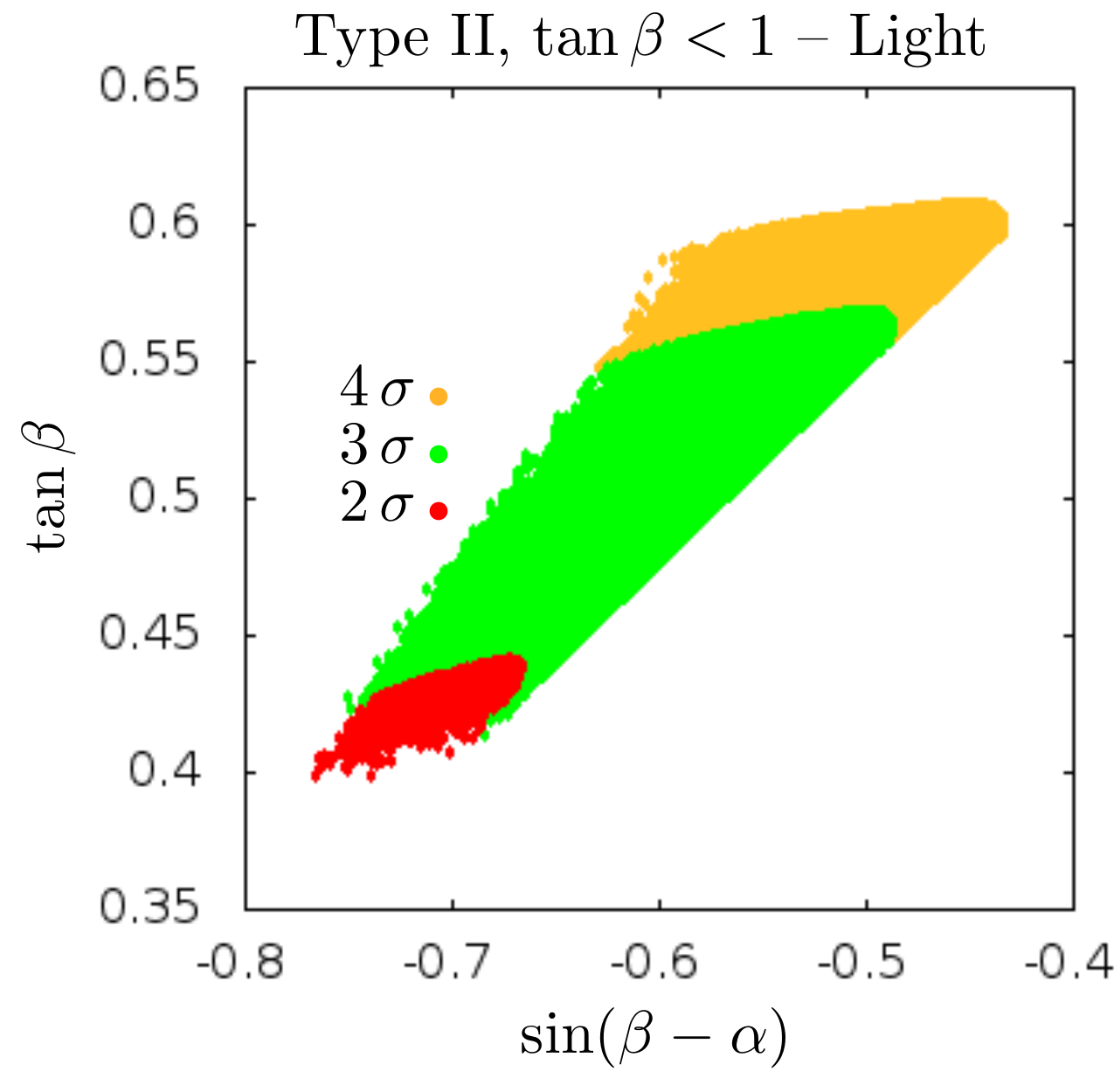}
\hspace{0.1\linewidth}
\includegraphics[width=0.35\linewidth]{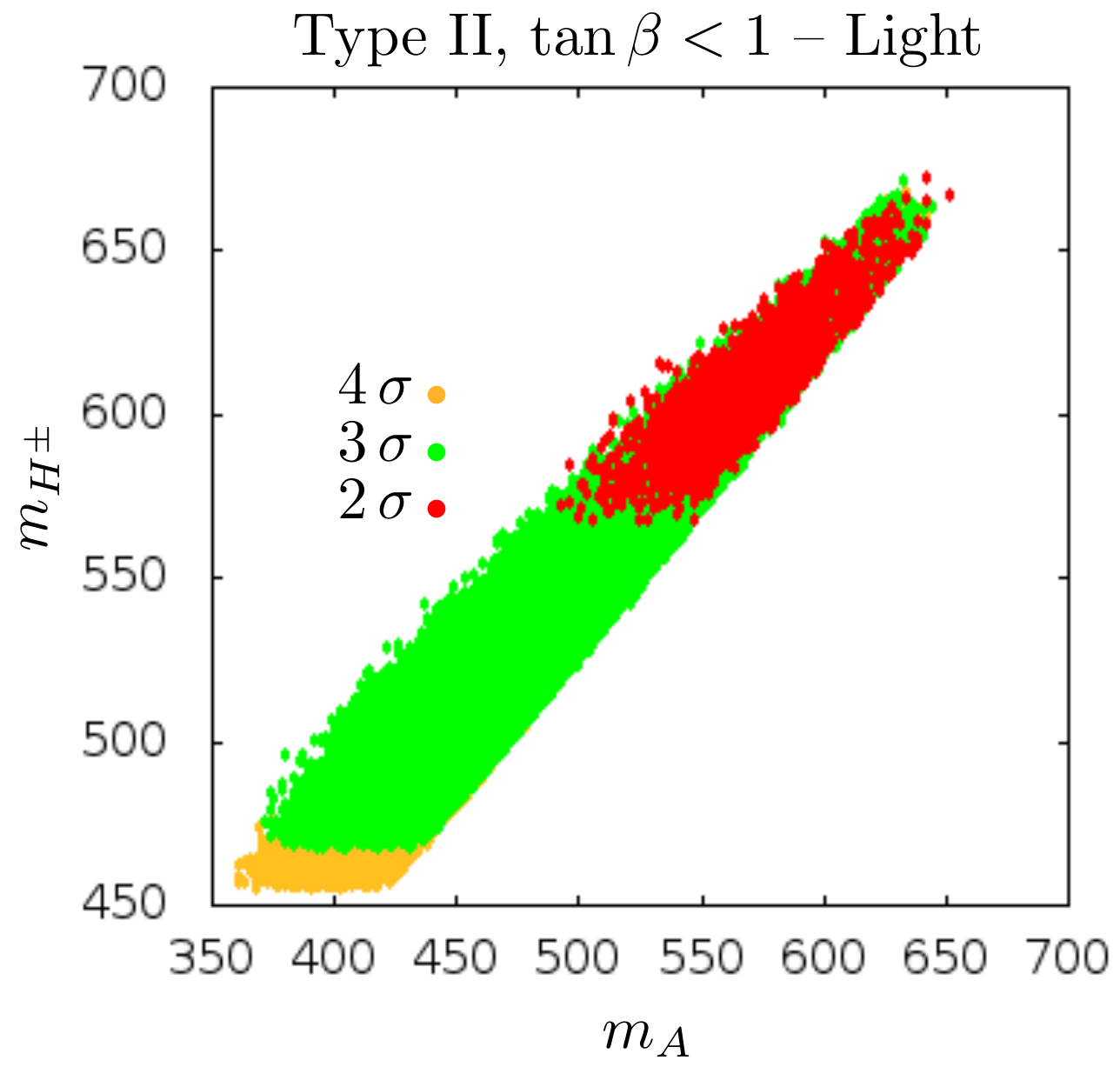}
\caption{Wrong sign scenario ($\kappa_V \, \kappa_U < 0$)
 for the lightest Higgs in the type II model for $\tan \beta <1$ with
all LHC data analysed so far taken into account at 2$\sigma$ (red), 3$\sigma$ (green) and 4$\sigma$ (yellow).
Left: $\tan \beta$ as a function of  $\sin (\beta - \alpha)$; right: $m_{H^{\pm}}$ as a function of $m_A$.}
\label{fig:f4}
\end{figure}
Let us now turn back to the first possibility $\kappa_V \, \kappa_U < 0$ and investigate the points that survive in the full scan\footnote{With all constraints taken into account, including collider data, $R_b$ and the $b\rightarrow s \gamma$ B-physics observable.}. In figure~\ref{fig:f4} (left) we present $\tan \beta$ as a function of $\sin (\beta - \alpha)$ 
for the type II model and $\tan \beta <1$ with all LHC data analysed so far taken into account at 2$\sigma$ (red), 3$\sigma$ (green) and 4$\sigma$ (yellow). Only
points with very low $\tan \beta$ survive and only from 2$\sigma$ onwards. In this figure~\ref{fig:f4} we have taken into account
all constraints, except the ones from the $B$--$\bar{B}$ mixing data.
In the right panel of figure~\ref{fig:f4} we present $m_{H^{\pm}}$ as a function of $m_A$. The main purpose is to  show that the values that give rise to the allowed 2$\sigma$ points require a large $m_{H^{\pm}}$ close to the unitarity limit.
This is a consequence of the structure of the vertex $tbH^{\pm}$ which gauges the new
physics contributions in loop processes where the $W$-loop is replaced by a charged Higgs loop. Hence, 
the constraints coming from $B$-physics are typically exclusion regions in the ($\tan \beta$, $m_{H^{\pm}}$)
plane. As previously discussed, because $\kappa_U$ is positive in our convention, we have $\kappa_V < 0$ and, as seen in the left panel of figure~\ref{fig:f4}, 
its value is well below $1$ meaning that it is therefore very hard to satisfy the LHC bounds on $\mu_{VV}$.

\begin{figure}
\centering
\includegraphics[width=3.5in,angle=0]{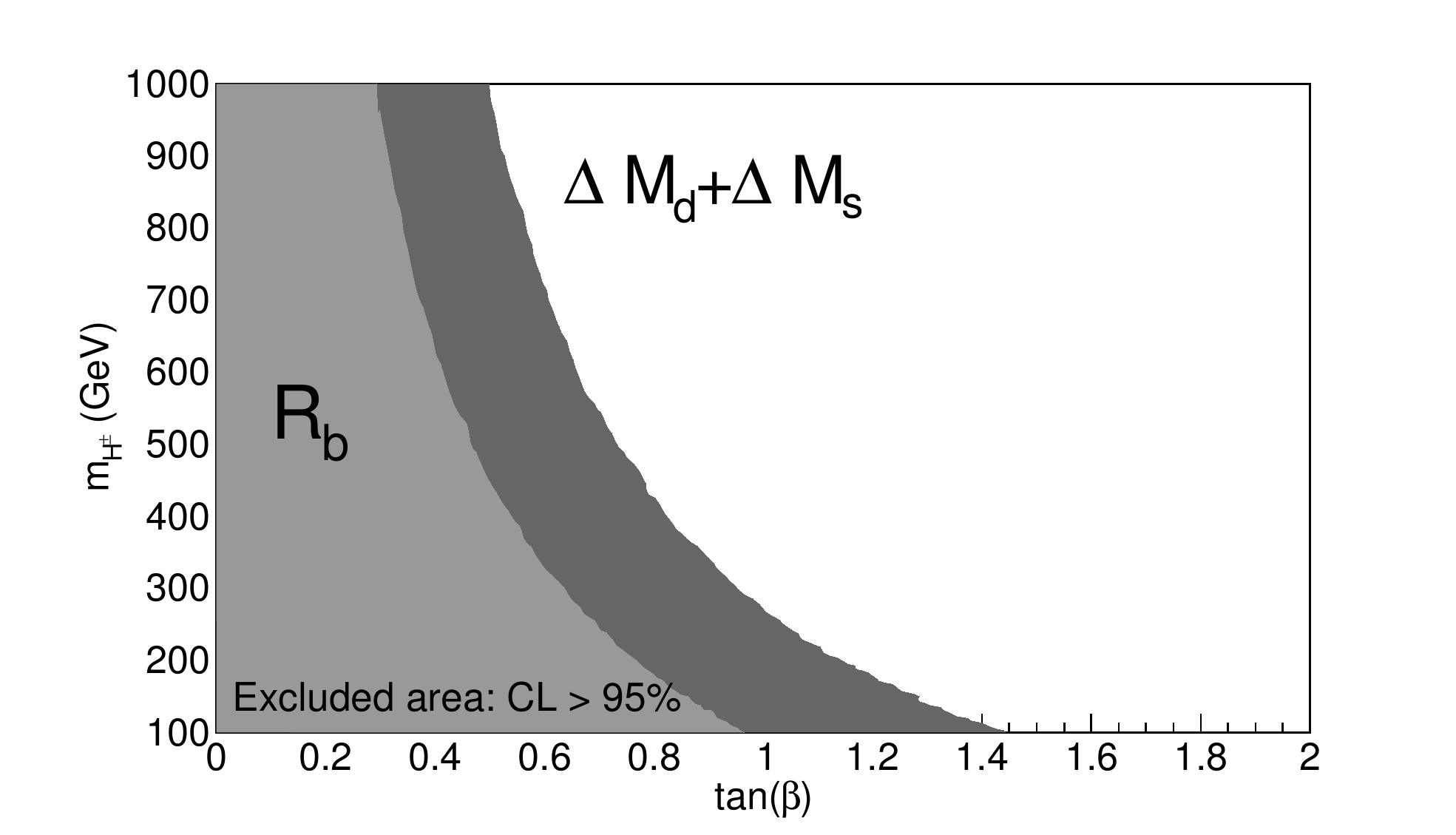}
\caption{Constraints on the $(m_{H^\pm}, \, \tan \beta)$  plane from $R_b$ and $B$--$\bar{B}$ mixing data at 95 $\%$ C.L. (data from~\cite{Deschamps:2009rh}).
}
\label{fig:f5}
\end{figure}

Let us now discuss in more detail the constraints available from B-physics. Contrary to type I and type LS, where $b \to s \gamma$ forces $\tan \beta > 1$, in type II (and type F) values slightly below $\tan \beta =1$
are still allowed for large charged Higgs masses. The main B-physics observables that provide an exclusion of the small $\tan\beta$ region are $R_b$, that was included as a filter at $95 \%$ in figure~\ref{fig:f4}, and the $B$--$\bar{B}$ mixing data, that was not included.
The constraint from $B$--$\bar{B}$ mixing is derived from the measurement of $\Delta m_d$ and $\Delta m_s$~\cite{Geng:1988bq}.
In the SM, neutral-meson mixing occurs due to a box diagram with $W$-boson exchange. In the 2HDM the box contains new contributions due to the charged Higgs bosons, which are obtained by replacing one or two $W$-boson lines by charged Higgs lines (expressions for such contributions at leading-order can be found in~\cite{Geng:1988bq}). The presence of the new diagrams implies that, in the 2HDM, the CKM matrix parameters should be determined from the data simultaneously with the 2HDM parameters in the diagrams (i.e. the charged Higgs mass $m_{H^{\pm}}$ and $\tan \beta$). This modifies the SM fit to fix the CKM matrix~\cite{Hocker:2001xe, Charles:2004jd}. When performing this simultaneous fit, the constraints on the $(m_{H^{\pm}}\tan \beta)$ plane become less restrictive as shown in~\cite{Deschamps:2009rh}. In figure~\ref{fig:f5}~\cite{Deschamps:2009rh} we present the exclusion lines  obtained from $R_b$ and from the $B$--$\bar{B}$ mixing data at 95 $\%$ C.L., on the $(m_{H^\pm}, \, \tan \beta)$ plane. 

Even if the simultaneous fit relaxes the bounds on $\tan \beta$, we have concluded that the inclusion of the $B$--$\bar{B}$ constraints  in the analysis at 95 $\%$ C.L., makes the entire region in the left plot of figure~\ref{fig:f4} vanish. Hence, the B-physics constraints exclude this particular wrong sign scenario at 95 $\%$ C.L.. However, it is clear that even without the constraints coming from loop processes, this scenario will be definitely excluded by the data obtained during the next run of the LHC, assuming that all measurements converge to the SM values with higher precision.

\subsection{Heavy Higgs scenario} 
\label{subsec:heavy}

We now return to the discussion of the wrong sign cases for the heavy Higgs scenario, focusing on $\tan\beta>1$. 
The case where the heaviest CP-even Higgs is the scalar state that was observed at the LHC was first analysed in~\cite{heavyh}, in the context of the 2HDM and it was discussed after the LHC 8 TeV run in~\cite{Wang:2014lta}.  
\begin{figure}
\centering
\mbox{\includegraphics[width=0.45\linewidth,clip=true]{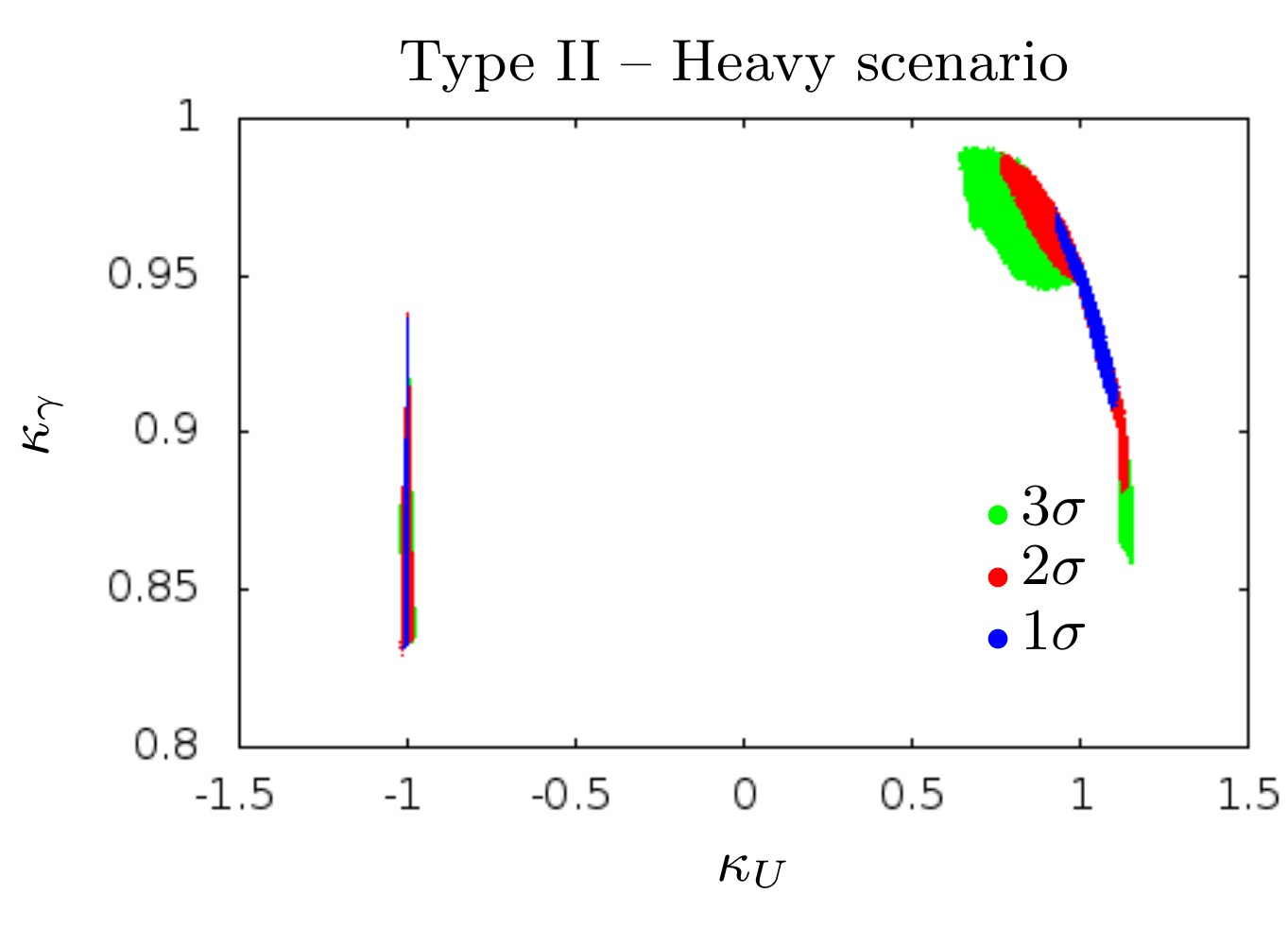}\hspace{6mm}\includegraphics[width=0.45\linewidth,clip=true]{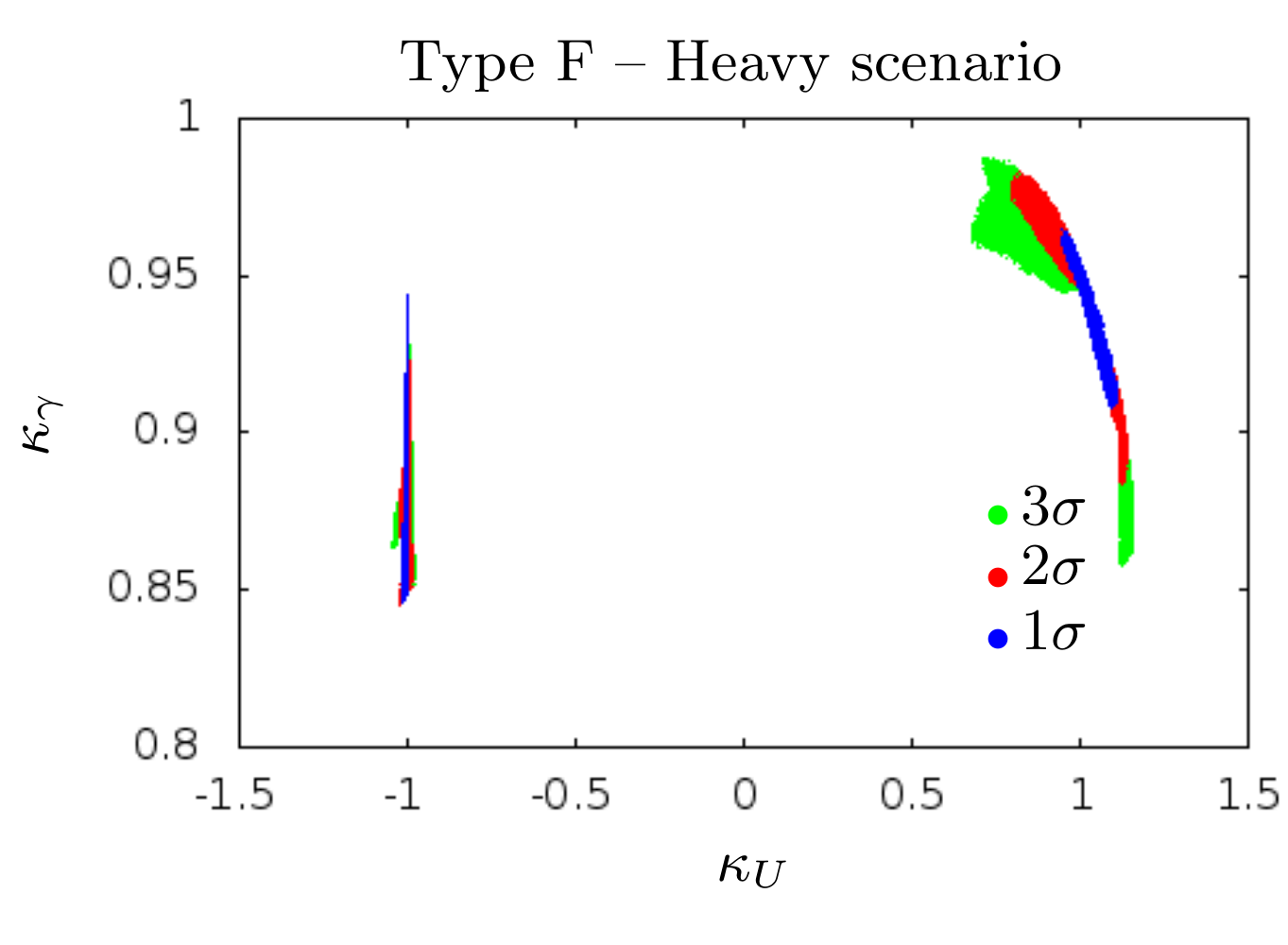}}
\mbox{\includegraphics[width=0.45\linewidth,clip=true]{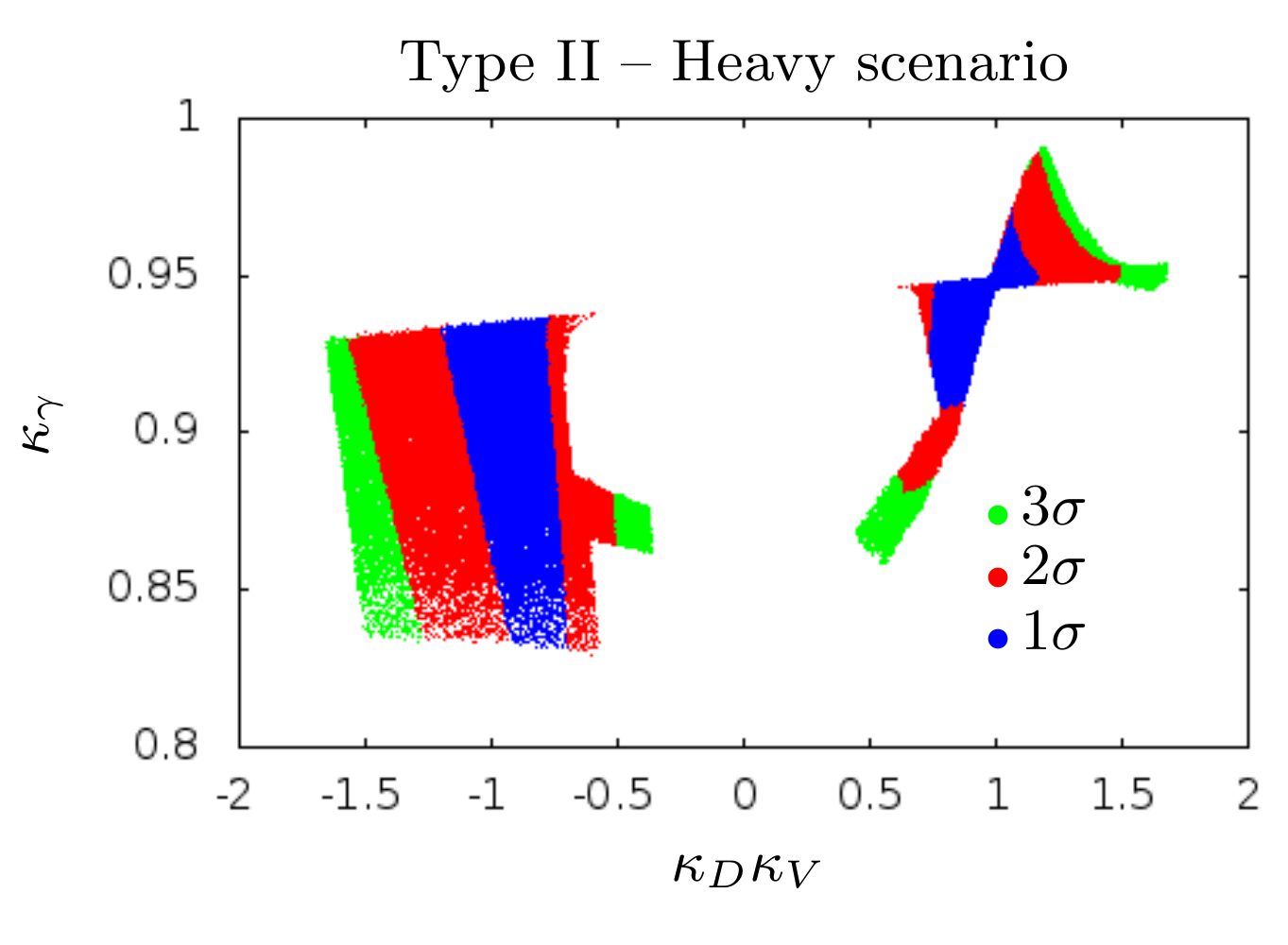}\hspace{6mm}\includegraphics[width=0.45\linewidth,clip=true]{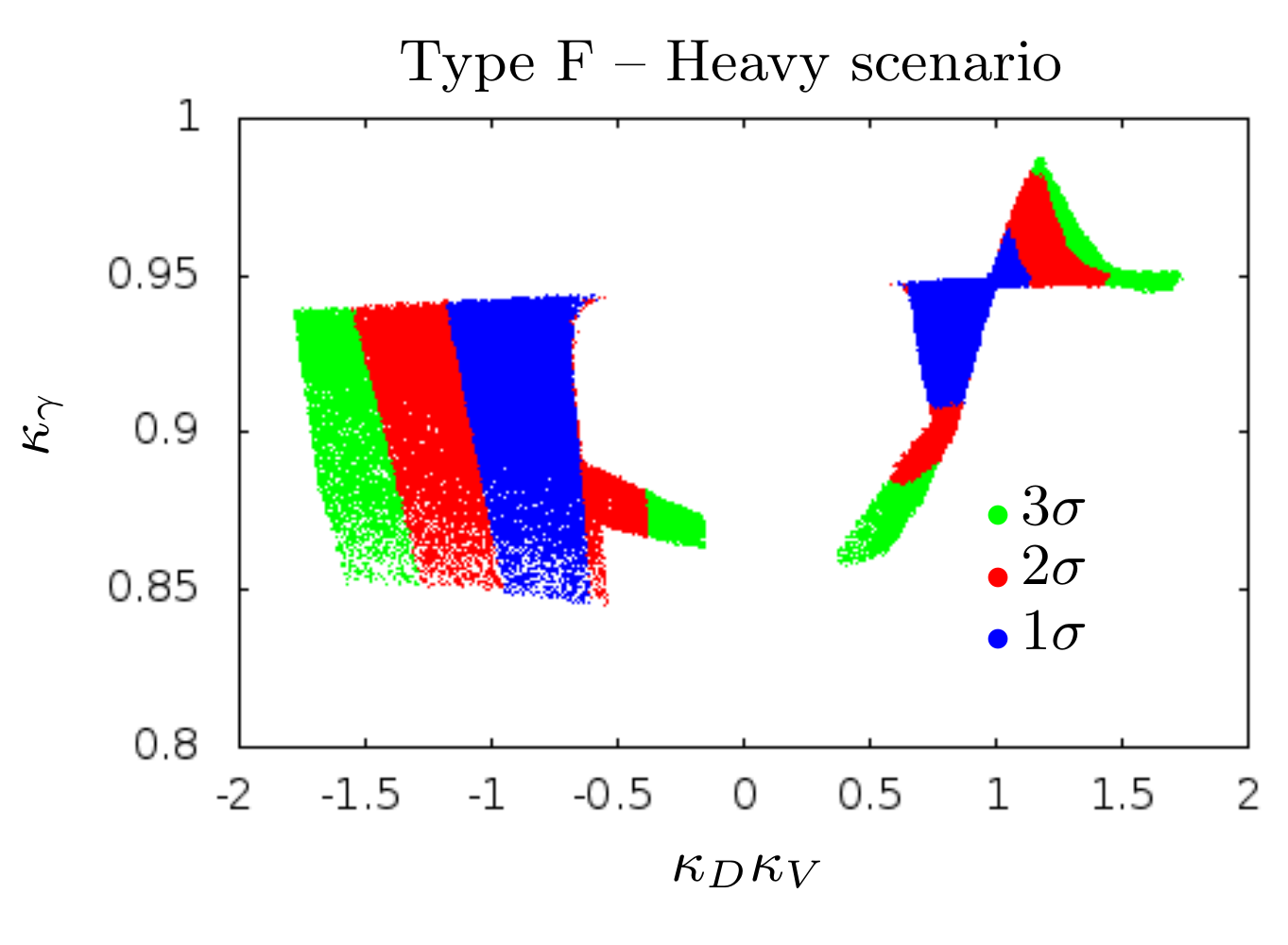}}
\caption{Left: type II; right: type F. $\kappa_\gamma$ as a function of $\kappa_D$ and of $\kappa_D \, \kappa_V$ with all points that have passed both the pre-LHC constraints 
and the 7/8 TeV LHC Higgs data at 1$\sigma$ (blue), 2$\sigma$ (red) and 3$\sigma$ (green). }
\label{fig:f11}
\end{figure}
As previously discussed, the type of wrong sign limits are exactly the same in the light and in the heavy Higgs scenario. Therefore, the only possible difference between the two scenarios could
only come from fact that the parameter spaces scanned are not exactly the same. In fact, in the heavy scenario there is a CP-even scalar with a mass below 125 GeV which alters
the conditions of the scan. However, the general trend is the same and in figure~\ref{fig:f11} we present $\kappa_\gamma$, for the heavy Higgs scenario, as a function of $\kappa_D$ and also as a function
of $\kappa_D \, \kappa_V$  with all points that have passed both the pre-LHC constraints and the 7 and 8 TeV LHC Higgs data at 1$\sigma$ (blue), 2$\sigma$ (red) and 3$\sigma$ (green),
for the type II model (left) and type F model (right). Qualitatively, there seems to be no major differences when we compare these results with the ones obtained for the light Higgs case. In the next section, while discussing
the non-decoupling nature of this scenario we will see that even the conclusions regarding the exclusion of the wrong sign scenario with a $5\%$ precision measurement of the rates also apply
to the heavy Higgs case. Moreover, the reason for this exclusion is exactly the same in the two scenarios - a decrease in $\Gamma (h \to \gamma \gamma)$ due to the charged Higgs loop contribution
in the wrong sign limit.

\section{The non-decoupling nature of the heavy scenario}
\label{sec:non-dec}

In this section we investigate what a measurement of the rates within $5$, $10$ and $20\%$
of the SM value could tell us about the heavy Higgs scenario. We start with type II (the results for type F are very similar)
and our first goal is to
understand if the light Higgs and heavy Higgs scenarios could be distinguished at the LHC. From now on we drop the superscript $H$ in $\kappa_i^H$.
\begin{figure}
\centering
\mbox{\includegraphics[width=0.45\linewidth,clip=true]{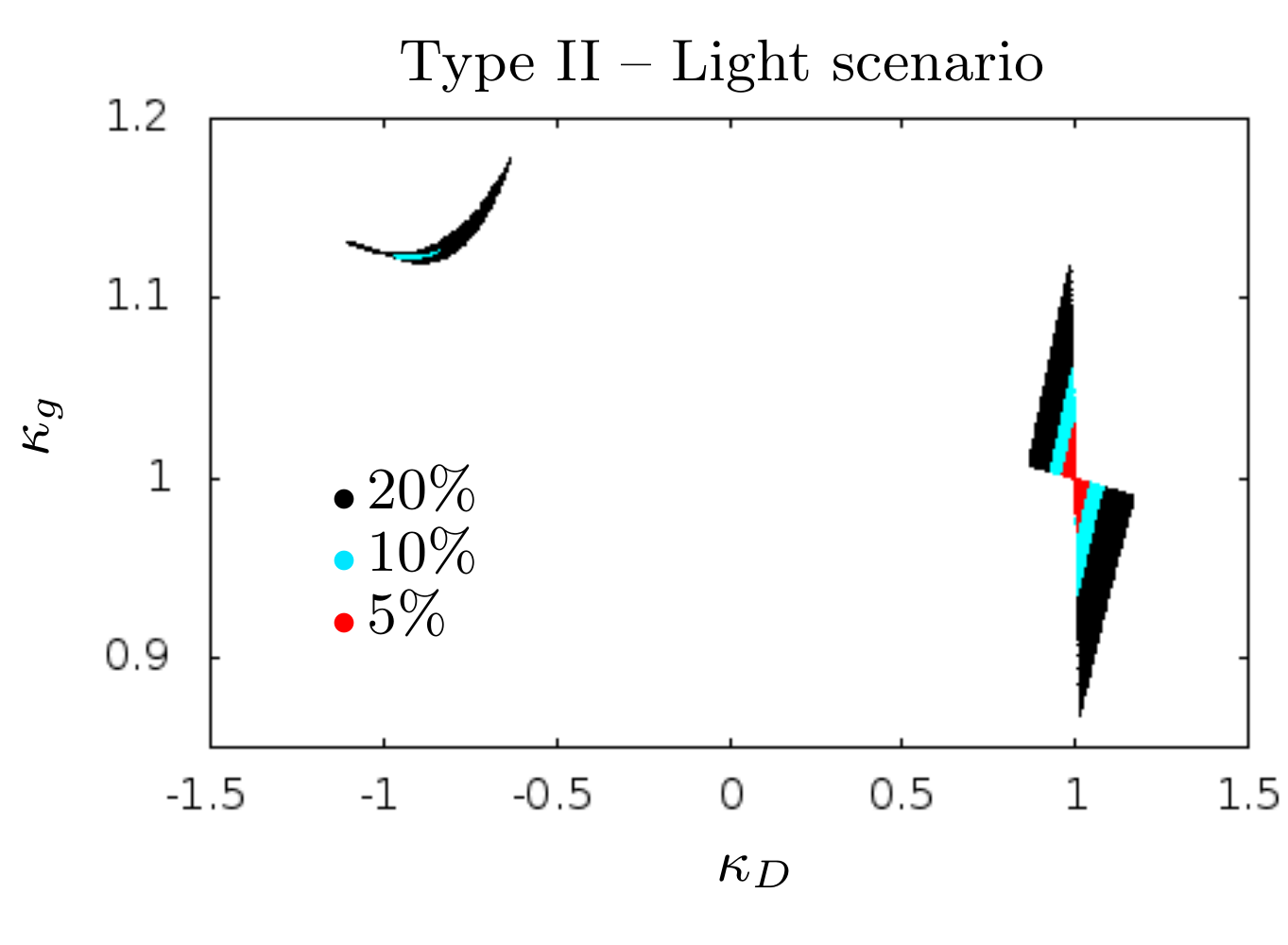}\hspace{6mm}\includegraphics[width=0.45\linewidth,clip=true]{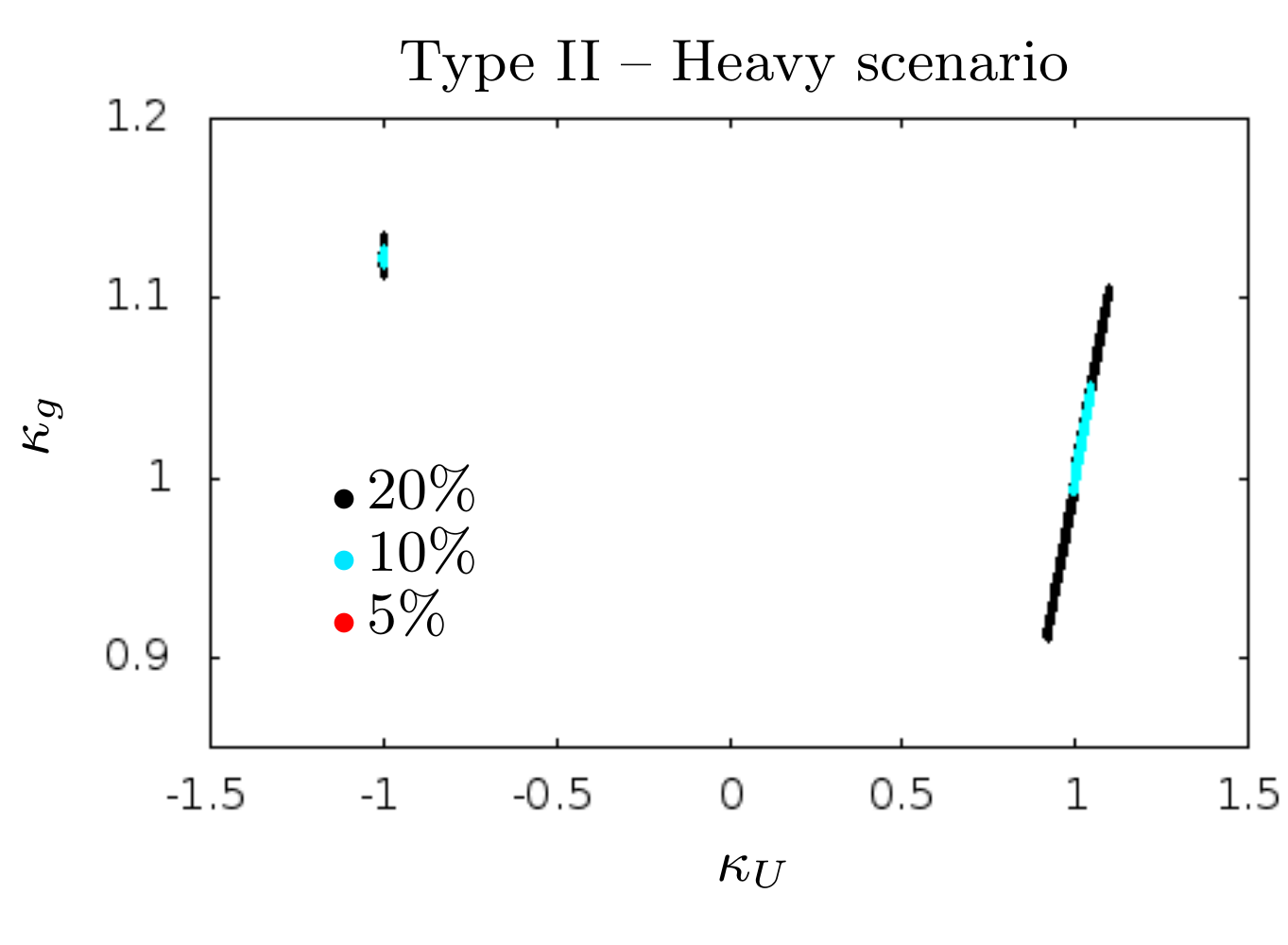}}
\mbox{\includegraphics[width=0.45\linewidth,clip=true]{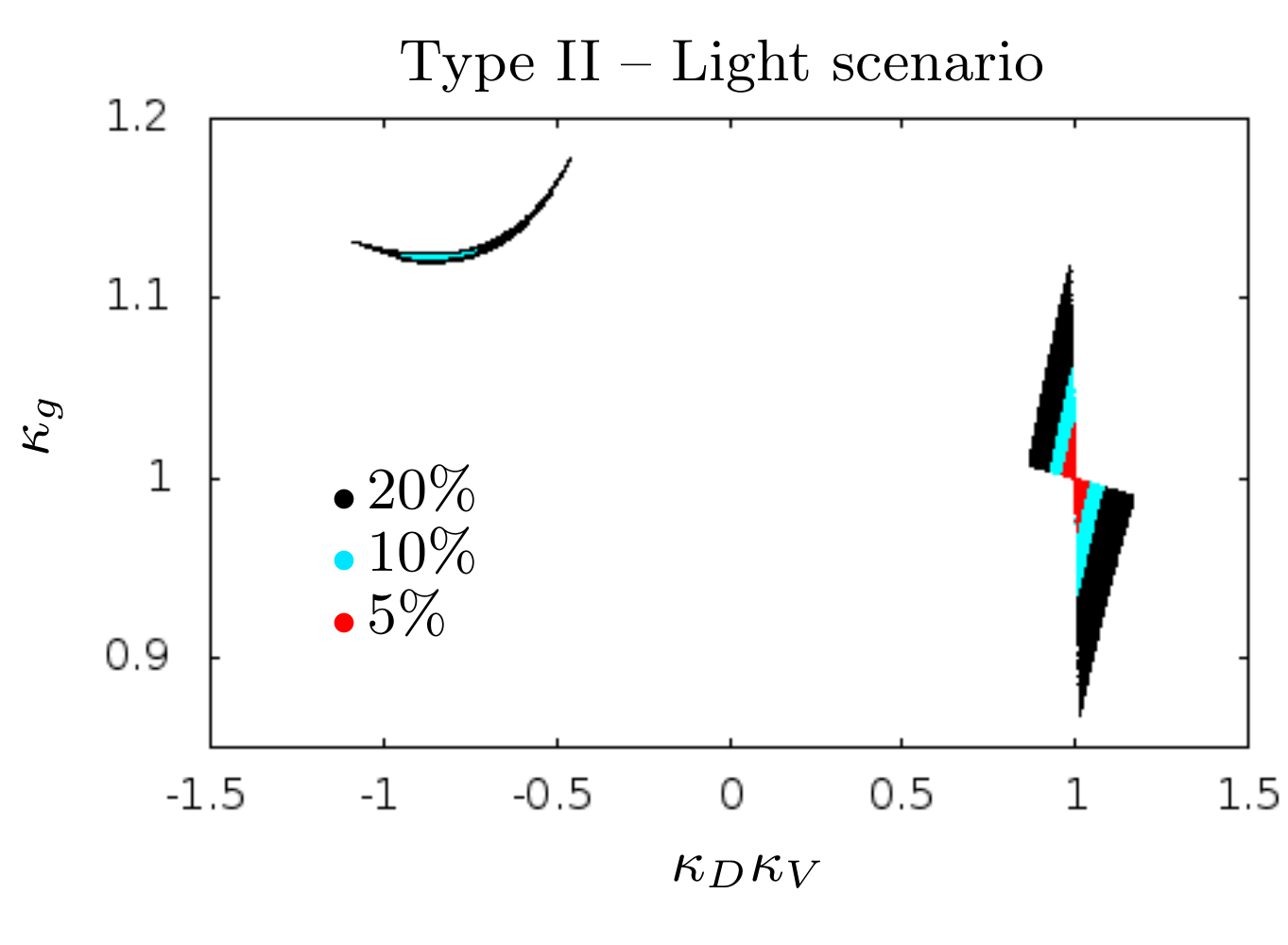}\hspace{6mm}\includegraphics[width=0.45\linewidth,clip=true]{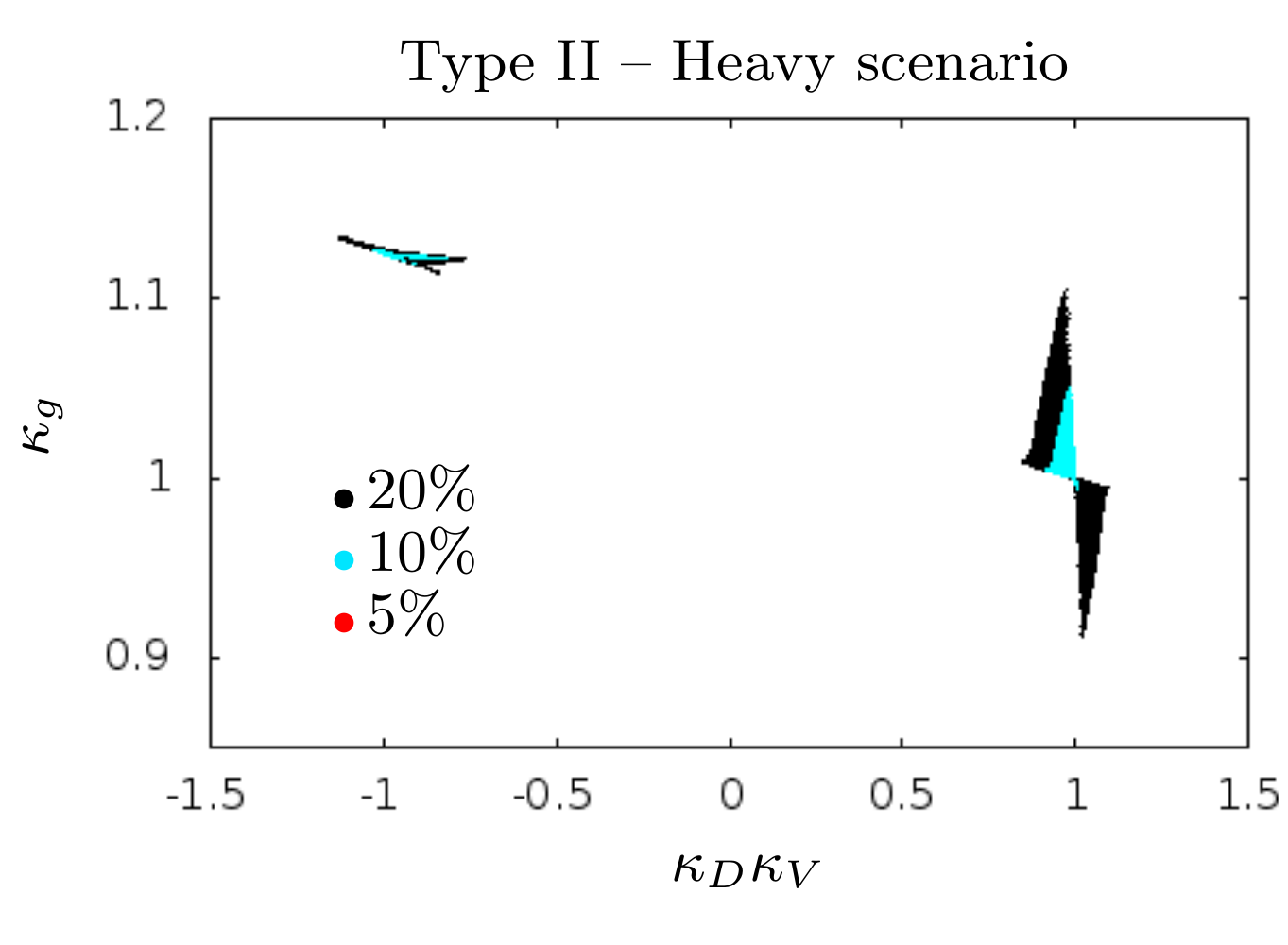}}
\caption{Predicted allowed parameter space for type II with all rates within $5$, $10$ and $20\%$
of the SM values. Left: lightest Higgs scenario; right: heaviest Higgs scenario.}
\label{fig:f10}
\end{figure}
In figure~\ref{fig:f10}  we show the predicted allowed space for type II with all rates within $5$, $10$ and $20\%$
of the SM values. In the left panel we present the light Higgs scenario while in the right panel one can see
the heavy Higgs case. Because the loop integrals are exactly the same in the two wrong sign limits  (heavy and light),
the values of $\kappa_g$ are both centred at $\approx$ $1.12$. The main difference between the two scenarios
is the shape of the allowed regions which is mainly due to the reduced size of the parameter space in the case of heavy Higgs which
implies smaller allowed regions. Hence, $\kappa_g$ can be used to distinguish between wrong and alignment scenarios
but not between the heavy and light cases.
\begin{figure}[h!]
\centering
\mbox{\includegraphics[width=0.45\linewidth,clip=true]{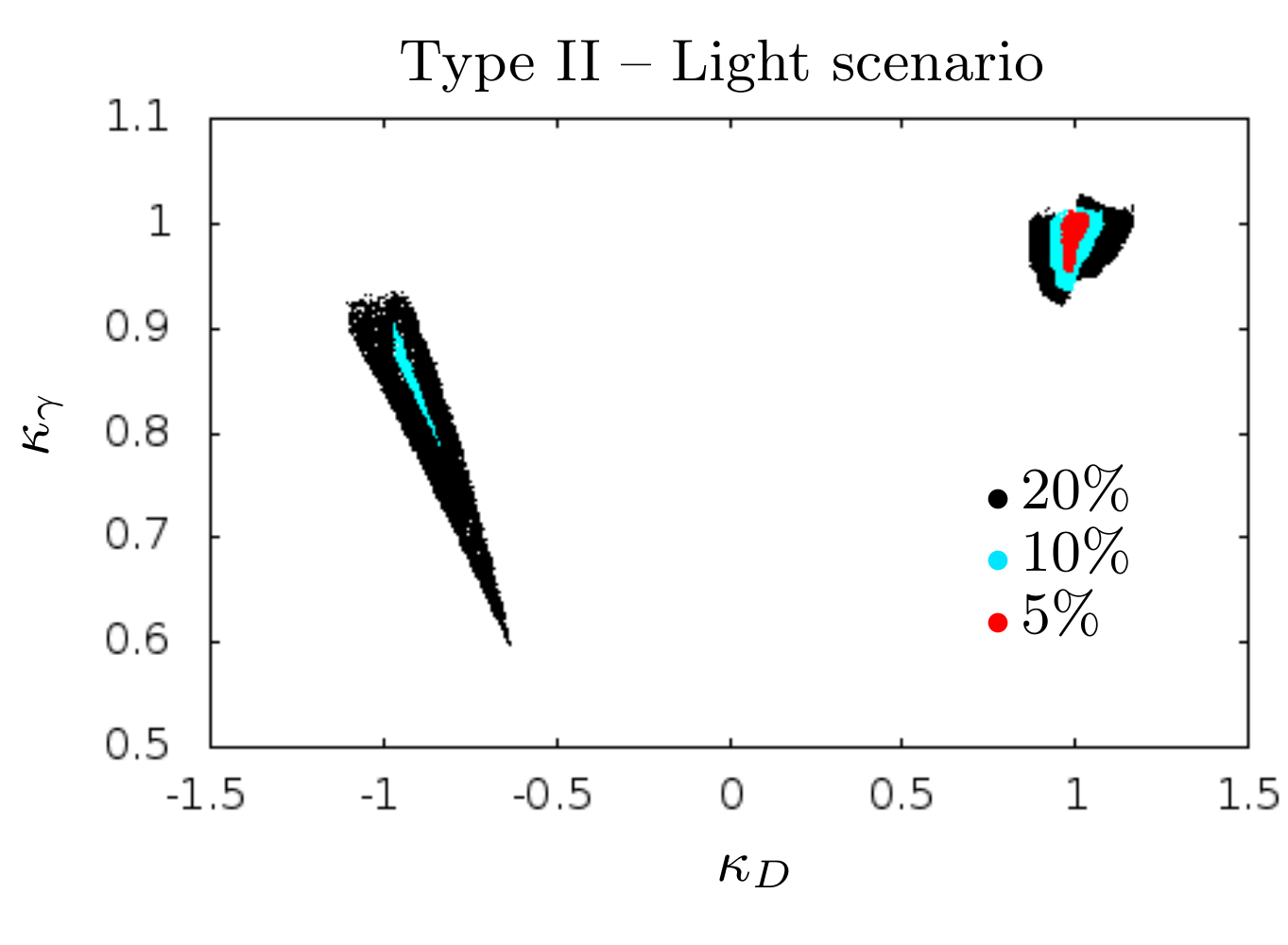}\hspace{6mm}\includegraphics[width=0.45\linewidth,clip=true]{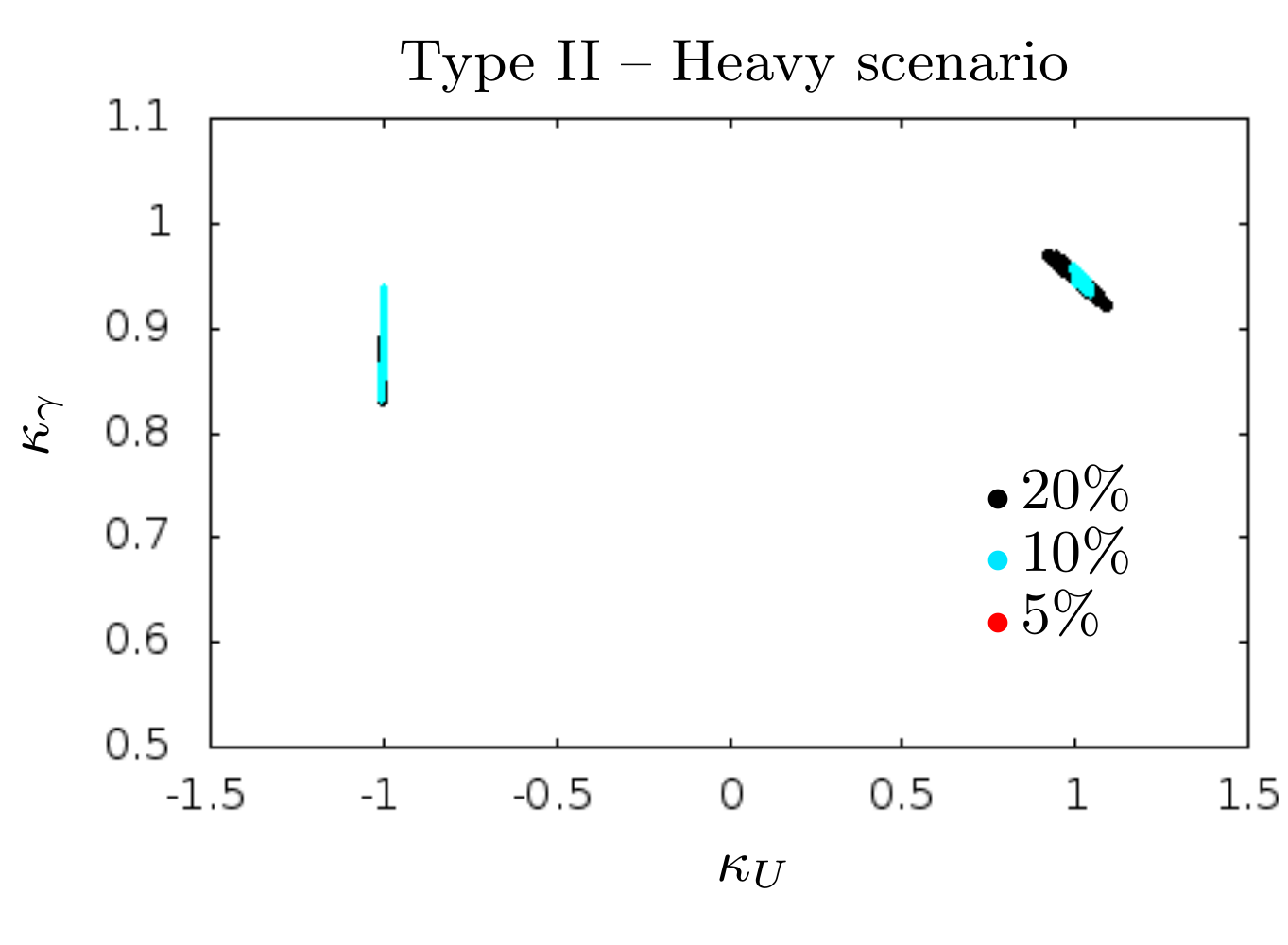}}
\mbox{\includegraphics[width=0.45\linewidth,clip=true]{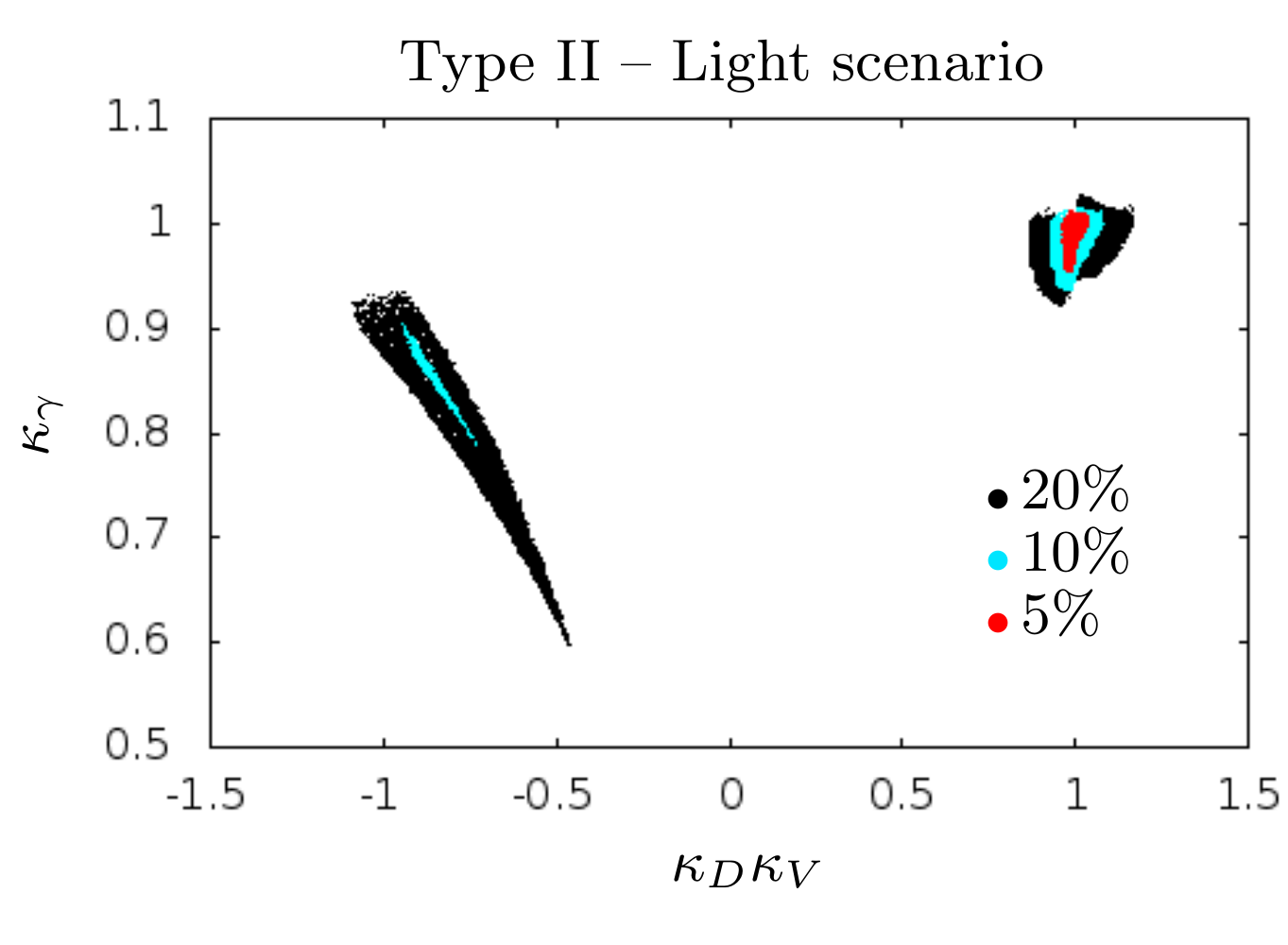}\hspace{6mm}\includegraphics[width=0.45\linewidth,clip=true]{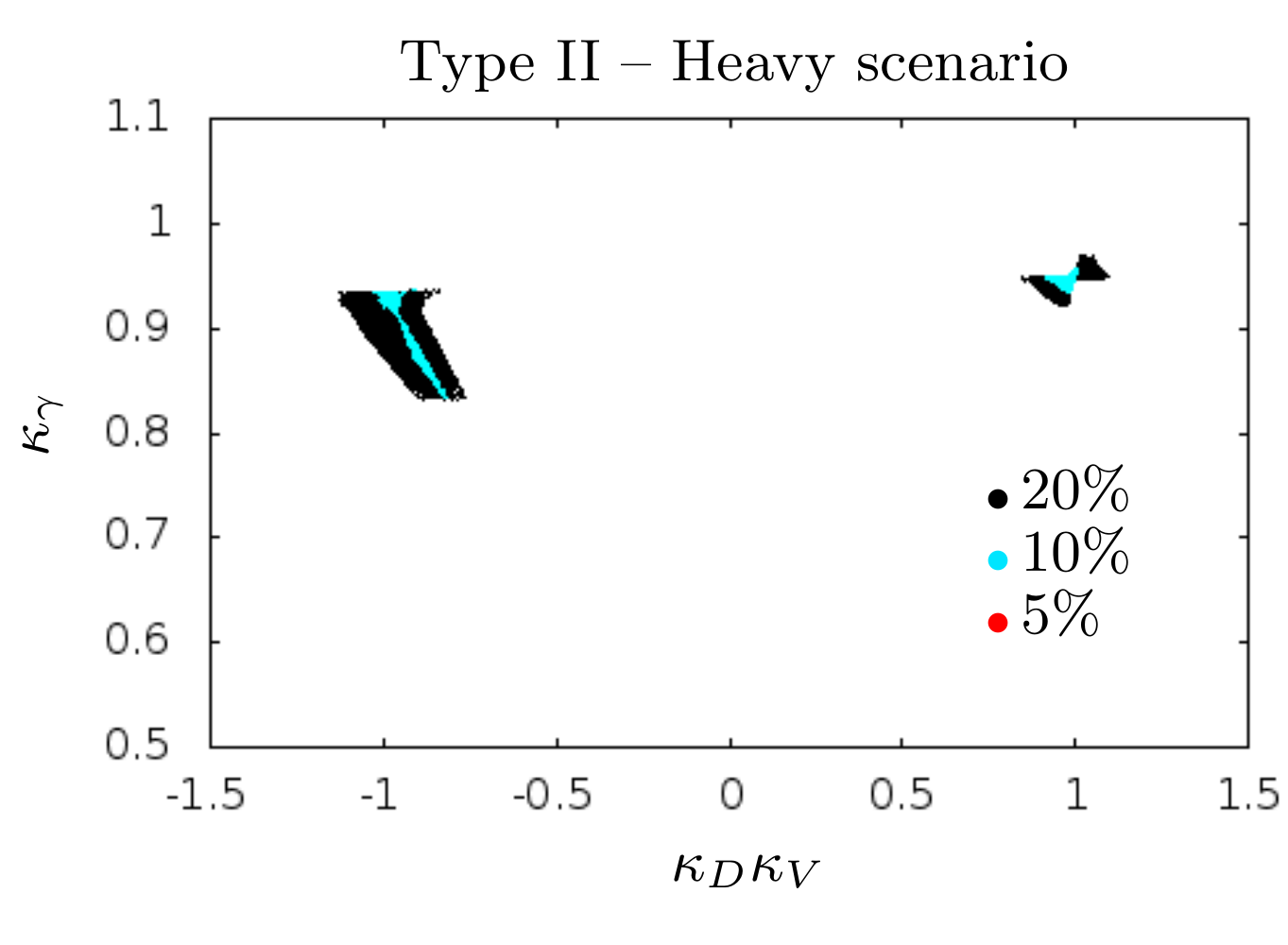}}
\caption{Comparing wrong sign scenario with $\kappa_\gamma$ as a function of $\kappa_U$ and $\kappa_D$. On the left we show the wrong sign
scenario for the lightest Higgs case and on the right for the heaviest Higgs scenario.}
\label{fig:f12}
\end{figure}
The same conclusion can be drawn from figure~\ref{fig:f12} where we compare  $\kappa_\gamma$ in the two wrong sign scenarios. 
On the left we show the light Higgs case and on the right  the heavy Higgs scenario. Clearly we see that a $5\%$ precision would allow us
to distinguish wrong sign from alignment scenarios both in the heavy and in the light Higgs case but not the two different wrong sign scenarios
from each other. We recall once again that it is the reduction of the width $\Gamma (h \to \gamma \gamma)$, which is due to the charged Higgs boson
loop contribution, that ultimately decreases $\mu_{\gamma \gamma}$ below $0.95$.

However, both figures~\ref{fig:f10} and~\ref{fig:f12}  reveal a much more interesting feature of the heavy Higgs scenario. In fact the alignment limit of the heavy
Higgs can be excluded with a measurement of the rates at $5\%$. Clearly this would not be possible for the light Higgs scenario due to the decoupling limit
of the 2HDM~\cite{Gunion:2002zf}. We now show that it is again the charged Higgs loop together with the theoretical and experimental constraints that
is responsible for a reduction in  $\mu_{\gamma \gamma}$ below $0.95$. We start by recalling that the couplings of the heavy Higgs to the charged 
Higgs bosons can be written in the form (for $\tan \beta > 1$)
\begin{equation}
g_{H H^\pm H^\mp} ^{\scriptscriptstyle {\rm Wrong Sign}}=-  \frac{2 m_{H^\pm}^2 - m_H^2}{v^2}   \, , \qquad  g_{H H^\pm H^\mp}^{\scriptscriptstyle {\rm Alignment}}    = - \, \frac{2 m_{H^\pm}^2 + m_H^2- 2 M^2}{v^2} 
\label{eq:eq10}
\end{equation}
for the wrong sign and alignment limit respectively. Now, what leads to the reduction of $\Gamma (h \to \gamma \gamma)$ in the wrong sign case is the almost constant negative value
of $v/m_{H^\pm}^2 \, g_{H H^\pm H^\mp} ^{\scriptscriptstyle {\rm Wrong Sign}}$. If we compare the two expressions in~\eqref{eq:eq10}, the difference is that the term
$-m_H^2$ is replaced by $m_H^2 - 2 M^2$. Hence, in order to show that a similar situation occurs in the alignment case, we have to prove that $|M|$ is of the order of $m_H$ 
and therefore small when compared to the charged Higgs mass. This is indeed the case -
when forcing the potential to be bounded from below, the condition $\lambda_1 > 0$, which in the alignment limit can be rewritten in the form
\begin{equation}
M^2 < m_H^2+m_h^2/\tan^2\beta
\label{eq:eq11}
\end{equation}
clearly shows that $m_H$ is indeed of the order of $|M|$ and $M^2 \ll m_{H^\pm}^2$ because, as discussed before, $m_{H^\pm} > 340$ GeV.
However, one should note that when removing the global minimum condition $M^2$ could also be negative. Therefore, it is in fact a combination of the theoretical conditions that leads to values of $M^2$ small enough to always keep $\mu_{\gamma \gamma}$ below $0.95$, even for $M^2 <0$.
Had we removed all theoretical conditions, points with $\mu_{\gamma \gamma}$ above $0.95$ would be allowed.
\begin{figure}[h!]
\centering
\mbox{\hspace{-8mm}\includegraphics[width=0.45\linewidth,clip=true]{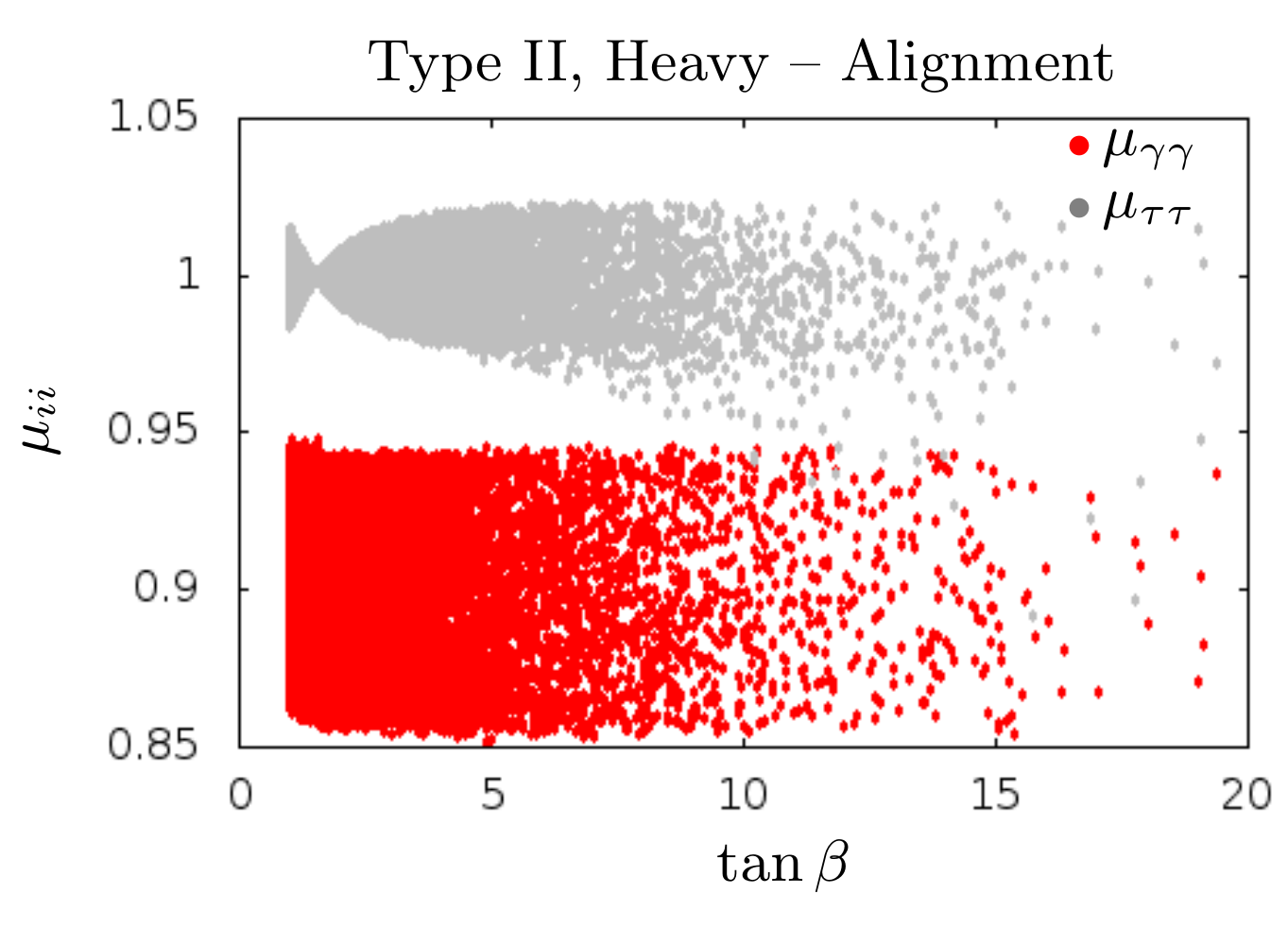}\hspace{20mm}\includegraphics[width=0.35\linewidth,clip=true]{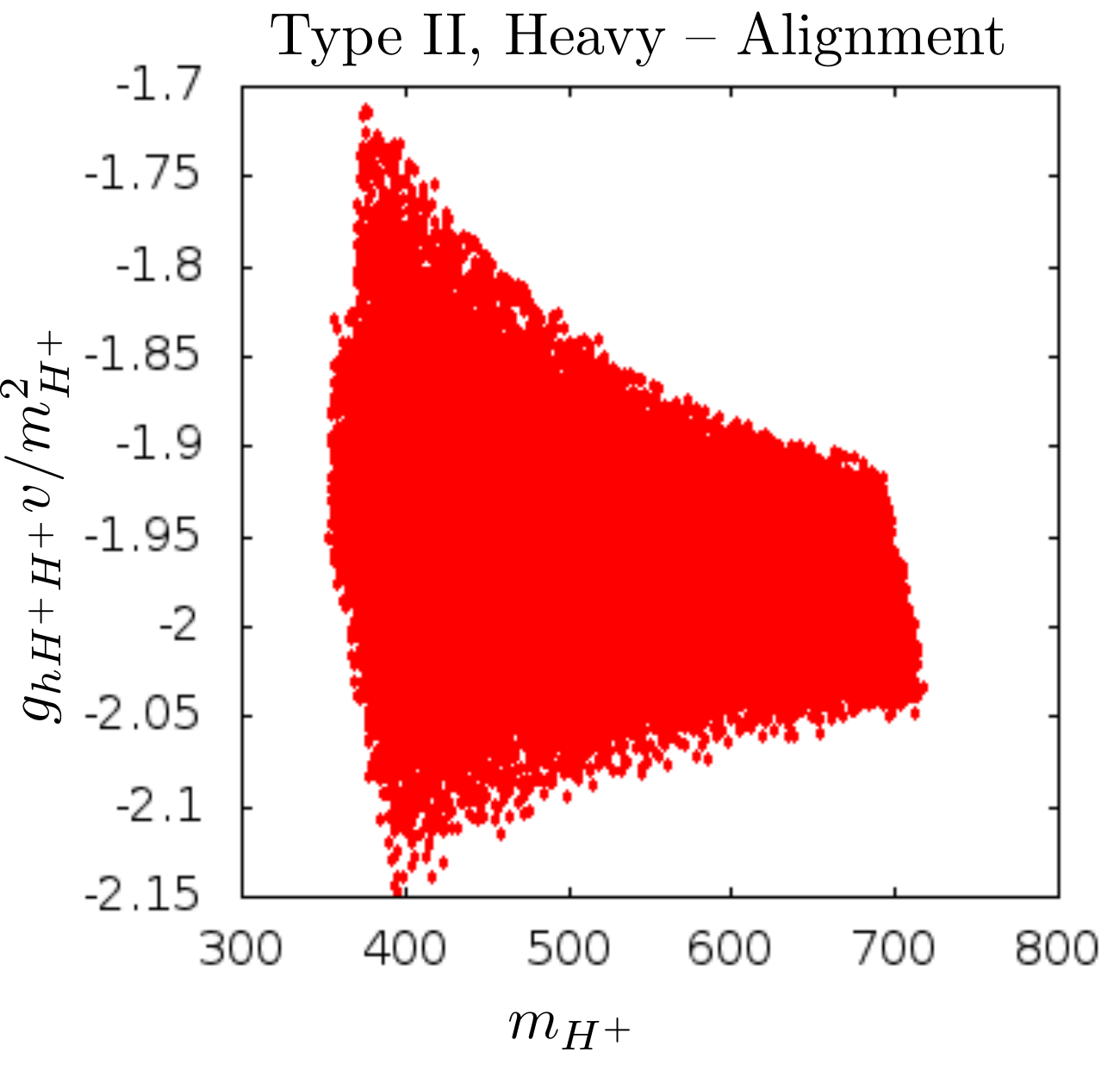}}
\caption{Left: $\mu_{\tau \tau}$ and $\mu_{\gamma \gamma}$ as a function of $\tan \beta$ in the alignment limit for the heavy scenario. 
Right: $v/m_{H^\pm}^2 \, g_{H H^\pm H^\mp} ^{\scriptscriptstyle {\rm Alignment}}$ as a function of the charged Higgs mass.}
\label{fig:f122}
\end{figure}
In the left panel of figure~\ref{fig:f122} we present the rates $\mu_{\tau \tau}$ and $\mu_{\gamma \gamma}$ as a function of $\tan \beta$ in the alignment limit for the heavy scenario
with $\mu_{VV}$ measured at $5 \%$. The decrease in $\mu_{\gamma \gamma}$ is clearly seen and explained by the plot on the right where 
$v/m_{H^\pm}^2 \, g_{H H^\pm H^\mp} ^{\scriptscriptstyle {\rm Alignment}}$ as a function of the charged Higgs mass is shown. As previously discussed
this coupling is always negative and almost constant which leads to a decrease in the Higgs to two photons width.

This result for the heavy scenario in the alignment limit is extremely interesting as it clearly shows the non-decoupling nature of the heavy scenario. 
In the next section we discuss the type I model that has no wrong sign limit for $\tan \beta > 1$. 

\section{The \textit{symmetric limit}}
\label{sec:sym}

In figure~\ref{fig:LHC8_2} we saw that in the heavy scenario there is a region analogous to the $\sin(\beta+\alpha)=1$ region of the light Higgs scenario. Such region is now centred on the line $\cos(\beta + \alpha) =1$. 
However, regardless of the scenario we are considering, the limits $\sin(\beta+\alpha)=1$ (light case) and $\cos(\beta+\alpha)=1$ (heavy case), are not \textit{a priori} wrong sign scenarios. That is,
if we consider the type I 2HDM, none of the Higgs couplings to the remaining SM particles changes sign relative to the SM one (modulo a global sign change in the heavy scenario). Nevertheless, the 
shift $\alpha \to - \alpha$ still changes the value of $\kappa_V$, which as previously seen is given by
\begin{equation}
\kappa_V^{h\, (H)}=(-)\frac{\tan^2 \beta -1}{\tan^2 \beta +1} 
\end{equation}
and we recall that in the heavy scenario there is a global minus sign change in the Higgs coupling  when $\cos(\beta + \alpha) =1$.
We call this limit \textit{Symmetric Limit} to distinguish it from the wrong sign limit because there is no sign change in the couplings
even though the shift $\alpha \to - \alpha$ occurs. 
\begin{figure}[h!]
\centering
\mbox{\includegraphics[width=0.45\linewidth,clip=true]{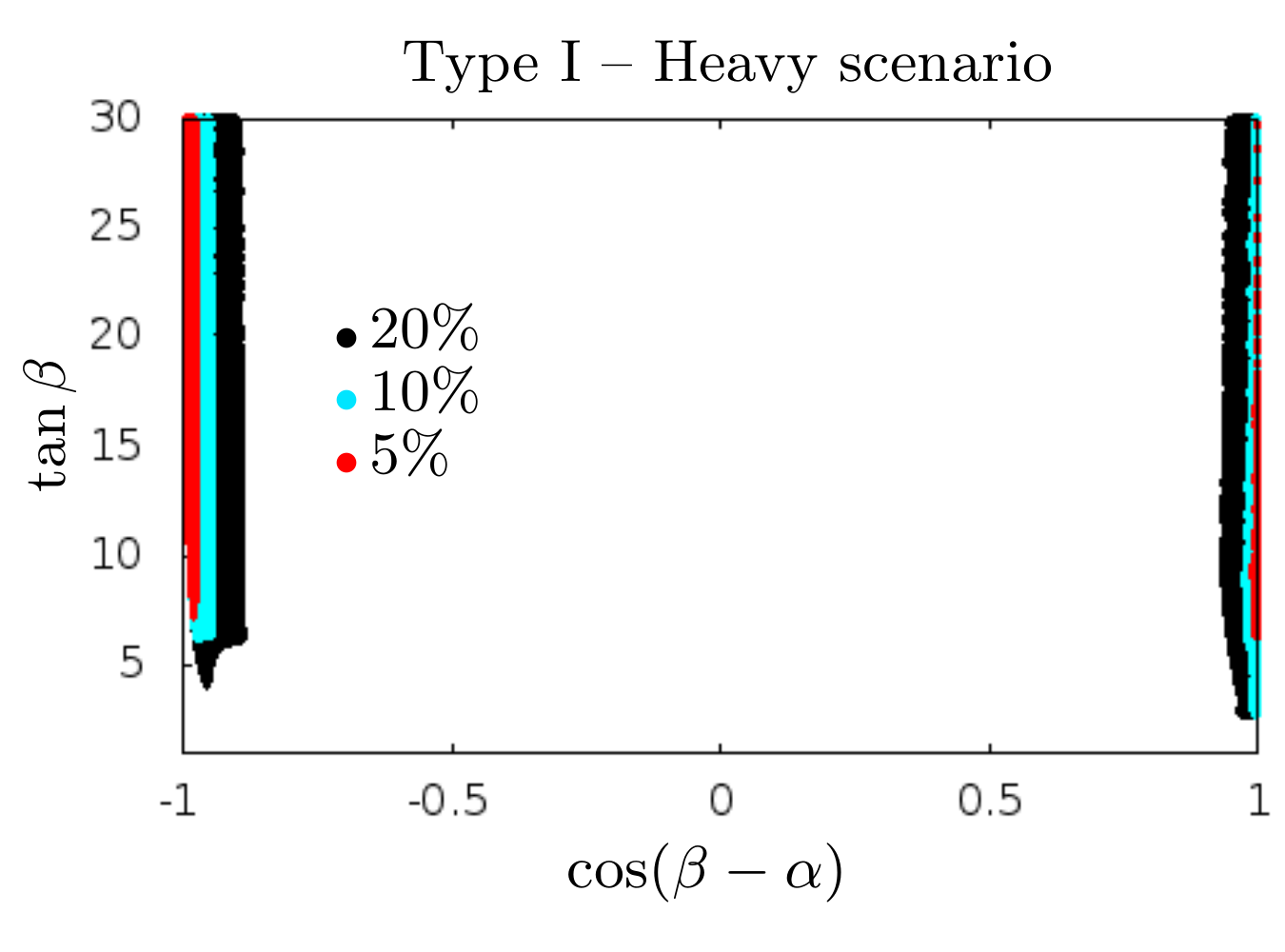}\hspace{6mm}\includegraphics[width=0.45\linewidth,clip=true]{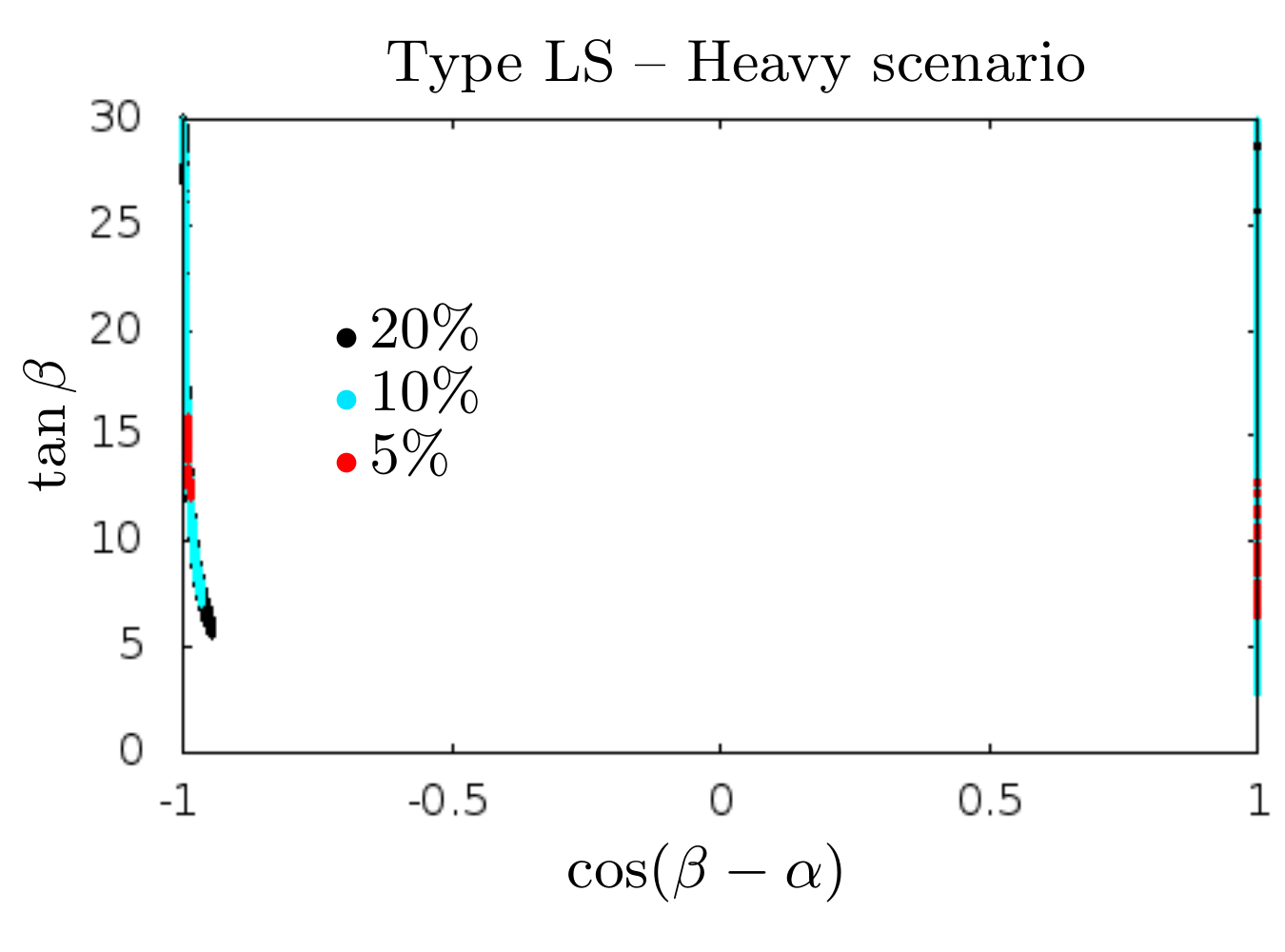}}
\caption{Predicted allowed parameter space with all rates measured at $5$ (red), $10$ (blue) and $20\%$ (black)
where the regions on the right correspond to the alignment limit and the ones on the left correspond to the \textit{symmetric} limit for type I (left panel)
and type LS (right panel).}
\label{fig:f13}
\end{figure}
For the case of type LS the only coupling that changes sign relative to the SM is $\kappa_L$ which plays no role in the discussion
given the predicted accuracy of future rate measurements at the LHC. For the remainder of this section we will focus on the heavy case (the discussion for the light case is similar). 
Since we are taking $\tan \beta > 1$, it is true for all $i,j=F,V$ that $\kappa_i^H \, \kappa_j^H >0$ ($i,j$ represent either a fermion or a massive gauge boson). Hence, for type I, 
and $\tan \beta > 1$ not only there are no sign changes but we recover exactly the alignment limit when $\tan \beta \to +\infty$.

In figure~\ref{fig:f13} we show the predicted allowed parameter space 
with all rates measured at $5$ (red), $10$ (blue) and $20\%$ (black) where the regions centred around $1$ correspond to the alignment 
limit and the ones around $-1$ correspond to the \textit{symmetric limit}. Type I is shown on the left panel while type LS
is in the right panel. 
There are two points worth discussing. First it is clear that, as the precision increases, the lower bound on $\tan \beta$
grows from about $4$ at $20\%$, to $6$ at $10\%$ and finally to $8$ at $5\%$. In type I this behaviour is also
present but it is not so striking. The second point is that even for the alignment limit there seems to appear again
some kind of non-decoupling effect that excludes the low $\tan \beta$ region.

\begin{figure}[h!]
\centering
\includegraphics[width=0.35\linewidth]{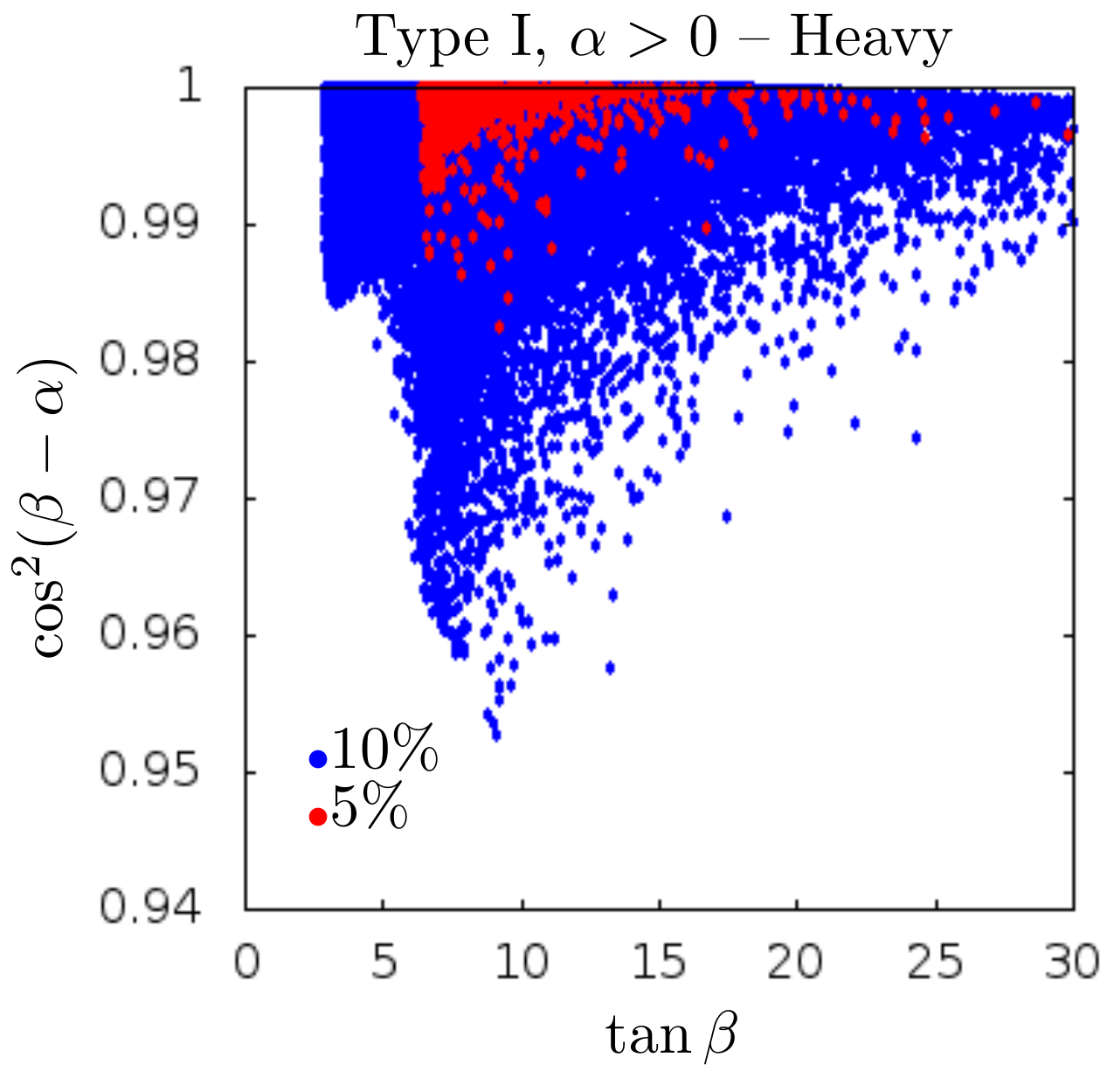}
\hspace{0.1\linewidth}
\includegraphics[width=0.35\linewidth]{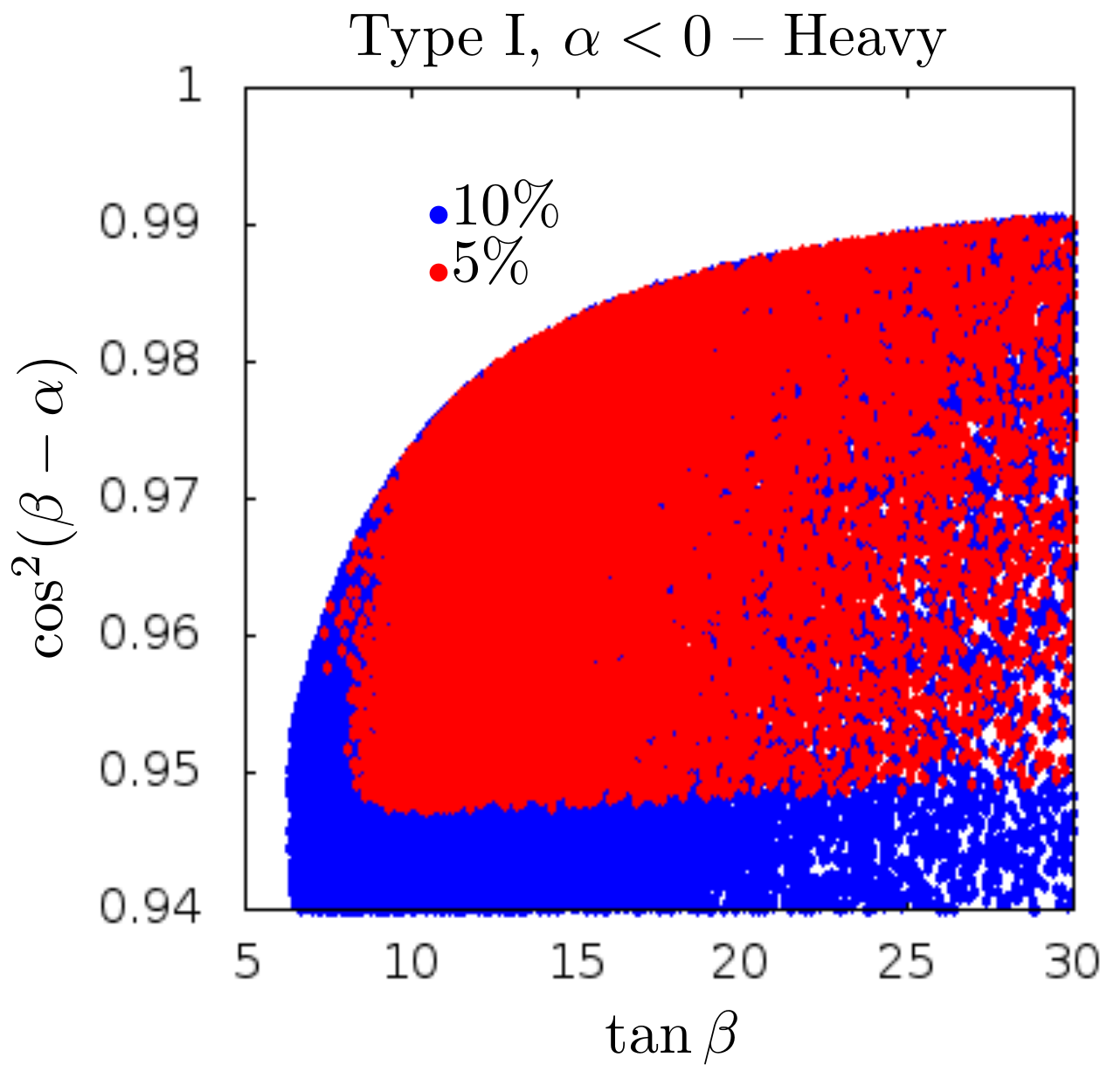}
\caption{Predicted allowed parameter space with all rates measured at $10\%$ (blue) and $5\%$ (red). Left: alignment limit; right: \textit{symmetric} limit.}
\label{fig:f9}
\end{figure}

Let us start with the first point. The \textit{symmetric limit} is clearly seen in type LS for low $\tan \beta$. Now one may ask if the \textit{symmetric limit} could be distinguished
from the alignment limit, for finite $\tan \beta$, given enough precision in type I.  In figure~\ref{fig:f9} we show the predicted allowed parameter 
space with all rates within $10\%$ (blue) and $5\%$ (red) of the SM predictions for the alignment limit - all points with $\alpha > 0$ (left panel) and for the \textit{symmetric limit} - all points with $\alpha < 0$
(right panel). Noting that $\mu_{VV} \approx \cos^ 2 (\beta - \alpha)$ it is going to be extremely hard to distinguish the two limits except for the very low $\tan \beta$ region
where they could both be excluded. In fact, this bring us to the second point. Why are values of low $\tan \beta$ excluded in the alignment limit when all rates are measured at $5\%$?
\begin{figure}
\centering
\mbox{\includegraphics[width=0.435\linewidth,clip=true]{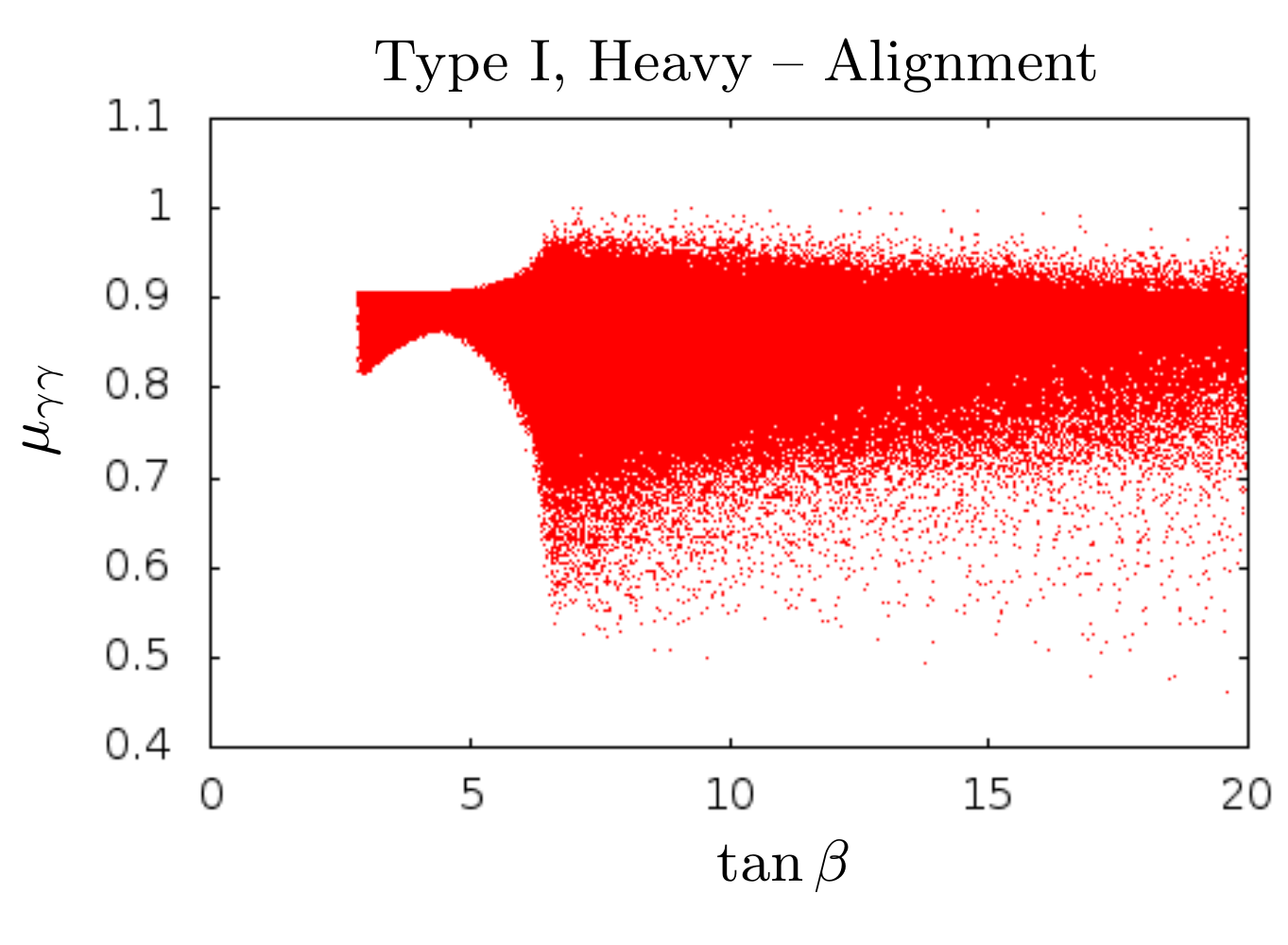}\hspace{6mm}\includegraphics[width=0.45\linewidth,clip=true]{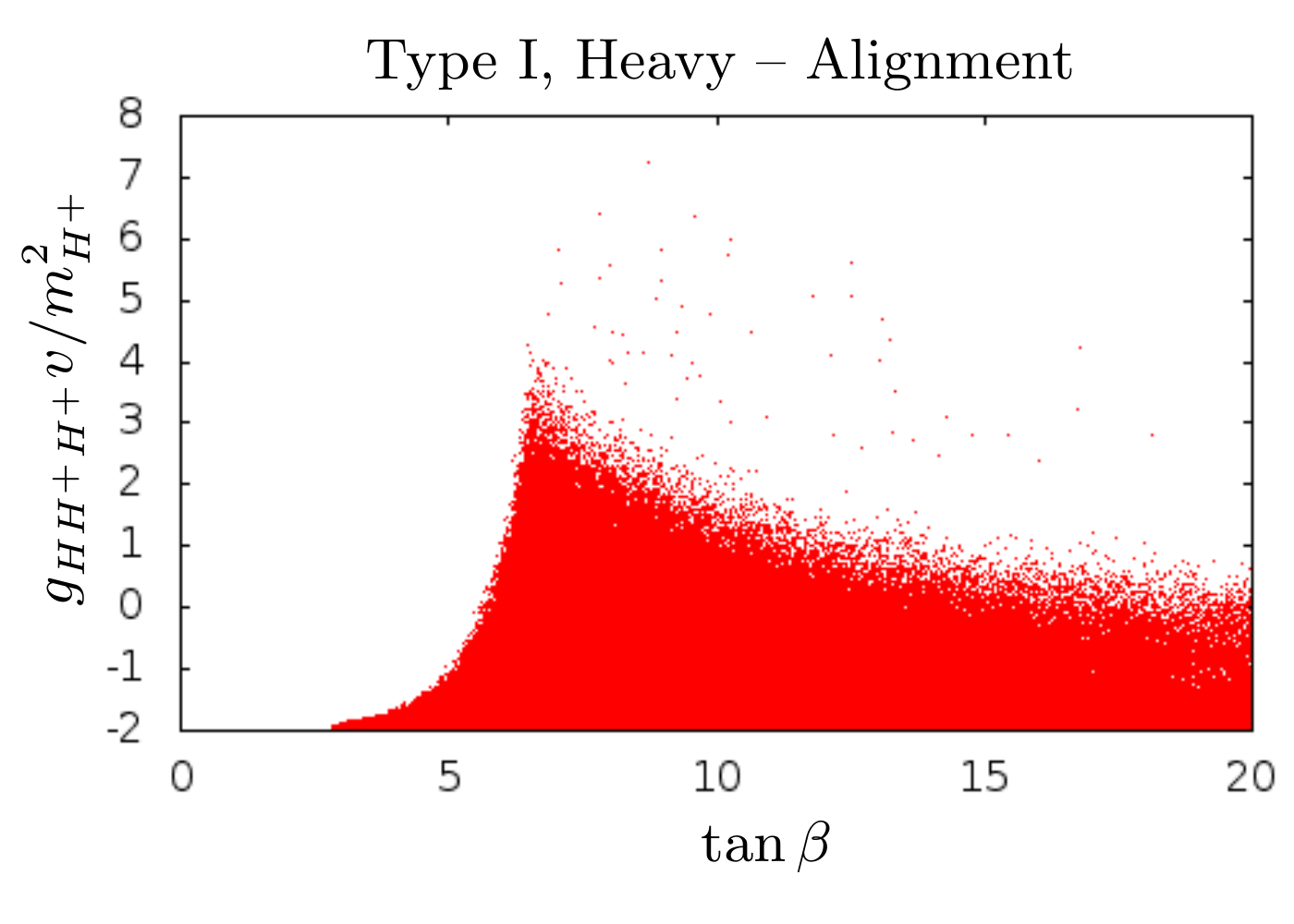}}
\caption{In the left panel we show $\mu_{\gamma \gamma}$ as a function of $\tan \beta$ while in the right panel we see $ g_{H H^\pm H^\mp} ^{\scriptscriptstyle {\rm Alignment}} \, v/m_{H^\pm}^2$ as a function
of $\tan \beta$.}
\label{fig:f131}
\end{figure}
The answer again lies in the behaviour of the $H$ coupling to the charged Higgs bosons together with the remaining theoretical and experimental constraints. In the left
panel of figure~\ref{fig:f131} we present $\mu_{\gamma \gamma}$ as a function of $\tan \beta$ with $\mu_{VV}$ measured at $5 \%$. The region where $\Gamma (h \to \gamma \gamma)$ is below $0.95$
is easily identifiable in the plot for the low $\tan \beta$ region. In the right panel we present the plot   $ g_{H H^\pm H^\mp} ^{\scriptscriptstyle {\rm Alignment}} \, v/m_{H^\pm}^2$ as a function
of $\tan \beta$. The low values of $\tan \beta$ correspond to negative values of the couplings and therefore to a decrease in the two photons width.
 As suggested by the shape of this plot, where one sees a sharp line cutting the low $\tan\beta$ region, this is mainly due to a combination of the theoretical constraints imposed on the model.
 Finally, we stress that $\kappa_g$ is the same in the symmetric and in the alignment limits while $\kappa_\gamma$ shows a negligible difference in the two limits.

\section{Conclusions}
\label{sec:conc}
We have discussed all the different possibilities for having a wrong sign limit in 2HDMs. A wrong sign limit is defined as a scenario where: (a) one or more Higgs couplings to SM particles change sign relative
to the corresponding SM couplings; (b) this difference has physical meaning, that is, it could in principle be measured experimentally.
Hence, each scenario is defined by a condition $\kappa_i \, \kappa_j < 0$. After listing all possible wrong sign scenario cases when the lightest Higgs is the alignment one, we have also discussed the case where it is the  heaviest CP-even Higgs.

We have shown that with all the 7/8 TeV data analysed so far, the wrong sign scenarios for both type II and type F
are still allowed even at $1\sigma$. This is true not only for the lightest Higgs case but also for the heaviest
Higgs scenario in the regime $\tan \beta > 1$. The light/heavy Higgs scenarios are very similar except
in the range of the parameter scan and in the non-decoupling nature of the heavy scenario.
Although we have concluded that each of the wrong sign scenarios can be distinguished from the respective alignment limit, we have also concluded that it is hard to differentiate between the two wrong sign scenarios (light or heavy). 

A possibility not previously discussed was the wrong sign scenarios for the case where $\tan \beta <1$ which is possible for all Yukawa types. Taking into account all constraints except the ones from $B$--$\bar{B}$ mixing data, we have shown that the LHC does allow this particular
wrong sign limit for types II and F at $2\sigma$. We have also shown that $B$--$\bar{B}$ mixing data at 95 $\%$ C.L. completely excludes this region. It is however possible that B-physics constraints taken at $3\sigma$ or $4\sigma$ would not exclude
this scenario. In the end, the next LHC run at 13 TeV will be able to definitely exclude this region.

We have then discussed the non-decoupling nature of the heavy Higgs case. In fact, we have shown that
due to a non-decoupling effect in the charged Higgs coupling to the heavy CP-even Higgs boson, there
is a reduction in  $\Gamma (h \to \gamma \gamma)$ in the alignment limit, similarly to what happens for
the wrong sign limit. We conclude that a measurement of $\mu_{\gamma \gamma}$ with a $5 \%$
precision would exclude the alignment limit scenario in the heavy type II case.

Finally, we have also discussed the \textit{symmetric} limit in the context of the heavy Higgs scenario. 
This is a limit that occurs in type I (also type LS if we disregard $\kappa_L$ which plays no major role in 
the LHC results) and corresponds to the flip $\alpha \to -\alpha$, and consequently to $\kappa_U^H = \kappa_D^H = \kappa_L^H = -1$
while $\kappa_V^H = (1 - \tan^2 \beta)/(1 + \tan^2 \beta)$. Although  $\kappa_V^H \to  - 1$ when $\tan \beta \to \infty$
this case could be in principle distinguishable from $-1$ for finite values of $\tan \beta$ (we again recall that 
in this case $-1$ corresponds to the SM $\kappa_V$ because there is a global sign change in the limit $\cos (\beta + \alpha) =1$).
We have shown that although possible, it will be extremely hard to differentiate between the  \textit{symmetric}  and the alignment limit.

\begin{acknowledgments}
We thank Maria Krawczyk and St\'ephane Monteil for discussions. 
RS also thanks Jo\~ao P.~Silva for discussions. 
P.M.F., R.G. and R.S. are supported in part by the Portuguese
\textit{Funda\c{c}\~{a}o para a Ci\^{e}ncia e Tecnologia} (FCT)
under contract PTDC/FIS/117951/2010 and PEst-OE/FIS/UI0618/2011.
R.G. is also supported by a FCT Grant SFRH/BPD/47348/2008.
M.S. is supported by a FCT Grant SFRH/BPD/69971/2010.
\end{acknowledgments}

\end{document}